# Spectroscopic experimental and theoretical study of Uranyl(VI) in an aqueous system – Molecular modelling meets environmental protection

Jakub Višňák[1,2,3], Lukáš Sobek[4], Nils Hoth[5]

[1] *Dept. of Nuclear Chemistry, FNSPE, Czech Technical University, Břehová 7, 115 19 Prague 1, Czech Rep.*
[2] *Dept. of Chemical Physics and Optics, Faculty of Mathematics and Physics, Charles Univ., Ke Karlovu 3, 121 16 Prague 2, Czech Rep.*
[3] *J. Heyrovský Institute of Physical Chemistry, Dolejškova 2155/3, 182 23 Prague 8, Czech Rep.*
[4] *Czech Physical Society, Na Slovance 1999/2, 182 21 Praha 8, Czech Republic*
[5] *Dept. of Mining and Special Constructions, Zeunerstr. 1A, 09599 Freiberg, Germany*

## Abstract

Time-Resolved Laser-Induced Fluorescence Spectroscopy (TRLFS) and cryo-TRLFS are sensitive tools for *in situ* speciation of low-level uranium in aqueous samples, including natural waters. To tailor, often ill-conditioned (i.e., leading to ambiguous interpretations), multi-linear analysis, first-principles based theoretical computation (Molecular Dynamics and Quantum Chemistry) of luminescence spectra would be beneficial. We present a methodology to simulate TRLFS and cryo-TRLFS spectra and present vibrationally resolved luminescence spectra for aquo complex $[UO_2(H_2O)_5]^{2+}$(aq). Comparison to experimental data, interpretation of spectra in the terms of a minimal non-redundant set of spectroscopic parameters (peak spacing $\omega$, the 0'→0 peak position $T_{00}$, envelope shape parameter $\Delta R$, average peak width $\sigma$ and luminescence life-times) and shifts due to ligand coordination are discussed. The minimum theory-experiment deviation in $T_{00}$ (exp. ~ 20 485 cm$^{-1}$) has been reported for SORECP/TD-DFT with LB$\alpha$ functional – 20 cm$^{-1}$, 90 cm$^{-1}$ and 100 cm$^{-1}$ for different models, similar level of agreement has been met for $\omega_{gs}$. The RMS along configuration space sampling CMD trajectory for $T_{00}$ corresponds well to $\sigma$. Preliminary predictivity study for a small set of uranyl complex species and comparison of pseudo-potential (SORECP) and all-electron results for $[UO_2(H_2O)_5]^{2+}$(g) are appended.

## Introduction

Chemical analysis of uranium containing aqueous solutions is of importance for environmental protection (monitoring, geochemical modelling studies and remediation), mining prospection, nuclear non-proliferation monitoring, for experimental speciation studies[1] and other applications. Not only total uranium concentration is of significance, but also the partitioning of uranium into various chemical forms (i.e., speciation). Uranium in different complex species has different chemical-physical properties predetermining different toxicities, mobilities and would require different means of remediation or extraction [1].

The Time-Resolved Laser-Induced Fluorescence Spectroscopy (TRLFS) [2] is "a unique tool for a direct low-level uranium speciation in aqueous samples" (to cite [3] exactly), including natural waters [4], [5]. Under oxidative conditions, dissolved uranium is present as uranyl(VI) cation, $UO_2^{2+}$. In neutral and alkalic *pH* natural water samples, carbonate ($CO_3^{2-}$) and bicarbonate ($HCO_3^{-}$) uranyl complex species often exhibit insufficient luminescence under ambient conditions and cryostatic measurements are therefore necessary (opposed e.g. to the $UO_2^{2+}$ - $XO_4^{2-}$ - $H_2O$ (X = S, Se) systems in acidic conditions) [6].

Sample spectra decomposition into individual chemical species contributions is [in the case of uranyl(VI) complexes] complicated by high similarity between individual components in spectral domain. The luminescence spectra of various different complex species have very similar shapes [5], [8]-[10], as the electronic transition responsible for all peaks in spectra is dominantly localized on uranyl central group and ligands contribute only by "fine-tuning" spectral parameters – 1. the main peak position (corresponding to the phonon-less excitation energy $T_{00}$), 2. cold- and hot-band peak spacing on wave-number scale (corresponding to the effective symmetric stretching vibrational frequency of uranyl group in the ground and excited electronic states, $\omega_{gs}$ and $\omega_{es}$, respectively) and 3. shape of whole spectrum envelope (corresponding to the effective shift between equilibria of above mentioned vibrational mode, $\Delta R$). For example, $T_{00}$ or $\omega_{gs}$ peak position parameters can differ by 20 $cm^{-1}$ (case of $UO_2^{2+}$ - $CO_3^{2-}$ - $Ca^{2+}$/$Mg^{2+}$ system, [5]), while peak-widths are around 200 $cm^{-1}$ [5]. In the time-domain, individual component spectra differ more significantly, but measurements corresponding to longer delays between laser-pulse-excitation and luminescence signal ICCD camera accumulation has worse signal : noise ratio.

Spectral decomposition methods should necessarily exploit information from both spectral and temporal domain simultaneously and, when reasonable, analyze data from more samples together. But even application of multi-linear methods faces problems steaming from ill-conditioning due to almost collinear spectra (such as multimodal and/or flat $\chi^2$ surfaces and corresponding ambiguity of possible interpretations). While spectral decomposition seems to be a purely mathematical problem (with solution quality strictly limited by signal : noise ratio and individual component similarities) an assignment of spectral components to particular chemical formulae is not.

For an experimental batch-of-artificial-solution speciation studies ("BoAS", realized, e.g., as spectroscopic measurements of a batch of artificial samples with varying metal : ligand (M/L) total concentration ratio $c_M$ : $c_L$) the assignment is usually clear as there is a speciation model – set of individual species concentrations $C_{mi}$ as a function $C_{mi} = C_m(\beta;\, c_M,\, c_{L,i})$, with $\beta$ being set of stability

[1] analysis of a batch of solutions differing by metal : ligand concentration ratio for 1 : 1, 1 : 2, … 1 : *n* complex stability constant determination, [5], [7]-[10])

constants as function parameters with values optimized in the $\chi^2$ minimization in process of multi-linear data analysis.

For set of natural samples, aforementioned routine can be approached, e.g. with geochemical modelling program usage (capable of reliable speciation calculation for a complex chemical system), but is limited by the need to independently experimentally determine total elemental concentrations in studied sample (and its other chemical-physical characteristic). On the top of that, geochemical database might not cover all possibly present species and might contain inaccurate or erroneous stability constants and ionic strength influence parameters. An important side note is that different complex species often greatly differ in „luminescence per unit concentration" – i.e., the total luminescence assigned to a given component is proportional to its concentration, but the proportionality constant (let us denote it $\mu_m$) is different for a different species (indexed *m*). The aforementioned problem is avoided in BoAS arrangement or in general in presence of a large number of samples with smoothly enough varying concentrations of a few studied species. That is in strict contrast to natural water analysis, with many species in few samples – in that case individual spectral component assignment to chemical species is often unclear and every possible independent information on studied complex species and their individual spectra can be crucial.

Possible remedies for assignment problem described in previous paragraph include

1. Large set of BoAS for all binary M : $L_1$, ternary, M : $L_1$ : $L_2$, … speciation studies determining $\mu_m$, individual component spectra and speciation constants completing and/or correcting geochemical databases.

2. TRLFS spectra library (with inputs from BoAS speciation studies).

3. Accurate first-principles based simulation of individual component spectra and/or speciation parameters.

The first two points (1., 2.) applicability is limited by the fact that luminescence life-times ($\tau_m$) but also the individual component spectra are often environment-dependent („matrix effect"). For life-times, matrix effect is easily seen through different quencher concentrations in different natural samples (Natural sample uranyl luminescence quenchers can be $Cl^-$ [11], $Fe^{2+}$ , $Mn^{2+}$ [12], … and many organic compounds [2], [13]-[15]). Matrix effect also limits Parallel Factor Analysis (PARAFAC, [16]-[22]) usage to rather qualitative, preliminary purposes as the PARAFAC decomposition relies on environment-independent individual component life-times and spectra.

Preliminary steps towards accurate first-principles based simulation (3.) have been taken by us in [5] following similar study of uranyl – sulphate system [10]. The aforementioned studies have been, however, only scalar quasi-relativistic and luminescence spectra have been approached through comparison of experimental and quantum chemistry determined spectral parameters ($T_{00}$, $\omega_{gs}$, $\omega_{es}$, $\Delta R$). The quantum chemical part was mostly based on TD-DFT / B3LYP [23]-[36] ground and excited electronic state geometry optimization. In this paper a methodology for simulation of vibrationally resolved luminescence spectra of various uranyl complex compounds in aqueous solution will be presented. The quantum chemical part includes spin-orbit splitting and TD-DFT [27-30] with the XALDA[2] approximation [S1], [42], larger variety of functionals have been used (including the Coulomb Attenuation Method, CAM-B3LYP [37], which better capture partly charge-transfer character of studied de-excitation process than B3LYP). The simulation of luminescence spectra shape has been based on Franck-Condon profile evaluation for both central group ($UO_2^{2+}$) only (in line with the methodology of „Focused" approaches [43]-[50]) and a whole complex species inside water solvent modelled at a lower level of theory. While QM/continuum-modelled-solvent approaches [51]-[54] are computationally less demanding, for directional solute-solvent interactions, in particular, for hydrogen bonding in aqueous solutions, discrete, usually atomistic, models should be used [43]-[50].

The simulation gives a proof that peak maxima spacing corresponds to vibrational frequency of normal mode connected dominantly with uranyl group symmetric stretching (while vibronic transitions corresponding to change of number of phonons in other vibrational modes contribute to rather wide peak widths, which cannot be significantly decreased even under cryogenic conditions). Wave-function methods, ΔSCF and Polarization Propagator Module (PPM, [55], [56]) are used for a comparison with TD-DFT excitation energy results.

## Theory

Based on detailed combined analysis of luminescence, absorption, Raman and IR spectra the characteristic luminescence spectrum of $[UO_2(H_2O)_5]^{2+}$(aq) can be assigned to single electronic transition X $^1\Sigma_{g,0}^+ \leftarrow$ a $^3\Delta_{g,1}$ , when using Focused approach and approximate λ-ω notation [5], [6], [10], [76].

The transition is localized on uranyl central group, corresponding dominantly to a single electron de-excitation $\sigma_{u,1/2}$ (HOMO) $\leftarrow$ $\delta_{u,3/2}$ (LUMO), with uranium-localized non-bonding LUMO of U$f_\delta$ character. An alternative Focused description X $0_g^+ \leftarrow$ A $1_g$ is based on double-group symmetry exclusively ($1/2_u \leftarrow 3/2_u$ for corresponding dominant orbital contribution).

[2] Adiabatic Local Density Approximation (ALDA) corresponds here to the use of the ground state reference calculation with a hybrid (non-local) functionals (B3LYP and asymptotically corrected variants – CAM-B3LYP [37], SAOP functionals [38]-[41] correcting B3LYP), but in response (excited state) computation, second functional derivatives of Exchange-Correlation (XC) functional have been approximated by second functional derivatives of LDA part of B3LYP (i.e., the SVWN functional [33]). The XALDA variant in addition keeps the respective fraction of Hartree-Fock exchange for exact second functional differentiation. The ALDA and XALDA variants are more accurate than Random Phase Approximation [29], [30], but less accurate than exact evaluation of second functional derivatives of XC functional as default in both Turbomole and DIRAC TD-DFT routines. However, error cancelation similar to that in Tamn-Dancoff Approximation (TDA) makes TD-DFT triplet and approximate triplet excitation energies obtained with XALDA approximation closer to experimental values than those from the exact second variational differentiation.

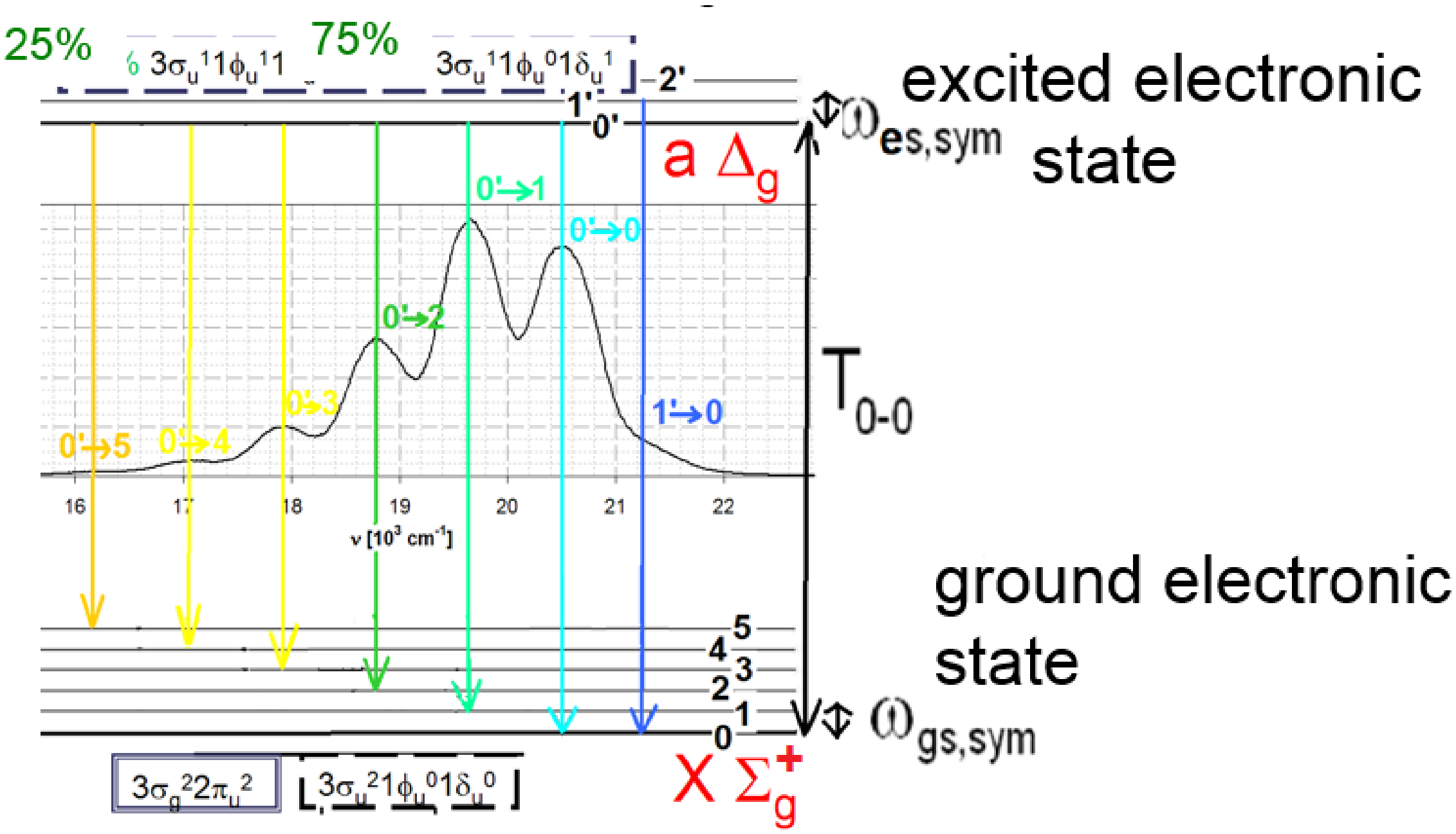

**Fig. 1':** Uranyl aquocomplex luminescence spectrum peaks assigned to 0'→*n* cold-band and 1'→0 hot-band vibronic progression. Figure adopted from [10].

For the $D_{5h}$-$[UO_2(H_2O)_5]^{2+}$(aq) model, the spectroscopic notation would be X $^1A_1$' ← a $^3E_2$' for scalar relativistic description and X $A_1$' ← a $E_1$" with spin-orbit splitting included. The transition is vibrationally resolved with most prominent progression 1'→0, 0'→0, 0'→1, 0'→2, …, 0'→*n* (the first peak being hot-band, Fig. 1') explaining sequence of equidistant (when using scale-proportional-to-energy, as the wave-numbers, $cm^{-1}$) peaks. The resolving vibrational mode is the symmetric stretching mode of (mostly linear) central uranyl group, which is highly harmonic, the U-O bond equilibrium bond length increase due to the excitation by ~ 4.6 pm.

However, it is not the only mode corresponding to spectrum. That is easily demonstrated by the fact that peak widths do not drop with decreasing temperature into cryogenic and deep cryogenic regimes as would be supposed for true single-mode contribution [76]. The vibrational modes connected to ligands and/or solvent water molecules are rather anhramonic and intercoupled. The Duschinsky effect [77]-[81] has to be therefore taken into account when the standard procedure of Franck-Condon profile computation for vibrationally resolved electronic spectra is performed – significantly rising computational demands for the „Franck-Condon profile part“ of spectral simulation.

Based on both similarity of experimental spectra and theoretical studies, luminescence spectra of uranyl complexes with other various inorganic ligands ($SO_4^{2-}$, $CO_3^{2-}$, $HCO_3^-$, $SeO_4^{2-}$, $OH^-$, $NO_3^-$, …) can be described through the same model as above, except different ligands fine-tune values of the characteristic parameters – $T_{00}$, $\omega_{gs}$ [82], $\Delta R$, $\mu$, $\tau$ – described in the Introduction section. Aside to inductive and electromeric effects, ligands affects the aforementioned parameters (and spectrum) due to the change of complex molecule point group.

However, practically more important is the „electronic part“ of spectral simulation, i.e. an accurate determination of the phonon-less transition energy. The reason is that while spectral shapes for different uranyl complex species can be very similar, tiny overall wave-number axis ($\omega$-axis) shift between them (down to 100 $cm^{-1}$ for $UO_2^{2+}$ - $SO_4^{2-}$, but only 10 $cm^{-1}$ for $UO_2^{2+}$ - $CO_3^{2-}$ - $Ca^{2+}/Mg^{2+}$ systems[3]) might be the key feature. The system in question contains heavy atom (relativistic effects), we are interested in small energy differences (the electron correlation as well as dispersive interaction has to be correctly described), and hydration has to be addressed explicitly.

## Experimental

For the TRLFS measurements 0.25 M $UO_2^{2+}$ in 0.42 M $HClO_4$ stock solution (prepared by dissolving solid $UO_3$ (p.a., depleted, n.p. Brno (Chemapol, Prague-Czechoslovakia), ≥ 98%) in conc. $HClO_4$ and subsequent demi-water dilution) have been used, $NaClO_4.H_2O$ (p.a., Sigma-Aldrich, ≥ 98%) to adjust ionic strength $I_m$, 1 M $HClO_4$ (by dilution of 60% p.a., Fluka) to adjust pH and demineralised water (Milipore, Direct-Q UV3, 18.2 MΩ.cm (25°C)). The measured solution has been prepared with total uranium concentration $c_U$ = $10^{-4}$ $mol.dm^{-3}$ (by gradual dilution of the stock solution and ionic strength and *pH* adjustment in the last step), $I_m = 4.0$ $mol{\cdot}kg_W^{-1}$, $pH = (2.0 \pm 0.2)$. The high ionic strength assures longer luminescence life-time ($\tau$ = (2.3 ± 0.2) μs) and therefore higher signal to noise ratio. Eventual perchlorate complexation has been ruled out by our experiments with varying perchlorate-adjusted $I_m$, and in agreement with $ClO_4^-$ considered as a bad complexing agent [83]). The cuvette with solution has been thermostated to $t$ = 25°C (Fig. 1)

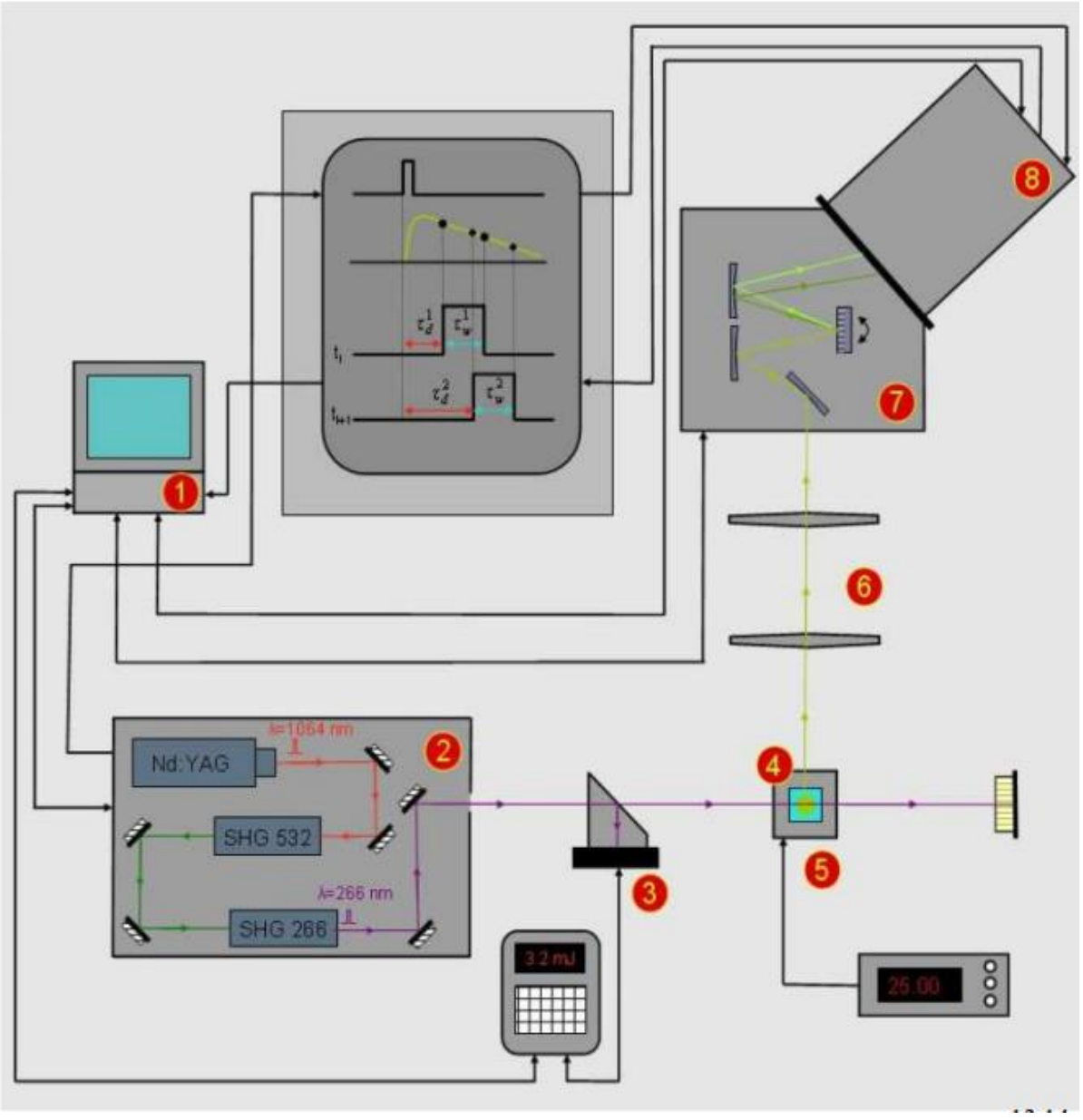

**Fig. 1.** The TRLFS measurement system used. 1 – PC control unit, 2 – tuneable Nd:YAG laser pumped laser system VIBRANT™ 355 II, 3 – beam splitter diverting defined part of the beam into digital time-resolved joule meter FieldMax II, which records individual pulses energies and the 1 synchronize the information with the CCD camera measurement results, 4 – quartz cuvette (SUPRASIL) with the

[3] in both cases smaller than gaussian variance-like peak width parameter $\sigma \sim 200$ $cm^{-1}$.

solution in thermostated holder (5). 6 – collimating lenses, 7 – Spectrograph MS257TM, dispersion element (optical grid), 8 – ICCD camera ANDOR iStar with time resolution [84].

Excitation pulses had 2 ns temporal width, 10 Hz repeating frequency, average[4] energy of (1.5 ± 0.1) mJ and mid wavelength of $\lambda_{exc}$ = 415 nm. Signal has been integrated in ICCD camera over $\Delta t$ = 10 μs. The time-resolved measurement (characterized by delays between excitation pulse and signal integration, $t_j$ ($j$ = 0, 1, …, 600)) corresponded to the collection of kinetic series with identical initial delay $t_0$ = 300 ns (long enough to avoid short-lived parasitic luminescence from sample impurities) and several different delay increments $dt$ = 100 ns, 200 ns and 400 ns. Up to 200 increments have been added to the $t_0$ (i.e. the longest delay has been $t_{600}$ = 80.3 μs).

## Experimental data analysis

The kinetic series of from TRLFS have been ordered into data matrix with spectral points as row index and temporal (delay) as a column index. The data matrix has been then analyzed by Singular Value Decomposition [63], [85-93] and Maximum Likeli-hood Estimation [94] – based FATS method [10], [5] using routine written by A. Vetešník in collaboration with J. Višňák in Matlab [S4]. The assumption of mono-exponential decay (in the temporal domain) of each spectral component has been consistent with the experimental data. Two spectral components have been, according to the speciation model, assigned to aquo complex $[UO_2(H_2O)_5]^{2+}$ (denoted also as $UO_2^{2+}$(aq) for short) and monohydroxo complex $[UO_2(OH)(H_2O)_3]^+$ (or $UO_2OH^+$(aq), luminescence life-time determined as $\tau$ = (70 ± 10) μs for the studied solution, minor component with $UO_2OH^+$ : $UO_2^{2+}$ ratio of luminescence rates ~ $7.7 \cdot 10^{-4}$) and only the former one investigated further (in comparison with computational chemistry simulated spectrum). In this study, temporal domain served for $UO_2^{2+}$(aq) discrimination from $UO_2OH^+$ component.

[4] but each detected spectrum has been correlated with individual pulse energy – measured continuously by joule-meter, allowing to remove noise due to variation in excitation beam energy

## Computationals

### *1. Work-flow diagram*

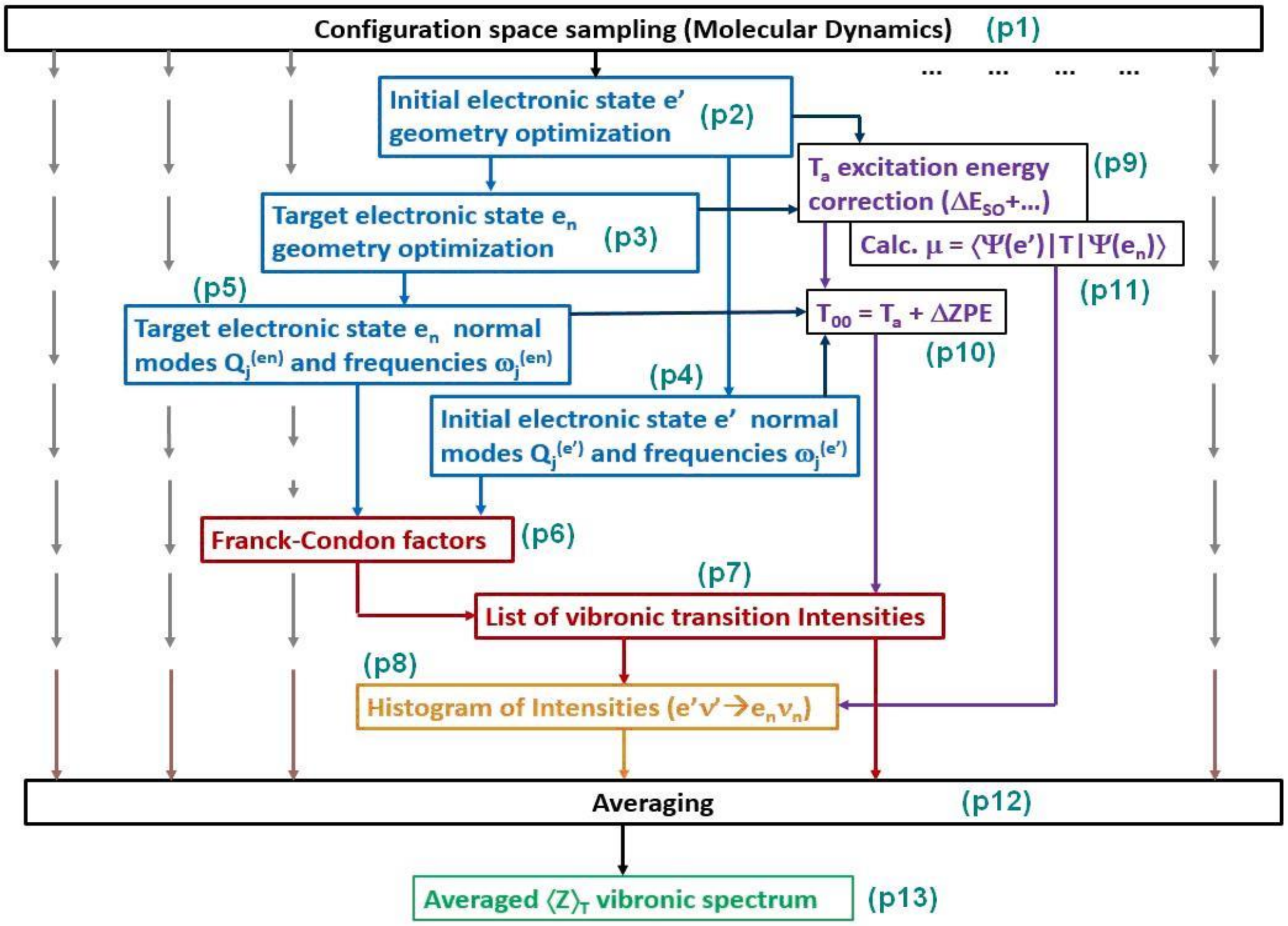

**Figure 3:** Workflow chart for vibronic spectrum simulation for case of a single electronic transition vibrationally resolved by arbitrary number of vibrational modes.

Methodology used is briefly summarized in Fig. 3. Several snapshots from CMD trajectory (p1, uppermost horizontal-oriented black-bordered box) are processed in the following way as expressed by different colour boxes between the two horizontal-oriented black-bordered boxes in Fig. 3 for one chosen example-snapshot (processing of others are depicted just by multiple grey arrows pointing downwards). The light blue boxes (p2-p5) correspond to scalar quasi-relativistic[5] ECP/(TD[6])DFT/B3LYP-D3[7] calculations (with $N_1$, $N_2$, $N_3$ partitioning as will be described in „System partitioning and Focused approach", local geometry optimization and normal mode analysis), def-

[5] Pseudo potential denoted ECPDS60MWBSF [143]-[146] in DIRAC [S1] has been used for the scalar quasi-relativistic computations. For spin-orbit resolved quasi-relativistic computation, ECPDS60MWBSO [143]-[146], which differ by spin-orbit part only has been used.
[6] Adiabatic TD-DFT, without XALDA approximation.
[7] For the semi-empirical D3 dispersion correction, please see [136].

TZVPP atomic basis set [146]-[148],[8] is assigned to central uranyl group and five water molecules directly ligated to it, def-SVP [165] to other atoms considered explicitly. The violet boxes (p9-p11) in the right part correspond to two single-point based computation of zero-phonon line $T_{00}$ (done according to eq. (2)) and possibly electronic transition moments $\mu$. The aforementioned single-point computations (p9-p11) are approached in greater detail in "Phonon-less excitation energy ($T_{00}$) decomposition" section later.

## 2. System partitioning and Focused approach

For a solvation description the Classical Molecular Dynamics (CMD, p1 in Fig. 3) study with force-field adapted from [95] (with SPC/E model for water molecule representation) has been done for $[UO_2(H_2O)_5]^{2+}$ surrounded by 4 000 water molecules (in one cubic cell, periodic boundary conditions has been used, two $Cl^-$ ions added to keep total electric charge zero, the minimum distance U-Cl have been checked to be large enough not to influence uranyl geometry, GROMACS V5 software has been used). First, 10 ns trajectory using NPT ensemble has been performed (pressure 1 bar, temperature 298 K, time step 2 fs) to achieve thermodynamic equilibrium. Next, 10 ns trajectory has been calculated using microcanonical ensemble (to correctly probe vibrational motion of atoms), snapshots have been collected every 50 ps and their representative subsets (Tab. 1) used for further quantum-chemical calculations (p2-p5 in Turbomole and p9-p11 in DIRAC).
For the aforementioned quantum chemical computations starting from each of chosen snapshots, the water molecules have been divided into three different groups (Fig. 2):

1. The inner-most shell of $N_1$ water molecules. These have been included explicitly for (TD)DFT calculation and coordinates of all their nuclei optimized together with uranyl central group.
2. The middle shell of $N_2$ water molecules – $H_2O$ in this group has been included explicitly but their coordinates fixed („frozen”) to values from CMD snapshot.
3. The outer-most shell of $N_3$ water solvent molecules – each atom has been included only as point-charge[9] and with fixed position adopted from CMD snapshot.

The reason to optimize (p2 and p3 in Fig. 3) solvated structure ($UO_2^{2+}$ and $N_1$ water molecules) comes from the fact that luminescence spectrum is vibrationally resolved and geometry optimization is necessary for the normal mode analysis (p4 and p5 in Fig. 3). The reason for $N_2 > 0$ is rather practical, as without this shell serving as kind of barrier, the point charges from the outer-most shell would „tear apart“ the solvated structure (an alternative would be to include higher multipoles and eventually polarizabilities to describe the outer-most shell).

Reasonable choices for $N_1$ are 0 (no vibrational resultion for ligand-connected vibrational modes, only uranyl central group coordinates optimized), 5 (only ligated water molecules [98], [61] included into inner-most shell) or up to 17 (the five coordinated water molecules plus two groups per six molecules – number adapted from *in Vacuo* optimized structure presented in Fig. 19 in [10]). The $N_2$ should be chosen at least one layer thick to truly work as „barrier“. As point-charge inclusion is computationaly cheap, $N_3$ can reach several thousands (Fig. 2).

[8] The def-SVP and def-TZVPP basis sets for hydrogen comes from Turbomole [S8] developer team and hasn't been published yet.
[9] For oxygen -0.834, for hydrogen +0.417, from TIP3P [96]. The difference with respect to SPC/E partial charges is rather small due to the explicit $N_1$+$N_2$ water molecule inner shell providing correct ligation and/or hydration of the central uranyl group. The outerlying $N_3$ water model influence is discussed in the Supplementary Information.

The $N_1$ parameter increase shouldn't be *a priori* expected as leading to more accurate results since spectral estimation is based (in our current work) on harmonic approximation. While uranyl stretching modes exhibit almost no anharmonicity [99]-[101], the same cannot be said about solvent molecule connected modes.

While the CMD trajectory calculation (p1) tailored for a luminescence spectrum prediction should be done with uranyl-excited-state force field (the emission comes from the "Thexi" state [2]), we used ground state parameters from [95]. The difference should be rather small as the only mode significantly affected by excitation (uranyl group symmetric stretching) is geometry optimized within computational protocol used anyway.

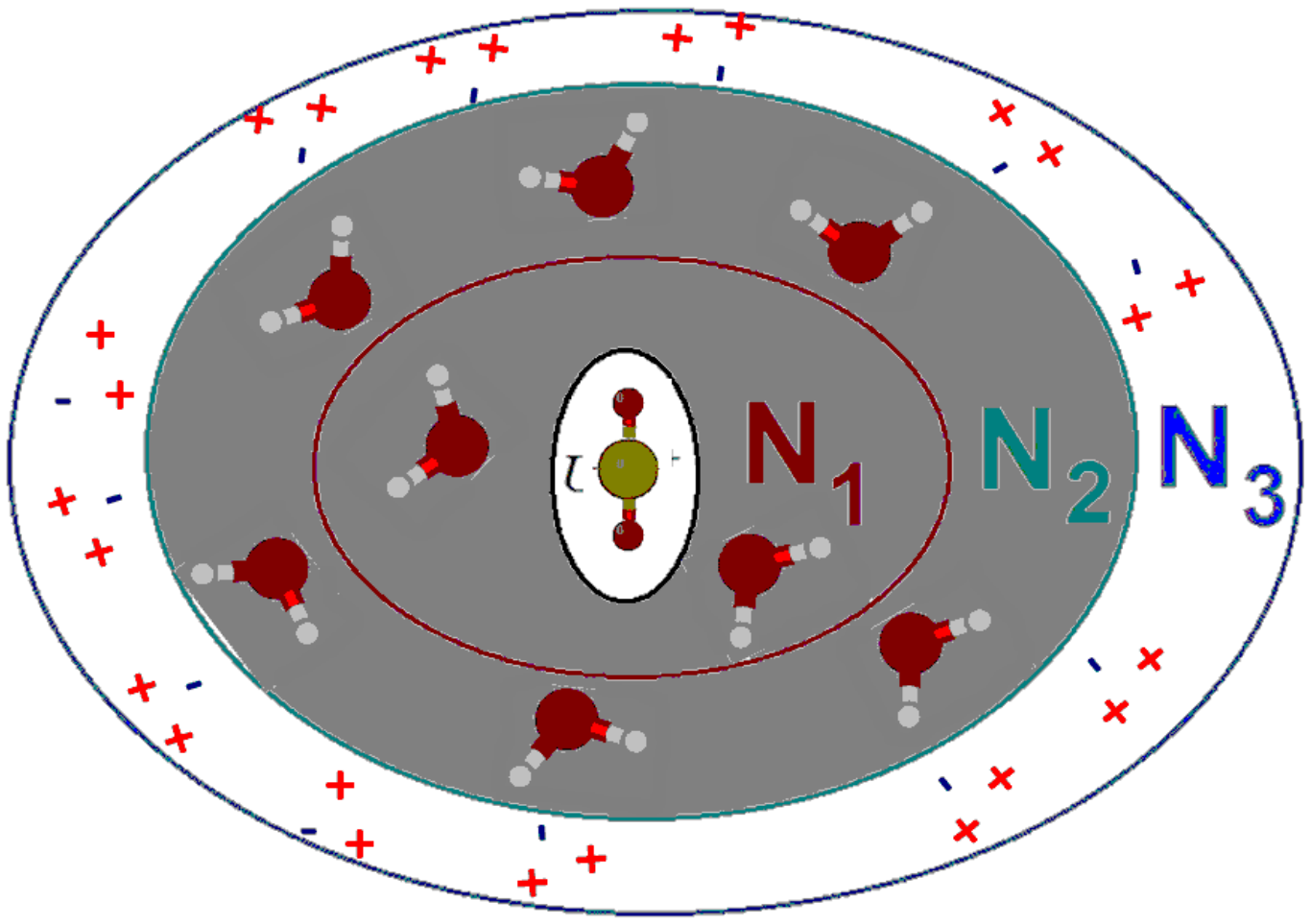

**Figure 2:** Studied system has been formally partitioned into four different regions. Starting from the innermost – uranyl central group (where electronic transition and "spectrum resolving" vibrational mode are dominantly localized) and the $N_1$ to uranium nearest water molecules are considered for Franck-Condon profile computation for each snapshot, $N_2$ next-to-nearest water molecules are accounted explicitly into QM computations, but their positions fixed as well as coordinates of the outhermost $N_3$ water molecules (modelled by point charges only).

| Settings name | $N_1$ | $N_2$ | $N_3$ |
|---|---|---|---|
| TT1 | 0 | 5 | 3260 |
| TT3 | 5 | 12 | 3248 |
| TT6 | 5 | 20 | 3140 |
| TT9 | 17 | 48 | 3165 |

**Tab. 1:** Possible settings for $N_1$, $N_2$ and $N_3$ parameters suggested

In the current simulation, only TT1 (33 snapshots) and TT3 (6 snapshots) settings (Tab. 1) have been used, the computational demands for TT6 and TT9 normal mode analysis in excited electronic state (using NumForce [102], [103] in Turbomole V7.1 [S8], [104], [105]) have been over our current accessiblity. Reasons mentioned in pre-previous paragraph together with wider statistics for TT1 (as each point is less computationaly demanding, more snapshots can be analyzed) favourize this settings over TT3. For TT1 frequencies scaling factor of 0.96 has been used [173], for the TT3 settings, no frequency scaling has been applied as the inclusion of more water molecules should facilatate a more correct vibrational frequency description.

## 3. Master formula for luminescence spectrum simulation

Assuming separation between electronic and nuclear motions (Born-Oppenheimer, [106], [107]), electronic transition moments independent of nuclear coordinates (Condon, [108]-[113]), harmonic approximation to vibrational motions in both initial and target electronic state (double-harmonic approximation) leads to luminescence spectrum estimated (via lowest order Dyson expansion [114]) as

$$Z(\nu) = \sum_{e'} \left|\mu^{(e',e)}\right|^2 W(e') \left\langle \sum_{n,n'} \rho_{e'}(n') \cdot \left|FCF(e'n' \to en)\right|^2 \cdot V\left(\nu - \left(T_{00} + E_{n'}^{vib,e'} - E_{n}^{vib,e}\right), \Gamma_{en}^{e'n'}\right) \right\rangle_T , \quad (1)$$

where $\nu$ stands for wavenumber ($cm^{-1}$), $\langle X \rangle_T$ for an average of *X* over Classical Molecular Dynamics (CMD) trajectory snapshots, $\rho_{e'}(n')$ for Boltzmann occupation factor for vibrational level described by set of vibrational quantum numbers *n*‘ of excited electronic state *e*‘, $FCF(e'n' \to en) = \langle \chi'_{e'n'} | \chi_{en} \rangle$ for Franck-Condon-Factor [115]-[119], central quantity for the spectral estimation (for their analytical evaluation - [120]-[123], or in special single- [97] or double- [124] mode cases), $e'n' \to en$ means transition from vibrational level *n*‘ of electronic state *e* to vibartional level *n* of electronic state *e*. The last term in angle braces, *V* is a Voigt peak profile centred at wavenumber corresponding to $e'n' \to en$ vibronic transition energy and with given width parameters, together denoted as $\Gamma_{en}^{e'n'}$. The sumation written inside the $\langle \cdot \rangle_T$ average is in principle over all possible initial and target vibrational quantum numbers combinations (i.e. 2(3*N*-6) individual sums). In practise it is necessary to restrict the calculation to few well chosen modes and only limited maximal vibrational quanta. The outermost sum is over excited (initial) electronic states, *W*(*e*‘) corresponds to *e*‘ electronic state initial probability and $|\mu^{(e',e)}|^2$ to *e*‘ → *e* electronic transition rate[10].

For $UO_2^{2+}$ in vacuum, $0_g^+ \leftarrow 1_g$ transitions are Lapport forbidden (and spin- and symmetry- allowed only due to the spin-orbit splitting of the excited state, as it originates dominantly from $^3\Delta_g$ state). After ligation, considering $D_{5h}$-$[UO_2(H_2O)_5]^{2+}$ in vacuum, the corresponding $A_1$’ $\leftarrow$ $E_1$“ transition is dipole forbidden as well.

In aqueous solution within *Focused approach* the non-symmetric position of water solvent molecules perturb dihedral angles of hydrogen atoms of ligated waters[11], U-$O_w$ bond lengths, and to a lesser extend other internal coordinates even for optimized solute structures. This leads to the $C_1$ point group instead

[10] Estimated from an average over few well-chosen geometries, under an assumption that $\mu^{(e',e)}$ will not vary with molecular geometry significantly. As spontaneous emission rate is proportional to $\nu^3$ [125], experimental data have to be divided by the third power of frequency before comparison with Franck-Condon profiles (as e.g. in [126]). The $\nu^3$ normalization, however, doesn’t seem to affect spectral parameters for our case significantly.

[11] angles between plane defined by ligated water (H-$O_w$-H) and uranyl-$O_w$ ($O_{yl}$-U-$O_w$) plane

of $D_{5h}$ (or $C_{2v}$, Fig. 9) for solute, and the electronic transition from the lowest lying excited electronic state (dominated[12] by $^3\Delta_g$ state of central uranyl group) to the ground electronic state is dipole allowed and its transition probability is by $10^7$ factor higher when compared to $A_1' \leftarrow E_1''$ quadrupole transition probability for $D_{5h}$-$[UO_2(H_2O)_5]^{2+}$ *in Vacuo*. Therefore, only dipole contribution to

## 4. *Phonon-less excitation energy ($T_{00}$) decomposition*

To allow for a more accurate computation (p9 and p10 in Fig. 3) of the phonon-less excitation energy $T_{00}$ (than by scalar quasi-relativistic methods used for geometry optimization and normal mode analysis, p2-p5 in Fig. 3) we suggest decomposition

$$T_{00}^{aq(pch),SO} = \underbrace{T_a^{pch,SO} - T_a^{pch,SF}}_{\Delta T_a^{SO(pch)}} + T_a^{aq,SF} + \Delta ZPE, \qquad (2)$$

where $T_{00}^{aq(pch),SO}$ stands for transition energy corresponding to $0 \leftarrow 0'$ vibronic transition, upper indices indicating „in **aq**ueous solution, with part of energy estimated by **p**oint-**ch**arge („pch“) solvation model, spin-orbit resolved („SO“) computation“, $T_a^{pch,SO}$ is spin-orbit resolved adiabatic excitation energy evaluated as an energy difference (5) – with solvent modelled by point-charges („pch“) only, $T_a^{pch,SF}$ is a scalar quasi-relativistic adiabatic excitation energy (8) with point-charges solvation model and $T_a^{aq,SF}$ is scalar quasi-relativistic adiabatic excitation energy (9), but evaluated with explicit inclusion of nearest *N* solvent water molecules into QM system (while remaining solvent molecules are included as point-charges). In addition, the difference of phononic **Z**ero-**P**oint-**E**nergy, $\Delta ZPE$ , between excited and ground electronic states is added (10). In formula (2) above, interpretation of „adiabatic part“ of $T_{00}^{aq(pch),SO}$ (i.e. $T_a^{aq(pch),SO}$ (3)) as a sum of scalar part (with explicit solvent model, $T_a^{aq,SF}$) and spin-orbit correction (however, only at a more approximate solvent model level of theory, $\Delta T_a^{SO(pch)}$) could be considered

$$T_a^{aq(pch),SO} \equiv T_{00}^{aq(pch),SO} - \Delta ZPE = E_{es}(R_{es}) - E_{gs}(R_{gs}), \qquad (3)$$

where $E_{es} = E_{es}(R)$ and $E_{gs} = E_{gs}(R)$ denote excited (**e**s) and ground (**g**s) electronic state Potential Energy surfaces respectively. $R_{es}$ and $R_{gs}$ being equilibrium geometries of the respective electronic states (i.e. coordinate of $E_{es}(R)$ and $E_{gs}(R)$ minima, $R_{es} = \text{argmin}\, E_{es}(R)$, $R_{gs} = \text{argmin}\, E_{gs}(R)$).

Alternatively, the formula (2) can be rewritten suggesting different partition of $T_a^{aq(pch),SO}$,

[12] but not limited to, as a result of state mixing due to spin-orbit interaction, ligation and solvent perturbations.

$$T_{00}^{aq(pch),SO} = T_a^{pch,SO} \underbrace{- T_a^{pch,SF} + T_a^{aq,SF}}_{H_{pch}^{SF}(T_a)} + \Delta ZPE, \quad (4)$$

i.e., as a sum of spin-orbit resolved adiabatic excitation energy $T_a^{pch,SO}$ and a solvation-correction term, $H_{pch}^{SF}(T_a)$ (evaluated at scalar quasi-relativistic level of theory only).

The first term in (2), (4), $T_a^{pch,SO}$ includes not only more accurate treatment of the spin-orbit interaction, but in principle could be based on different DFT functional/approach, or on *ab initio* (Wave-Function-Theory) method. In case of DFT (and therefore TD-DFT for excited states), corrections (XALDA), approximating XC-Kernel integrals in Casida equations have been included here.
The XALDA [S1], [70] results have been compared here with the ordinary ATD-DFT (Tab. 3[13] and Tab. 5[14]), the corresponding difference is up to +1300 cm$^{-1}$ and corrects otherwise underestimated ATD-DFT excitation energies. The first term in (4) can be also evaluated within an all-electron computation based on many-body Dirac-Coulomb(-Gaunt) Hamiltonian [135]. General expression for the first term in (4) is

$$T_a^{pch,SO} = E_{es}^{pch,SO}(R_{es}) - E_{gs}^{pch,SO}\left(R_{gs}\right), \quad (5)$$

where meaning of the symbols on the right side is analogical to (3). The $R_{gs}$ and $R_{es}$ does include coordinates of complex molecule and solvent molecules (but the optimization is done with respect to only $N_1$ of them, with additional $N_2$ being explicit, but with fixed coordinates) and have been determined only at the scalar quasi-relativistic B3LYP-D3[15]/ECP level of theory. However, as DFT/B3LYP-D3 usually determines geometries and normal modes (the ΔZPE term will be discussed later on) accurately and as simple-model study of $UO_2^{2+}$ revealed that PES for given electronic multiplet (arising from spin-orbit splitting) have similar shapes in equilibrium region, this methodology is, at least in principle, justified. The adiabatic excitation energy should not be confused with the vertical excitation energy

$$T_v^{pch,SO} = E_{es}^{pch,SO}(R_{gs}) - E_{gs}^{pch,SO}\left(R_{gs}\right), \quad (6)$$

or vertical de-excitation energy

$$T_{de}^{pch,SO} = E_{es}^{pch,SO}(R_{es}) - E_{gs}^{pch,SO}\left(R_{es}\right), \quad (7)$$

where electronic state differences are evaluated in the same point of the configurational space ($R_{gs}$ and $R_{es}$ for $T_v$ and $T_{de}$ respectively (upper indices dropped for brevity)).

The second and third terms on right side of (4), denoted as solvation-correction $H_{pch}^{SF}(T_a)$ includes also the D3-dispersion interaction correction [136], used to improve B3LYP performance. The terms can be defined as

$$T_a^{pch,SF} \;=\; E_{es}^{pch,SF}(R_{es}) \;-\; E_{gs}^{pch,SF}\left(R_{gs}\right), \tag{8}$$

$$T_a^{aq,SF} \;=\; E_{es}^{aq,SF}(R_{es}) \;-\; E_{gs}^{aq,SF}\left(R_{gs}\right). \tag{9}$$

---

[13] For the $D_{5h}$-$[UO_2(H_2O)_5]^{2+}$ *In Vacuo* results, the corrections are denoted $X_v$ for vertical excitation energy, $X_a$ for adiabatic excitation energy, $T_v$ and $T_a$ have been computed with XALDA in Tab. 3.
[14] the XALDA based computation is written under given functional
[15] D3 dispersion interaction correction [136].

Partitioning of $T_a^{aq(pch),SO}$ is consistent with the „*Focused approach*“ common in computational spectroscopy. The spin-orbit (and electron correlation, level of theory in inclusion of relativistic effects, …) correction for the complex alone and solvation shift, however, are not additive and therefore, at least as point-charges, the solvent should be added for a more accurate electronic study of complex (performed here in DIRAC program package[16]). To quantify the aforementioned non-additivity, „pch“ in round parentheses are included into upper index in $T_{00}^{aq(pch),SO}$ to account for the approximate level of solvent model for the spin-orbit resolved computation. The comparison to analogical quantity $T_{00}^{aq(vac),SO}$ is of interest (where „vac“ stand for vacuum, i.e. solvent excluded from the spin-orbit resolved computation) as well as difference between $H_{pch}^{SF}(T_a)$ and $H_{vac}^{SF}(T_a)$ solvation corrections[17] (and are discussed in the last paragraphs of 3.2 section of Results).

The difference of phononic **Z**ero-**P**oint-**E**nergy, $\Delta ZPE$ , between excited and ground electronic states, in double-harmonic approximation

$$\Delta ZPE \;=\; \frac{1}{2}\hbar \sum_j \left(\omega_{j,es} - \omega_{j,gs}\right), \tag{10}$$

has been evaluated at scalar quasi-relativistic B3LYP-D3 level of theory in Turbomole V7.1 [S8]. The $N_1$ and $N_2$ settings for ΔZPE evaluation should be the same as for $T_a$ computations. However, the $T_{00} = T_a + \Delta ZPE$ could depend on $N_1$, $N_2$ values. As the vibrational analysis has been performed with „frozen nuclei“ (in $N_2$ and $N_3$ solvation shells) of some or all solvent molecules, six modes (usually the first six in increasing-frequency ordering) corresponds to overall translation and tangential rotation of uranyl and $N_1$ water molecules and there is question whether to include them in the sum on the right side of (10). We have chosen to exclude them, but the difference will be discussed later.

**DFT functionals used** in $T_a^{pch,SO}$ computations: Functional B3LYP has been called „golden standard of quantum chemistry” [31]-[36]. But its application to charge-transfer systems is problematic due to underestimated exchange energy for long-separated electrons, Coulomb Attenuated Method B3LYP (CAM-B3LYP) [37] and Statistical Average of Orbital Potentials (SAOP) [38]-[41] have been tested here as a potential remedies.

**Note on isotopic effects:** Through later text, only the most abundant isotopes have been considered for all elements (i.e. $^{238}$U, $^{16}$O and $^{1}$H). Eventual $^{A}$O- uranyl group oxygen substitution (for A ≠ 16) would modify symmetric stretching frequency $\omega_{gs}$ and thus luminescence peak spacing (and to a lesser extend affect coupling with other vibrational modes). The $^{235}U$ presence would have much lesser impact as $\omega_{gs}$ does not depend on uranium atom mass, but might affect electronic structure [140], [141] through different nuclear spin-electronic angular momenta interaction – however, those effects have not been studied in this paper.

[16] The point-charges in DIRAC calculations as been included through Polarized Embedding module (PEQM, [137]-[139]), where only charges have been included in the embedding potential.

[17] Step for a more correct description should be inclusion of point-like dipole and quadrupole moments and dipole-dipole anisotropic polarizibilities to point-modelled solvent atoms (instead of electric charges only). Or on the other side, at least approximate inclusion of spin-orbit splitting for solvation-correction computation. Both are further prospects for our research (as well as a more accurate methodology for the *Focused approach*).

**Franck-Condon factors/profiles** (FCF, red boxes p6 and p7 in Fig. 3) have been evaluated using ezSpectrum [S2]. To process the Turbomole [S8], [104] NumForce script [101], [102] outputs for the case of frozen $N_2$ water molecules, the six lowest lying frequencies (connected to translations and rotations of uranyl and $N_1$ water molecules together) have been artificially zeroed and frozen atoms removed before further processing by tm2ezspectrum script and ezSpectrum [S2]. For TT1 settings (Tab. 1), the maximal number of vibrational quanta in the initial (lowest lying excited) and target (ground) electronic states has been set to 7 and 10 respectively. Larger number of vibrational modes in TT3 settings ($N_1$ = 5 implying 48 vibartional modes as compared to $N_1$ = 0 with 4 modes for TT1) forced us to limit maximal quanta number in initial and target states to 1 and 5 respectively. Duschinsky effect [77]-[81] has been included in both settings (Tab. 1).

**Histograms** (of vibronic transition wave-numbers) **with weights** (FCF, orange box p8 in Fig. 3) have been created from ezSpectrum outputs in Matlab [S4] using function „histwc" [142]. Aforementioned histograms with weights represent simulated vibrationally resolved luminescence spectra. The bin widths have been chosen as $d\nu$ = 10 cm$^{-1}$, slightly wider than approximate FWHM of envelope of peaks from rotationally resolved luminescence spectrum of bare uranyl (not presented here), which was estimated as $\Gamma$ = 4 cm$^{-1}$. On the other side, typical experimental resolution for wave-number scale was 20 cm$^{-1}$.

## Results and discussion

### *1. $UO_2^{2+}$ in Vacuo*

The quasi-relativistic ATD-B3LYP study on $UO_2^{2+}$ *in Vacuo* revealed that inclusion of spin-orbit part in pseudo potential has significant effect on excitation energies (1 400 cm$^{-1}$ difference between $T_a$ values for scalar $^3\Delta_g$ state and its spin-orbit resolved $1_g$ component corresponds to almost twice peak spacing in experimental spectrum!), but rather small effect on equilibrium distances $R$ (Tab. 2) and vibrational frequencies ($\omega_{sym}$).

| State | $R$ [A°] | $T_v$ [ $10^3$ cm$^{-1}$] | $T_a$ [$10^3$ cm$^{-1}$] | $T_a - T_v$ [cm$^{-1}$] | $\omega_{sym}$ [cm$^{-1}$] |
|---|---|---|---|---|---|
| **X $^1\Sigma_g^+$** | **1.6949** | **vertical** | **adiabatic** | **difference** | **1044** |
| X $0_g^+$ | 1.6965 | excitation | excitation | | 1056 |
| **A $^3\Delta_g$** | **1.7409** | **18.6** | **17.6** | **-938** | **925** |
| **$1_g(^3\Delta_g)$** | **1.7424** | **17.2** | **16.2** | **-985** | **925** |
| $2_g(^3\Delta_g)$ | 1.7432 | 18.7 | 17.7 | -1016 | 909 |
| $3_g(^3\Delta_g)$ | 1.7448 | 20.8 | 19.7 | -1061 | 915 |
| **A $^3\Phi_g$** | **1.7546** | **17.0** | **15.4** | **-1647** | **903** |
| $2_g(^3\Phi_g)$ | 1.7568 | 14.4 | 12.8 | -1637 | 903 |
| $3_g(^3\Phi_g)$ | 1.7564 | 15.7 | 14.0 | -1625 | 905 |
| $4_g(^3\Phi_g)$ | 1.7559 | 19.4 | 17.8 | -1614 | 927 |
| **A $^1\Phi_g$** | **1.7590** | **22.3** | **20.4** | **-1918** | **903** |
| **A $^1\Delta_g$** | **1.7594** | **27.7** | **26.0** | **-1774** | **865** |

**Table 2:** Quasi-relativistic SORECP ATD-DFT/B3LYP-D3/def-TZVPP computation for bare uranyl ($UO_2^{2+}$) spectroscopic properties and comparison with scalar-quasi-relativistic approximation.

Different spin-orbit components arising from given scalar-relativistic pure triplet (Fig. 5) are highlighted with the same colour in Tab. 2 ($^3\Delta_g$ (luminescence active) in yellow, $^3\Phi_g$ in greenish blue, last two lines corresponds to higher lying scalar-quasi-relativistic singlets) With certain level of uncertainty this observation justify the $T_{00}$ separation (2) even for condensed phase and application of scalar quasi-relativistic methods for normal mode analysis in this study.

For comparison, ECP/HF method used by Tsushima et al [172] predicts $\omega_{gs,sym}$ = 1195 cm$^{-1}$ for uranyl ground electronic state in gaseous phase and 1058 cm$^{-1}$ in aqueous solution. For the ground state we have reproduced Pomogaev et al [95] ECP/B3LYP values within error typical for normal mode computation machine/software-related uncertainities.

Luminescence directly involved molecular orbitals are plotted (Fig. 4) as an energy levels The 5f uranium orbitals after splitting by molecular-axis field of oxygen atoms combine into occupied $5\sigma_u$ (HOMO, bonding, from the one *f* orbital with $j_z = 0$) shared molecular orbitals, virtual pairs of uranium-based (non-bonding) $1\phi_u$ and $1\delta_u$ (which are quasi-degenerated, roughly corresponding to the original four 5f with $|j_z| > 1$) and shared anti-bonding $4\pi_u$ (formed from 5f with $|j_z| = 1$). Right to scalar-(quasi)relativistic case, spin-orbit resolved levels are plotted. While the LUMO is the $1\phi_{u,5/2}$ spin orbital (for bare uranyl, as opposed to the aquo complex, where $\phi_u$ and $\delta_u$ energy levels has reversed ordering with $\delta_u$ being lower), the luminescence-important virtual orbital is $1\delta_{u,3/2}$ as the molecular axis projected total angular momentum difference $|\Delta j_z|$ to $5\sigma_{u,1/2}$ HOMO is $|\Delta j_z|=1$.

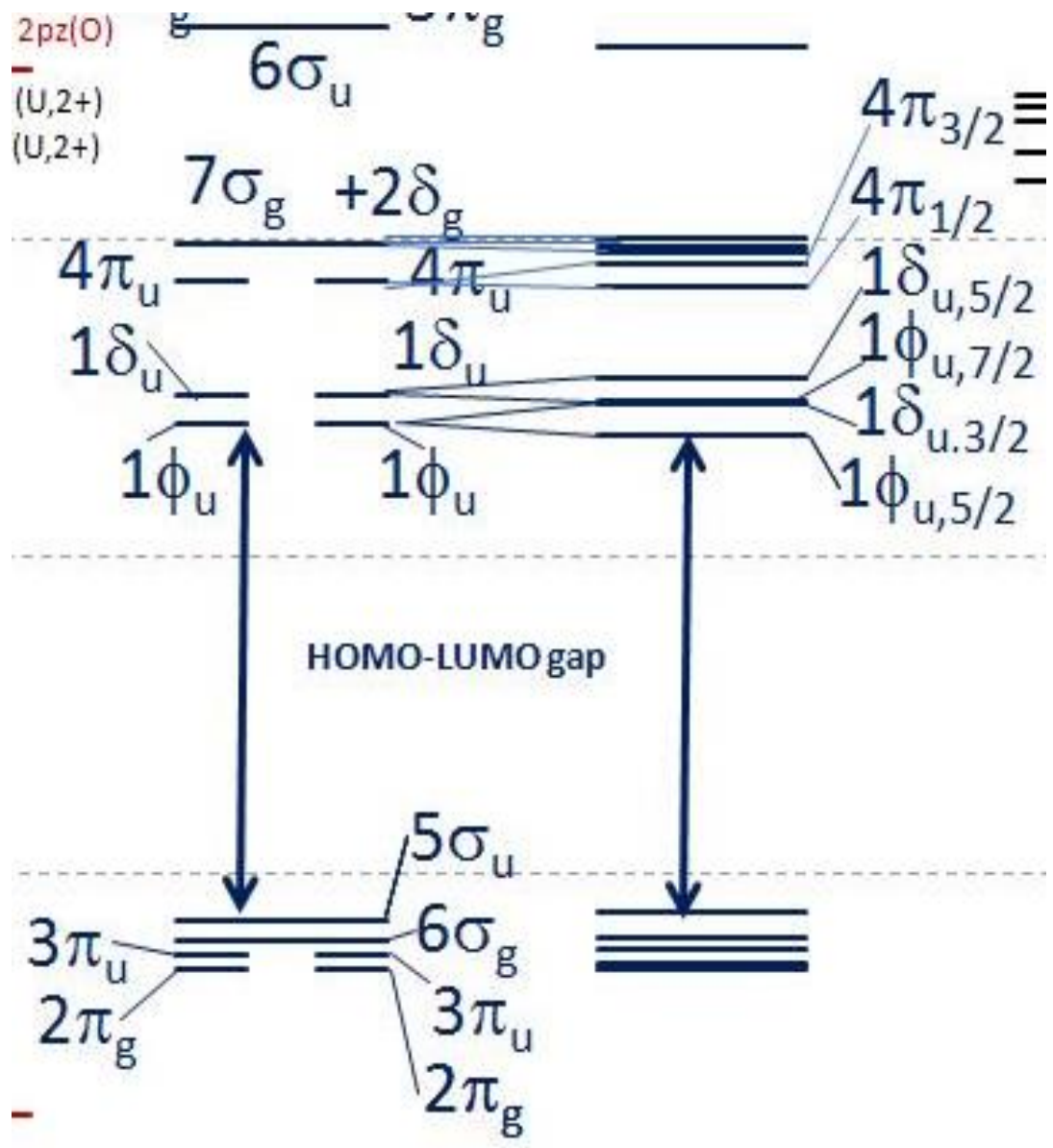

**Figure 4.** Molecular orbital level scheme for $UO_2^{2+}$ (U-O distance of 175 pm corresponds roughly to $1_g(^3\Delta_g)$ excited state equilibrium geometry or ground state equilibrium distance for *hydrated* uranyl) within RECP/CAM-B3LYP, expanded TZVPP basis set[18]), the columns on left and right correspond to SF (spin-free) and SO (spin-orbit resolved via SORECP [143]-[146]), respectively. Uranium atom partial charge has been determined as +2.8 for this geometry and method (SO).

Please note that spin-orbit splitting of both $\delta_u$ and $\phi_u$ is comparable to $\phi_u - \delta_u$ energy separation. Under relativistic double-group symmetry, the orbital angular momentum labels $\sigma$, $\pi$, $\delta$, $\phi$ are no longer exact and in particular for the $1\phi_{u,5/2}$ and $1\delta_{u,5/2}$ the $\delta$-$\phi$ mixing is almost 1 : 1.

For an easier identification and approximate spin-labelling of spin-orbit resolved state, „spin-orbit-scaling“ (or „spin-scaling“ for short) computations has been done within spin-orbit relativistic pseudo potential (SORECP) framework. The literature-preferred SORECP, implemented in Dirac [S1], [143-146], $U^{SORECP}$ has been divided into scalar part ($U^{AREP}$) and spin-orbit ($U^{SO}$) term and a scaling parameter $p \in [0;1]$ has been inserted as a multiplicative prefactor for the latter,

---

[18] Augmented def-TZVPP [147], [148] and aug-cc-pVTZ [149] (for cc-pvTZ see [150], aug-T addition consists of single gaussian added to s, p, d and f functions, corresponding exponents 0.07376, 0.05974, 0.214 and 0.5, no reference in Turbomole [S8], [104]), for oxygen and two different triple dzeta basis for uranium (together (28s26p20d16f9g) / [16s15p10d8f6g], ref. [146] and [151]).

$$U^{SOREP} = U^{AREP} + p \cdot U^{SO}, \tag{11}$$

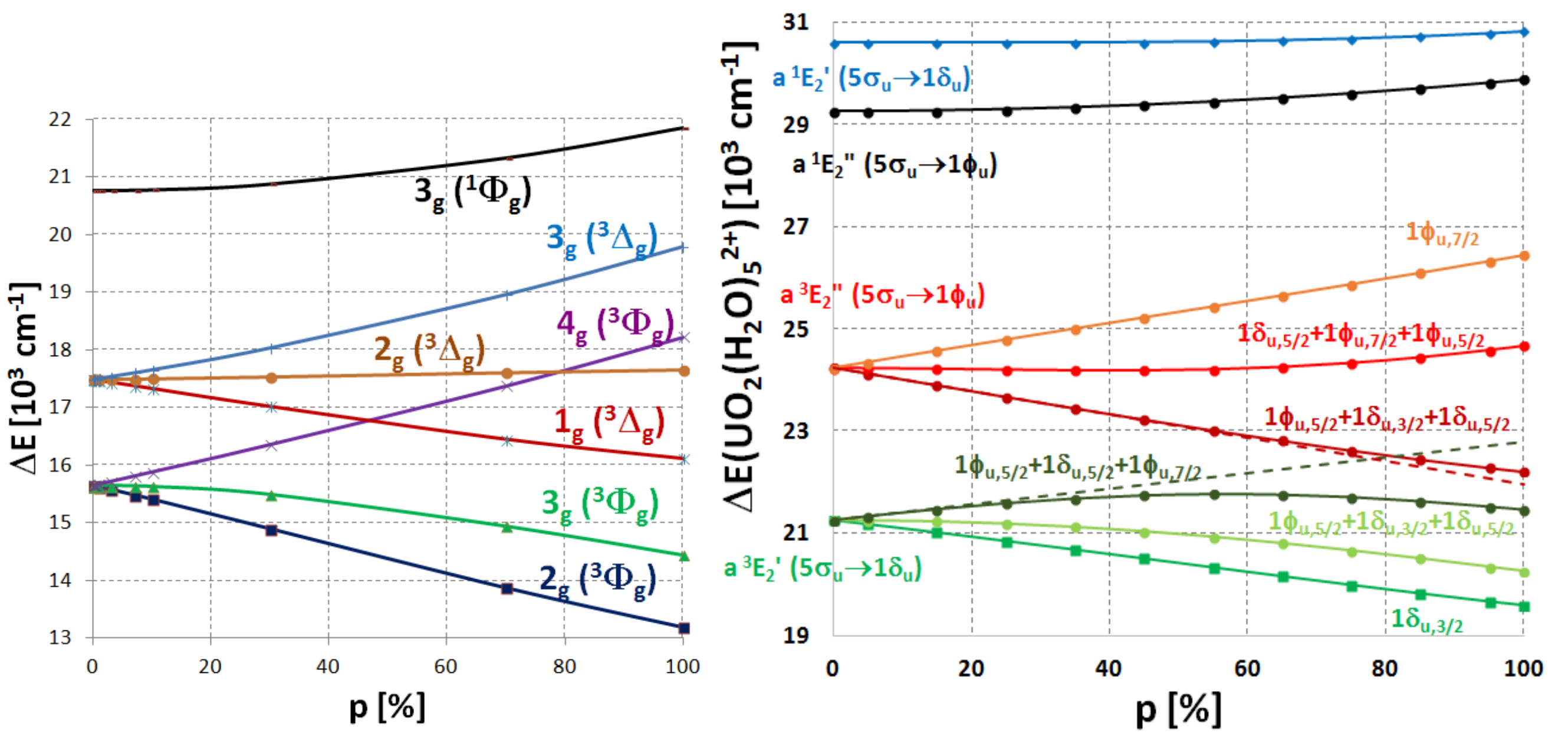

**Fig. 5 (left), 6 (right):** The spin-orbit splitting diagram showing the spin-orbit ECP parameter *p* dependence (11) of vertical excitation energies Δ*E* (labeled as $T_v$ in Tab. 3) for the lowest lying states of $UO_2^{2+}$ (left, U-O bond length R = 175 pm) and $[UO_2(H_2O)_5]^{2+}$ (right, corresponds to PES ground state minimum for B3LYP functional). The ATD-DFT/CAM-B3LYP/triple dzeta basis[19] method has been used.

in figures (Fig. 5, 6) plotting excited electronic state energies (the ground state is dominantly closed-shell singlet, $0_g^+(^1\Sigma_g^+)$) the parameter *p* is given in percents.

As bare uranyl (the molecular ion *in Vacuo*) luminescence spectra, to our best knowledge, have never been measured yet (but gas phase experimental study on uranyl complex species are available [152]), we report the respective theoretical simulation in the Supplementary Information only.

For a single electronic transition $X^1\Sigma_{g,0}^+ \leftarrow a^3\Delta_{g,1}$ no Duschinsky effect has been observed within simulation protocol (unlike for complex species in aqueous solution, where symmetric and anti-symmetric stretching rather well distinguished in the ground electronic state mix in the excited) and vibrationally resolved luminescence spectrum consists dominantly of peak progression 0'→0, 0'→1, …, 0'→6, … corresponding to different vibrational levels of symmetric stretching mode.

For comparison with all electron Dirac-Coulomb Hamiltonian [135] based ATD-DFT computations of vertical excitation energies ($T_v$) for $R$ = 1.683 A and 1.708 A please see [70]. Studies [71] and [72] investigate uranyl electronic structure as well.

[19] For the $UO_2^{2+}$ case, extended basis set as for computations presented in Fig. 4 has been used. However, neither computations for $UO_2^{2+}$ nor $[UO_2(H_2O)_5]^{2+}$ in Fig 5. and Fig. 6 has been done with XALDA. Aside to ca. 1200 cm$^{-1}$ shift of all excitation energies upwards no changes due to XALDA approximation are expected.

## 2. $D_{5h}$-$[UO_2(H_2O)_5]^{2+}$ *in Vacuo and with COSMO model*

Tab. 3 demonstrates the shift of excitation energies for $D_{5h}$-$[UO_2(H_2O)_5]^{2+}$ *in Vacuo* due to water molecule ligation and difference between TD-HF, TD-DFT/XALDA and Generalized Active Space Configuration Interaction (GAS-CI). As excitation energy values in Tab. 2 correspond to ATD-B3LYP without XALDA approximation, it is important, for direct comparison, to subtract $X_v$ or $X_a$ from Tab. 3 from the respective $T_v$ or $T_a$ and consider "B3LYP" as a method. For the SORECP/$\Delta SCF_n$ method (where $n$ = 0 stands for HOMO→LUMO, $n$ = 1 for average-of-configurations HOMO→{LUMO,LUMO+1}) adiabatic excitation energies $T_a$ (5) has been determined as $T_a(\Delta SCF_0)$ = 22263 cm$^{-1}$ and $T_a(\Delta SCF_1)$ = 22591 cm$^{-1}$. After resolving the open-shell excited state configurations by small-scale CI, the energies lowered to $T_a'(\Delta SCF_0)$ = 19470 cm$^{-1}$ and $T_a'(\Delta SCF_1)$ = 19673 cm$^{-1}$.

The geometries $R_{gs}$ and $R_{es}$ (5), (6) used for the electronic ground and excited state equilibria have been determined at the scalar quasi-relativistic ECP/ATD-B3LYP-D3/def-TZVPP level in Turbomole within the following protocol:

1. Ground state optimization with point group restriction to $D_{5h}$ *in Vacuo*

2. Freezing ligated water molecule internal coordinates and scanning PES with respect to U-$O_{yl}$ uranyl bond length and distance between uranium and each ligand U-$O_w$ ($D_{5h}$ assumed) distance.

The respective change of U-$O_w$ in second point has been smaller than scanning sampling period and neglected, for other parameters, please see Tab. 3 (first and third row).

In last three columns (Tab. 3) the All electron results (based on Dirac-Coulomb Hamiltonian [135] and four-component Dirac-Hartree-Fock and Dirac-Kohn-Sham TD-HF and TD-DFT/XALDA [42], [41], [153]) are given for comparison with spin-orbit resolved quasi-relativistic (with pseudo potential) results (values in the first five columns, denoted SORECP in Tab. 3). In All-electron computations, Dyall's triple dzeta basis [154], [155] has been used for all elements, integrals[20] (SS|SS), have been approximated [156] and the shift with respect to the Dirac-Coulomb-Gaunt Hamiltonian[21] [135] has been added, yet computed on Dyall's double dzeta [154], [155] basis set level only[22]. For $T_a$ the aforementioned shift is accounting for - 682 cm$^{-1}$ (TD-HF), - 93 cm$^{-1}$, -98 cm$^{-1}$ and -98 cm$^{-1}$ for TD-DFT/XALDA with functionals B3LYP, CAM-B3LYP, LB94 and LB$\alpha$ respectively. The (SS|SS)-integrals-approximated [156] $T_a$ decreased with respect to Dyall.v2z→Dyall.v3z basis set change by 150 cm$^{-1}$ for TD-HF and between 70 cm$^{-1}$ to 100 cm$^{-1}$ for all DFT functionals used. However, for the ground state energy in equilibrium geometry alone, the respective shifts have been -16 and -14 *thousands* cm$^{-1}$.

---

[20] where S stands for the smaller two components of four-component bispinor
[21] with all (SS|SS) integrals taken accurately
[22] The triple-dzeta all electron computations are in progress.

| | SORECP | | | | | All electron | | |
|---|---|---|---|---|---|---|---|---|
| Method | $T_v$ | $T_{de}$ | $T_a$ | $X_v$ | $X_a$ | $T_v$ | $T_{de}$ | $T_a$ |
| HF | 19195 | 17454 | 21128 | | | | 16883 | 20480 |
| B3LYP | 19969 | 18119 | 18983 | 1395 | 1337 | 19595 | 17763 | 18513 |
| **CAM** | **20752** | **18974** | **20556** | **1154** | **1110** | **20346** | **18587** | **20054** |
| **LB94** | **21417** | **19558** | **20446** | **1370** | **1141** | **21036** | **19202** | **19973** |
| LBα | 21969 | 20092 | 20993 | 1364 | 1301 | 21506 | 19656 | 20444 |
| GASCI[23] | 24325 | 21391 | 23571 | | | | | |
| **Exp.-H** | | | **20120** | | | | | |

**Table 3:** Excitation energies ($T_v$ vertical, $T_{de}$ vertical in excited state equilibrium geometry, $T_a$ adiabatic), the XALDA part ($X_v$ for vertical $T_v$, $X_a$ for adiabatic $T_a$) and All electron results for $D_{5h}$-$[UO_2(H_2O)_5]^{2+}$ *in Vacuo*. The "Exp.-H" stands for the experimental $T_{00}$ = (20485 ± 20) $cm^{-1}$ (see Tab. 6) minus hydration shift $H^{SF}_{vac}(T_a)$ = (420 ± 140) $cm^{-1}$ and ΔZPE = (-57 ± 240) $cm^{-1}$ determined for TT3 settings (Tab. 1, resulting uncertainty should be 280 $cm^{-1}$). All values are in $cm^{-1}$.

In Subsection 3 of Computationals it is mentioned that LB94 and LBα asymptiotic corrections have been applied to B3LYP (within XALDA approximation in Dirac) functionals and, e.g., LBα/B3LYP is further simplified in LBα in tables and text. For the $D_{5h}$-$[UO_2(H_2O)_5]^{2+}$ *in Vacuo*, comparison with a "more traditional" SAOP correction (with SORECP pseudopotential), the LBα/LDA and LB94/LDA is given, resulting in $T_a$(LBα/LDA) = 19648 $cm^{-1}$ and $T_a$(LB94/LDA) = 18702 $cm^{-1}$. Both LBα/LDA and LB94/LDA underestimates luminescence transition energy (equivalent $T_{a,exp}$ = 20 120 $cm^{-1}$), however the former only sligthly, while the latter similarly to B3LYP.

As the molecular orbitals directly involved into luminescence phenomenon are to a greater extend localized on the uranyl central group, following correlation molecular orbital diagrams show bare uranyl ($UO_2^{2+}$) frontier molecular orbital levels (Fig. 7), together with a $D_{5h}$ point group model of $[UO_2(H_2O)_5]^{2+}$ (the PES true global minimum for aquo complex *in Vacuo* is of $C_2$ symmetry, with ligands slightly rotated from perpendicular position to the equatorial coordination plane, but as the ground electronic state energy difference is just 70 $cm^{-1}$, the $D_{5h}$ model will be favourized for qualitative investigation)

[23] Generalized Active Space Configuration Interaction with SORECP, in Dirac software. Out of total 96 electrons, 14 have been active in spaces with 4,7,2 and 20 spin orbitals. Minimum 6, 11 and 13 electrons have been accumulated below given active space. Before active space generation (but after Hartree-Fock equations solution), spin orbitals have been permuted according to (1..41,43,44,45,47,42,46,48,49,51,52,54,50,53) new order (Kramers-restricted spin orbitals, one number correspond to the whole pair) – the reasoning has been based on the largest TD-DFT amplitudes for the orbitals permuted into lower active spaces. Total 18 995 803 Slater determinants have been used for A irreducible representation of $C_2$ computational subgroup of $D_{5h}$ and 18 995 306 for B irrep. states. The particular GAS choice hasn't been extensively optimized, but it is prospect for future study.

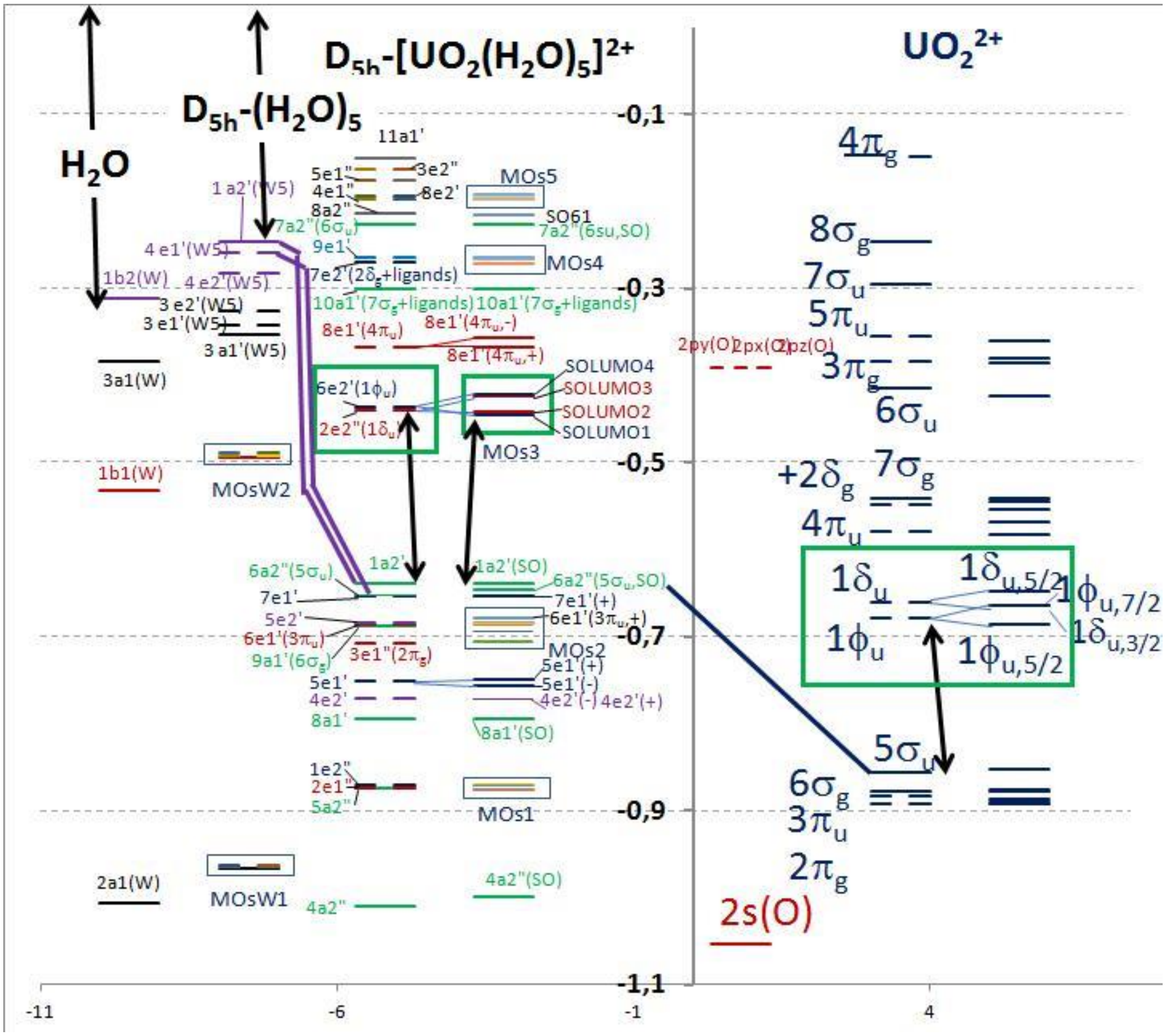

**Fig. 7:** Kohn-Sham molecular orbital diagram (RECP/ B3LYP, Turbomole-def-TZVPP basis set), the columns from left to right correspond to water molecule ($H_2O$) ligand, five ligand fragment, the complex $D_{5h}$-$[UO_2(H_2O)_5]^{2+}$ (SF), the complex, but with spin-orbit splitting (SORECP included [143]-[146]), oxygen atom (red levels right after vertical axis) and $UO_2^{2+}$ central group (the left part correspond to SF and right part to SO computation, spin-orbit splitting of $1\phi_u$ LUMO and $1\delta_u$ LUMO+1 molecular orbitals of $UO_2^{2+}$ into $1\phi_{u,5/2} << 1\delta_{u,3/2} < 1\phi_{u,7/2} << 1\delta_{u,5/2}$ is highlighted by green rectangle). Vertical energy axis is in Hartree atomic units.

The luminescence active excited state, a $E_1$“ (corresponding to $1_g(^3\Delta_g)$ for bare uranyl) has transition probability to X $A_1$‘ ground state non-zero within quadrupole order only (Fig. 8), similarly, the lowest lying excited state deexcitation is dipole-forbidden for $C_{2v}$* case (the $C_{2v}$ geometry, preferred for $[UO_2(H_2O)_5]^{2+}$ in COSMO modelled aqueous solution is presented in Fig. 9). However, the dipole order transition probability rates for ($C_1$) geometries distorted by the presence of water solvent has been by 7 orders of magnitude larger than quadrupole order contribution for the same geometry (using formulae derived in [157]).

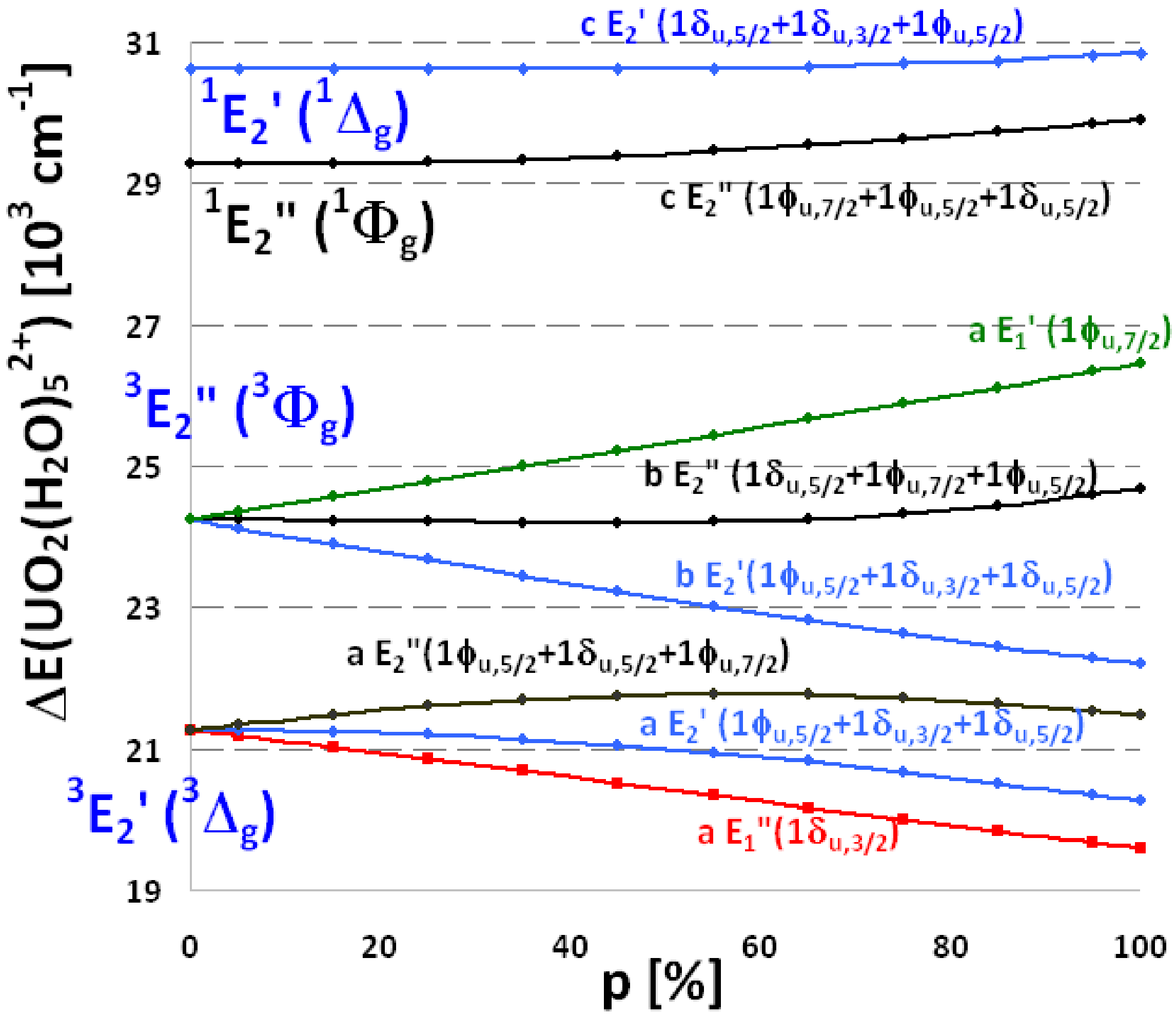

**Fig. 7:** Spin-scaling diagram for aquo complex $D_{5h}$ model (as in Fig. 6), lines corresponding to spin-orbit resolved states are equipped with $D_{5h}$* double-group irreducible representation labels and dominant virtual orbital contributions.

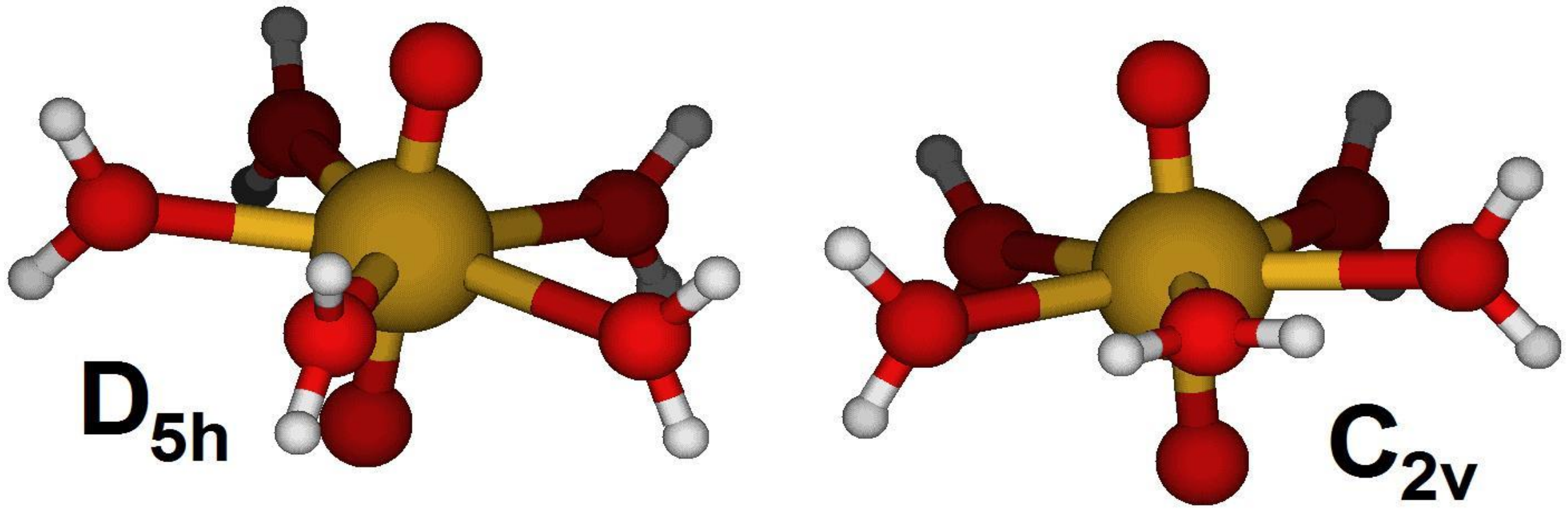

**Fig. 9:** Two most stable configurations for $[UO_2(H_2O)_5]^{2+}$ complex species, corresponding to the $D_{5h}$ (left) point group and $C_{2v}$ point group (right).

The TD-DFT computations in Tab. 2, 4 and for spin-scaling diagrams in Fig. 5-7 have been done without XALDA, used in all other simulations, except specified.

## *3. Uranyl aquo complex in aqueous solution, excitation energy*

### 3.1. Continuum model, COSMO

Hydration shift for $[UO_2(H_2O)_5]^{2+}$ phonon-less excitation energy $T_{00}$ can be estimated by COSMO model ($\varepsilon$ = 78.1) as +491 $cm^{-1}$ for $D_{5h}$ model and scalar quasi-relativistic B3LYP-D3/def-TZVPP description. Influence on equilibrium geometry and uranyl group vibrational frequencies has been estimated with COSMO model (Tab. 4).

| state | envir. | $U\text{-}O_{yl}$ | $U\text{-}O_w$ | O-H | HOH | $E_{min}\text{-}E_{ref}'$ | $\omega_{sym}$ | $\omega_{asym}$ | $\omega_{bend}$ | OUO |
|---|---|---|---|---|---|---|---|---|---|---|
| X | In Vac. | 174.1 | 249.2 | 96.7 | 106.9 | 0.308 | 945.2 | 1029.1 | 238 | 179.9 |
| X | W | 175.5 | 244.2 | 96.5 | 107.1 | 0.000 | 911.0 | 959.7 | 244-327 | 177.6 |
| $T_1$ | In Vac. | 178.7 | 249.8 | 96.7 | 106.9 | 0.397 | 835.7 | 822.9 | 211-328 | 179.2 |
| $H_2O$ | In Vac. | | | 96.1 | 104.9 | | | | | |
| $H_2O$ | W | | | 96.3 | 104.4 | | | | | |

**Table 4.** Non-constrained geometry optimization results for the ground (X) and first excited ($T_1(^3E_2')$) electronic states, *In Vacuo* values (In Vac.) compared for X with values from COSMO model (W). Ground state of water molecule *In Vacuo* and In COSMO ("water in water continuum") equilibrium geometry supplemented. $U\text{-}O_{yl}$ is U-O bond length inside the uranyl group, $U\text{-}O_w$ is an average bond length uranium-oxygen from water ligand, O-H is an average bond length O-H in coordinated water ligands (or single water molecule, for $H_2O$ rows), all in pm. HOH is an average bending angle of water molecule, OUO bending angle of uranyl central group, all in degrees (°). $\omega_{sym}$, $\omega_{asym}$ and $\omega_{bend}$ are symmetric and anti-symmetric stretching and bending vibrational frequencies of the central $UO_2^{2+}$ group in $cm^{-1}$. $E_{min} - E_{ref}'$ is energy level in Hartree (a.u.), with reference value $E_{ref}'$ = -1009.509283 a.u.

As aqueous solution experimental value ($\omega_{sym}$ = (870 ± 20) $cm^{-1}$) for symmetric stretching mode vibrational frequency differ, the COSMO model isn't accurate enough for quantitative spectroscopic study. On the other side, explicit inclusion of solvent water molecules does improve computational results on both structural and vibrational properties [6,57] and has been preferred in spectral simulation here as well.

### 3.2. Discretized solvent model data

Following table summarize adiabatic ($T_a$) and phonon-less ($T_{00}$) excitation energies (the former for TT3 settings (Tab. 1) also separately for point charges-modelled solvent ($T_a^{(pch)}$) and with explicit solvent inclusion shift ($T_a^{(pch,hyd)}$)).

| | | TT3 | | | | | TT1 | | | |
|---|---|---|---|---|---|---|---|---|---|---|
| | $T_a^{(pch)}$ | | $T_a^{(pch,hyd)}$ | | $T_{00}$ | | $T_a^{(pch,hyd)}$ | | $T_{00}$ | |
| Method | $<X>_T$ | $\sigma_T$ | $<X>_T$ | $\sigma_T$ | $<X>_T$ | $\sigma_T$ | $<X>_T$ | $\sigma_T$ | $<X>_T$ | $\sigma_T$ |
| B3LYP | 17.25 | 0.12 | 17.41 | 0.11 | 17.29 | | | | | |
| .XALDA | 18.49 | 0.14 | 18.63 | 0.13 | 18.58 | 0.34 | 18.53 | 0.29 | 18.38 | 0.29 |
| CAM | 19.08 | 0.10 | 19.25 | 0.10 | 19.12 | | | | | |
| **.XALDA** | **20.18** | **0.11** | **20.33** | **0.10** | **20.27** | **0.30** | **20.23** | **0.22** | **20.07** | **0.22** |
| **LB94** | **20.15** | **0.10** | **20.29** | **0.11** | **20.24** | **0.30** | **20.22** | **0.20** | **20.06** | **0.20** |
| **LBα** | **20.64** | **0.11** | **20.79** | **0.11** | **20.73** | **0.31** | **20.70** | **0.22** | **20.55** | **0.22** |
| **TDHF** | **20.77** | **0.84** | **20.92** | **0.09** | **20.85** | **0.30** | **20.83** | **0.17** | **20.67** | **0.17** |
| $\Delta SCF_0$ | 21.43 | | 21.88 | | 21.76 | | | | | |
| $\Delta SCF_1$ | 21.28 | | | | | | | | | |
| *PPM* | 20.56 | 0.36 | 21.02 | 0.47 | 21.03 | | | | | |
| ***Exper.*** | | | | | **20.49** | **0.07** | | | **20.49** | **0.07** |

**Table 5:** Excitation energies for $[UO_2(H_2O)_5]^{2+}$ in aqueous solution. $<X>_T$ stands for average over chosen snapshots (37 for TT1 and 7 for TT3 settings, Tab. 1) from CMD trajectory, $\sigma_T$ is the respective standard deviation. For definitions of $T_a^{(pch)}$, $T_a^{(pch,hyd)}$ and $T_{00}$ see eq. (5), (3) and (4) respectively.

The ".XALDA" row under B3LYP and CAM(-B3LYP) functional gives comparison for inclusion of approximation into Cassida equations [23]. For Statistical Average of Orbital Potentials (SAOP) with asymptotic potentials LB94 and LBα (functional is B3LYP), [38]-[41] the .XALDA option in DIRAC has been activated as well. $\Delta SCF_0$ stands for HOMO→LUMO, while $\Delta SCF_1$ for HOMO→LUMO+1, *PPM* for Polarization Propagator Method [55], [56], here used with cut-off to only 26 virtual orbitals, ***Exper.*** for experimental $T_{00}$ value (average from four different experimental spectra fit, Tab. 6).

The difference between vibrational zero-point-energies for electronic excited and ground state ΔZPE for TT1 settings (Tab. 1) has been computed from four highest frequency modes and accounted for (-156 ± 4) $cm^{-1}$. For TT3 settings the ΔZPE has been computed from 48 modes, but only 7 snapshots with result (-57 ± 240) $cm^{-1}$ clearly reflecting the insufficient statistics.

While correction $H^{SF}_{pch}(T_a)$ (4) counting difference in $T_a$ for explicit and point-charge solvent model in scalar quasi-relativistic case is $H^{SF}_{pch}(T_a)$ = (150 ± 120) $cm^{-1}$ (for TT3 settings, the analogical values for TT1 are (20 ± 130) $cm^{-1}$ and (70 ± 90) $cm^{-1}$ for 12 and 55 explicit solvent water molecules respectively), solvent shift based on difference between $T_a$ for explicit and no solvent inclusion (case where solvent molecules have been completely deleted for subsequent spin-orbit resolved computations) has been (420 ± 140) $cm^{-1}$ (for TT3 settings).

The average difference between methodology computing spin-orbit resolved energies with deletion of solvent and with point-charges has been only 28 $cm^{-1}$, but the latter is certainly more correct with practically no computational overhead.

### 3.3. Peak bimodality, smoothing, graphical comparison to experimental data

As neither the 6, nor 33 snapshots used for whole spectrum computation (Fig. 3) in TT3 and TT1 settings (Tab. 1) respectively were enough for good statistics, histograms with 10 $cm^{-1}$ wide bins (the peak widths deduced from bare uranyl rotationally resolved luminescence spectrum simulation would be 4 $cm^{-1}$) were too noisy for further analysis. Smoothing by discrete convolution with gaussian width variance parameter $\sigma$ has been used. Results for $\sigma$ = 40 $cm^{-1}$ (Fig. 10), $\sigma$ = 100 $cm^{-1}$ (Fig. 11) and $\sigma$ = 200 $cm^{-1}$ (Fig. 12) for selected methods (see Fig. 10 caption) are plotted below together with experimental (done in this study, denoted "(i)" in Tab. 6) luminescence spectrum $I(\omega)$ divided by $(\omega/\omega_0)^3$ ($\omega_0$ = 20 000 $cm^{-1}$) and renormalized to unit maximum value.

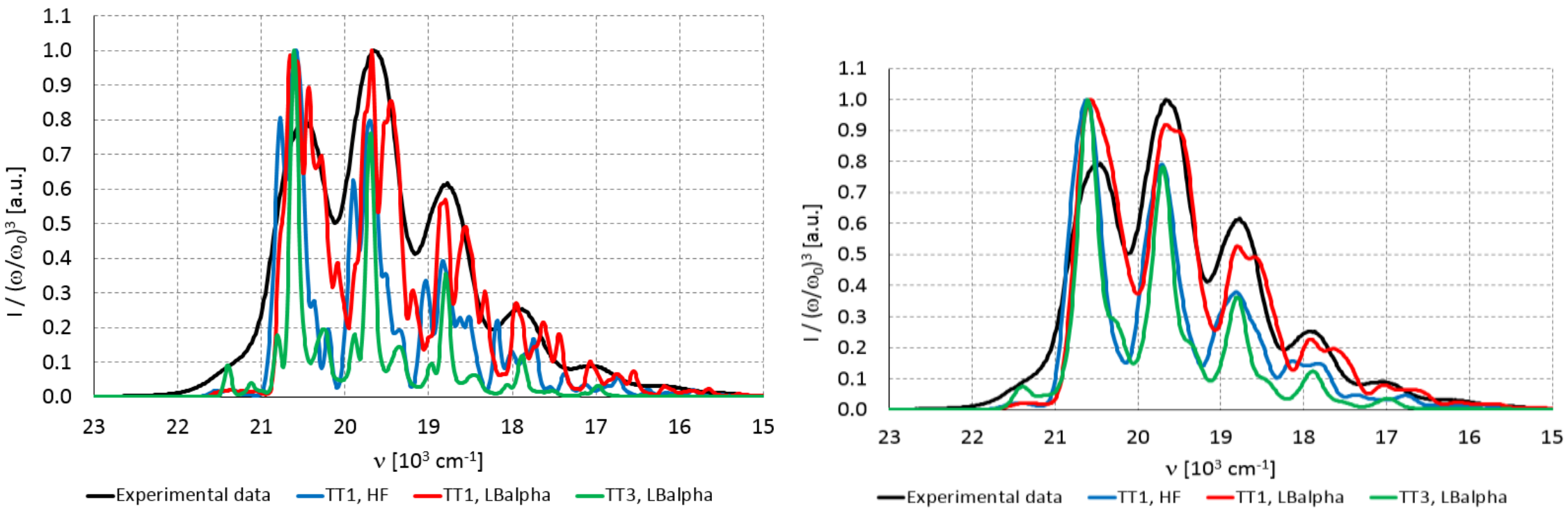

**Fig. 10,11:** The TT1 settings with TD-HF (blue) and TD-LBα (red) and TT3/TD-LBα (green) plotted predict the experimental spectrum (black) most accuartely. Electronic tranistion coefficient change with snapshot haven't been included here. Left (Figure 8) – smoothing parameter $\sigma$ = 40 $cm^{-1}$, right (Figure 9) $\sigma$ = 100 $cm^{-1}$.

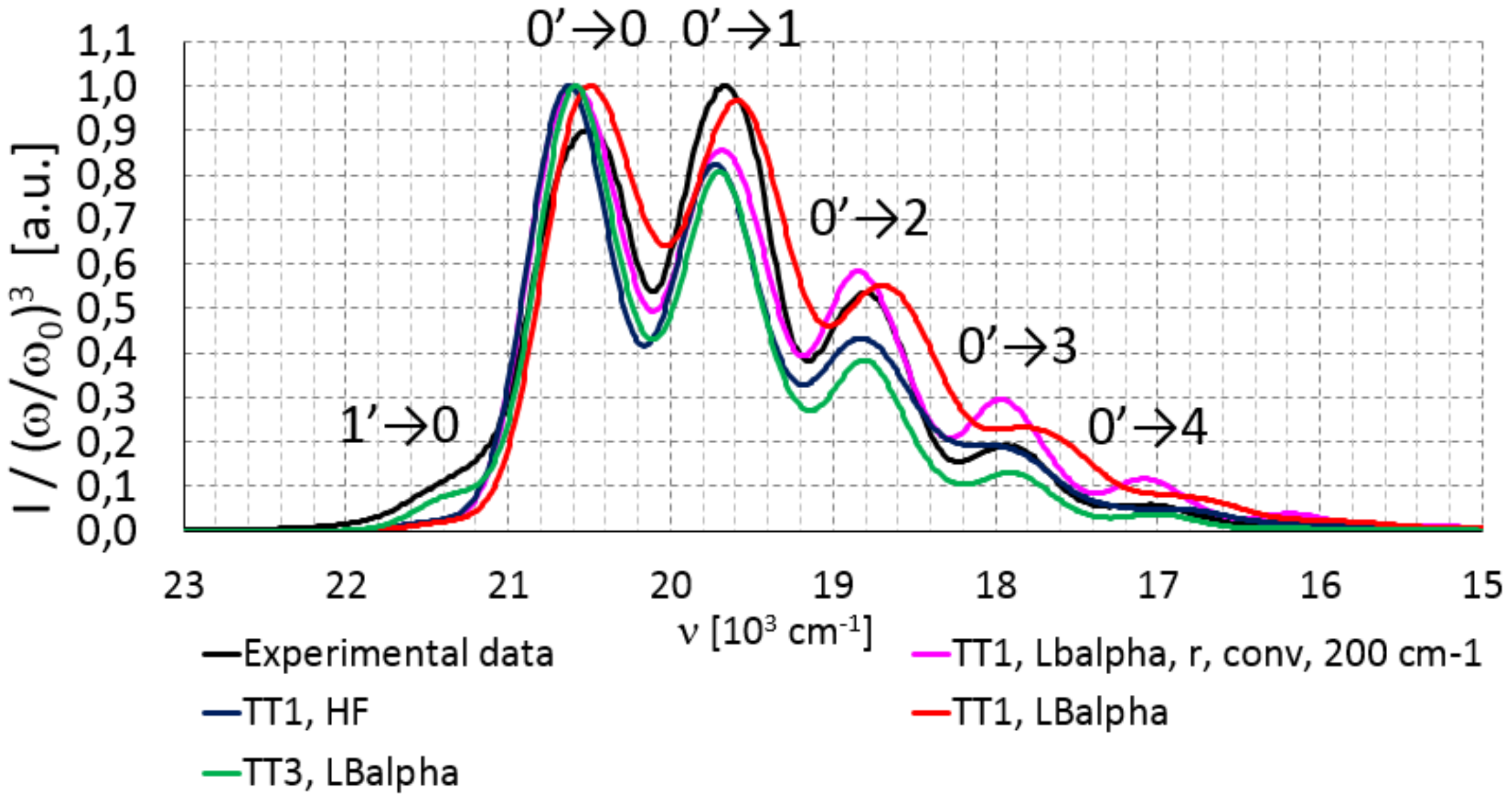

**Fig. 12:** The predictions from Fig. 10, 11 with smoothing parameter $\sigma$ = 200 $cm^{-1}$. TT1/rLBα method (magenta curve) corresponds to $|\mu|^2$-weighted sum over snapshot Franck-Condon profiles ($\mu$ is the electronic transition moment in excited state equilibrium geometry). For logaritmic scale intensity axis see Supplementary Information.

When smoothing parameter $\sigma$ reach the approximate experimental variance parameter for fitted gaussian peaks (~ 200 cm$^{-1}$), the agreement between experimental and theoretically predicted spectra is the best, in particular for the LBα SAOP and TD-HF within TT1 settings (Fig. 12). For CAM-B3LYP or B3LYP functionals, the spectra are noticeably shifted to lower frequencies (as $T_{00}$ values in Tab. 5 suggest) and haven't been plotted.

For comparison an approximation where "average shape" spectrum (computed from Franck-Condon profiles from all snapshots, yet with one fixed $T_{00}$ energy) is convoluted with $T_{00}$ histogram (which is of interest alone as well, Fig. 14) has been compared with experimental spectrum (Fig. 13).

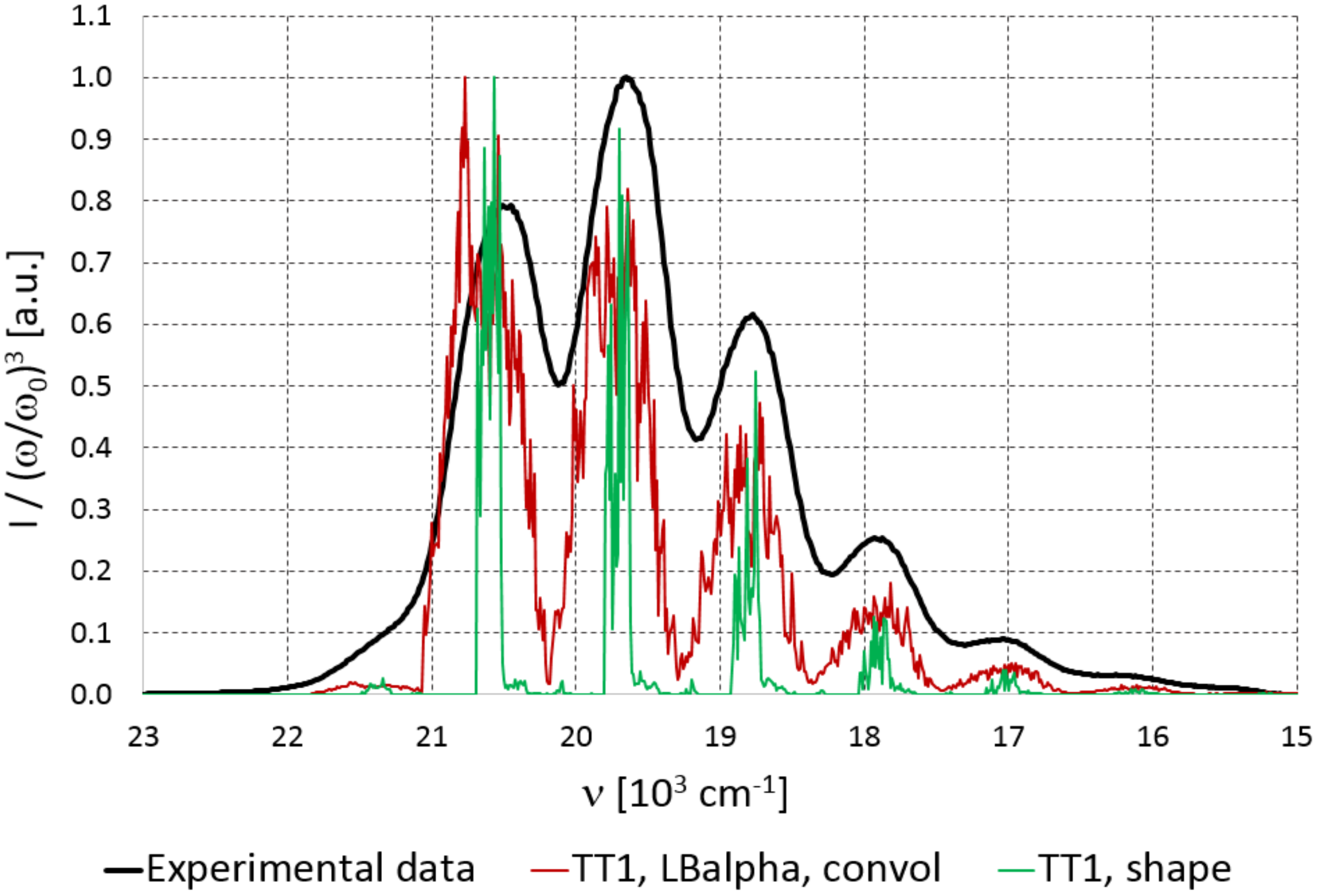

**Fig. 13:** Spectrum theoretical estimate based on averaged spectral shape (with TT1 settings, green) and convolution with $T_{00}$ histogram from snapshots (red) for TD-LBα

The peak splitting for TT1 settings (Tab. 1), visible for smaller smoothing parameter $\sigma$ in Fig. 10 and 13, in lower energy part of spectra in Fig. 9 and in red curve in Fig. 13 is a computational artefact, not present in any experimental spectra and steams from classical (not quantized) description of nuclear motions in $N_2$ and $N_3$ part of studied system. As the classical harmonic oscillator has bimodal distribution of positions, this translates into bimodal distribution of $T_{00}$ (Fig. 14) and any other studied quantity.

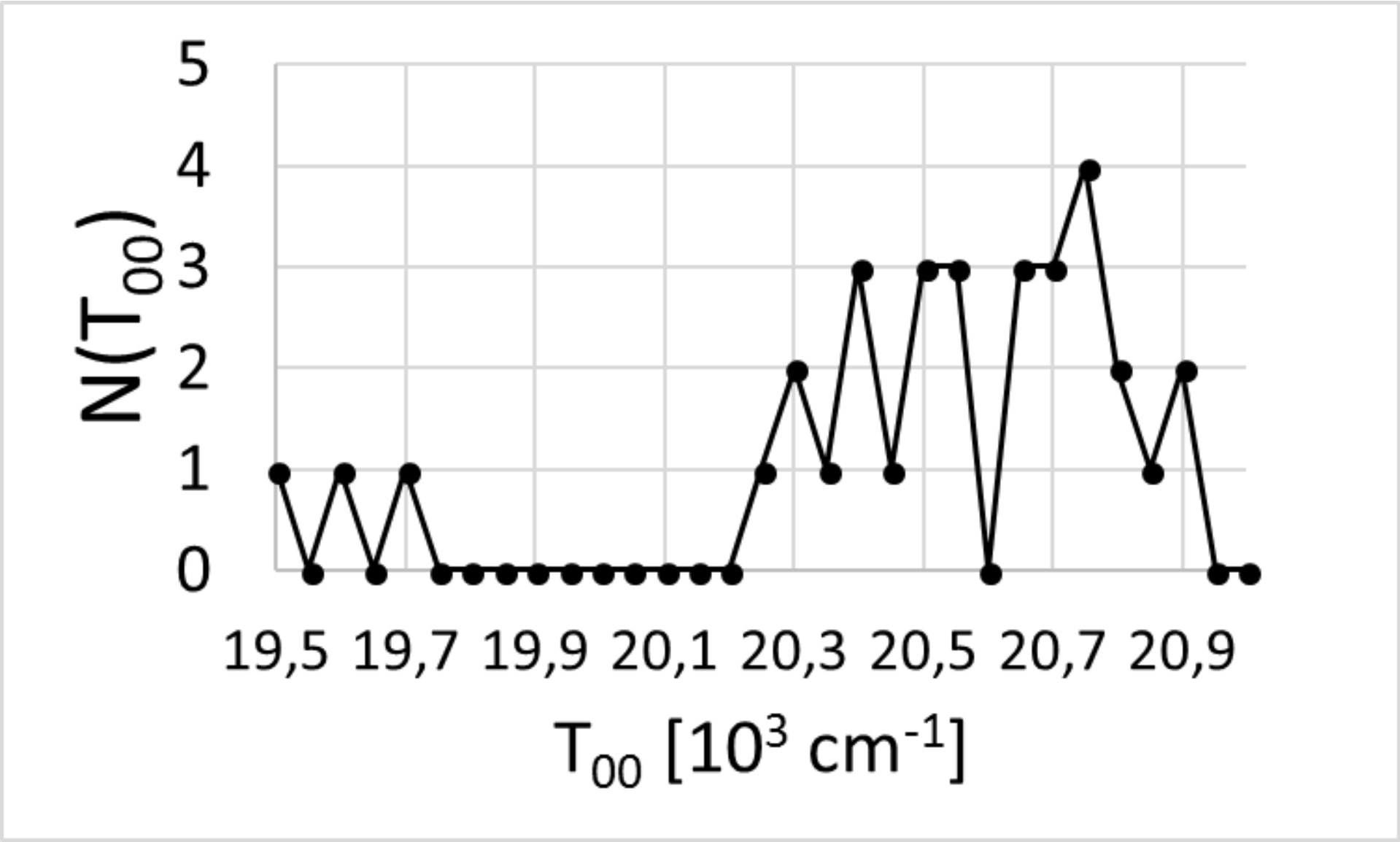

**Fig. 14:** Bimodal shaped histogram of phonon-less excitation energies, $T_{00}$, for LBα/TT1, wide bin size 50 $cm^{-1}$ has been used for plotting purposes here only.

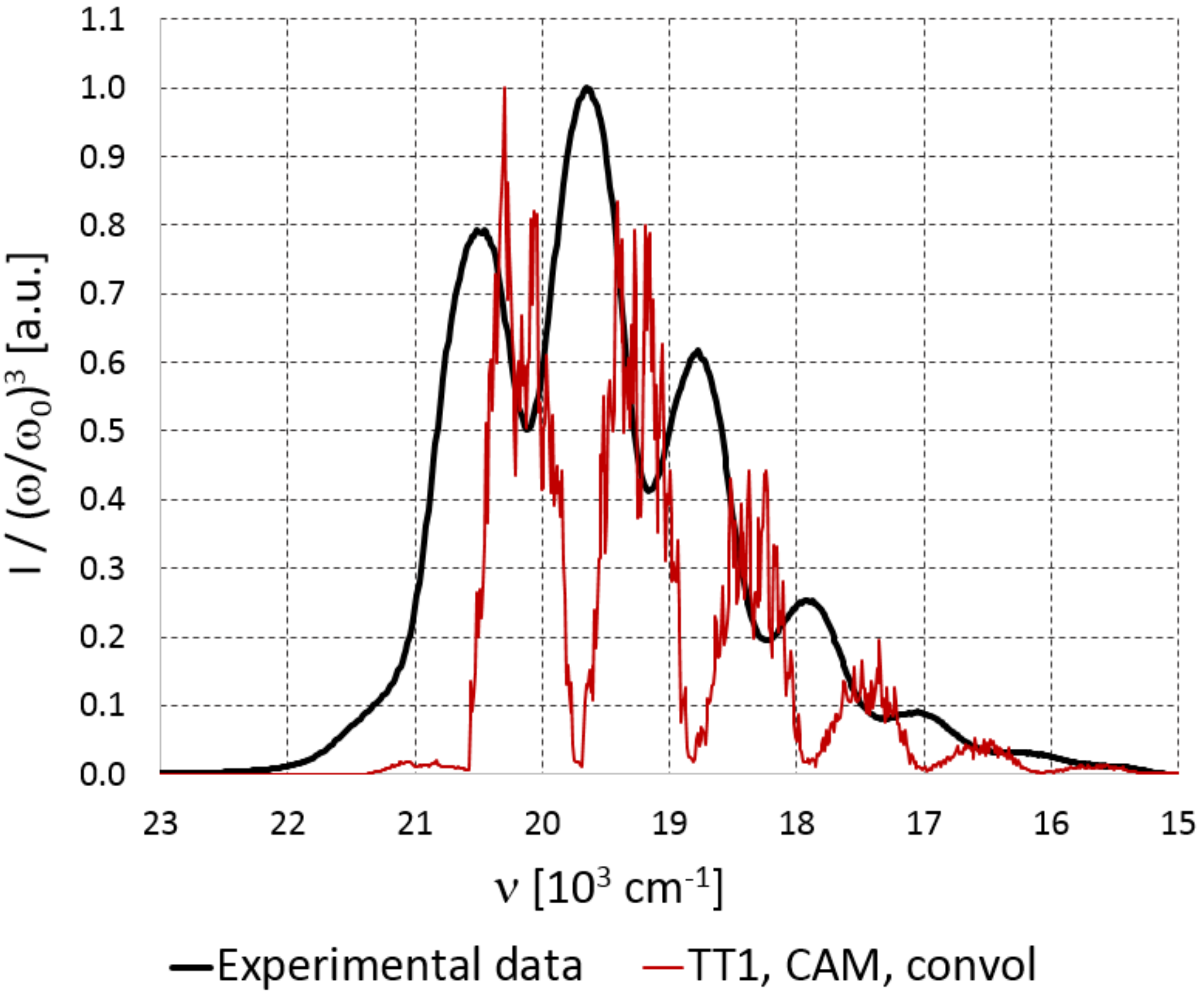

**Fig. 15:** Spectrum theoretical estimate based on averaged spectral shape (with TT1 settings (Tab. 1)) and convolution with $T_{00}$ histogram from snapshots (red) for TD-CAM-B3LYP.

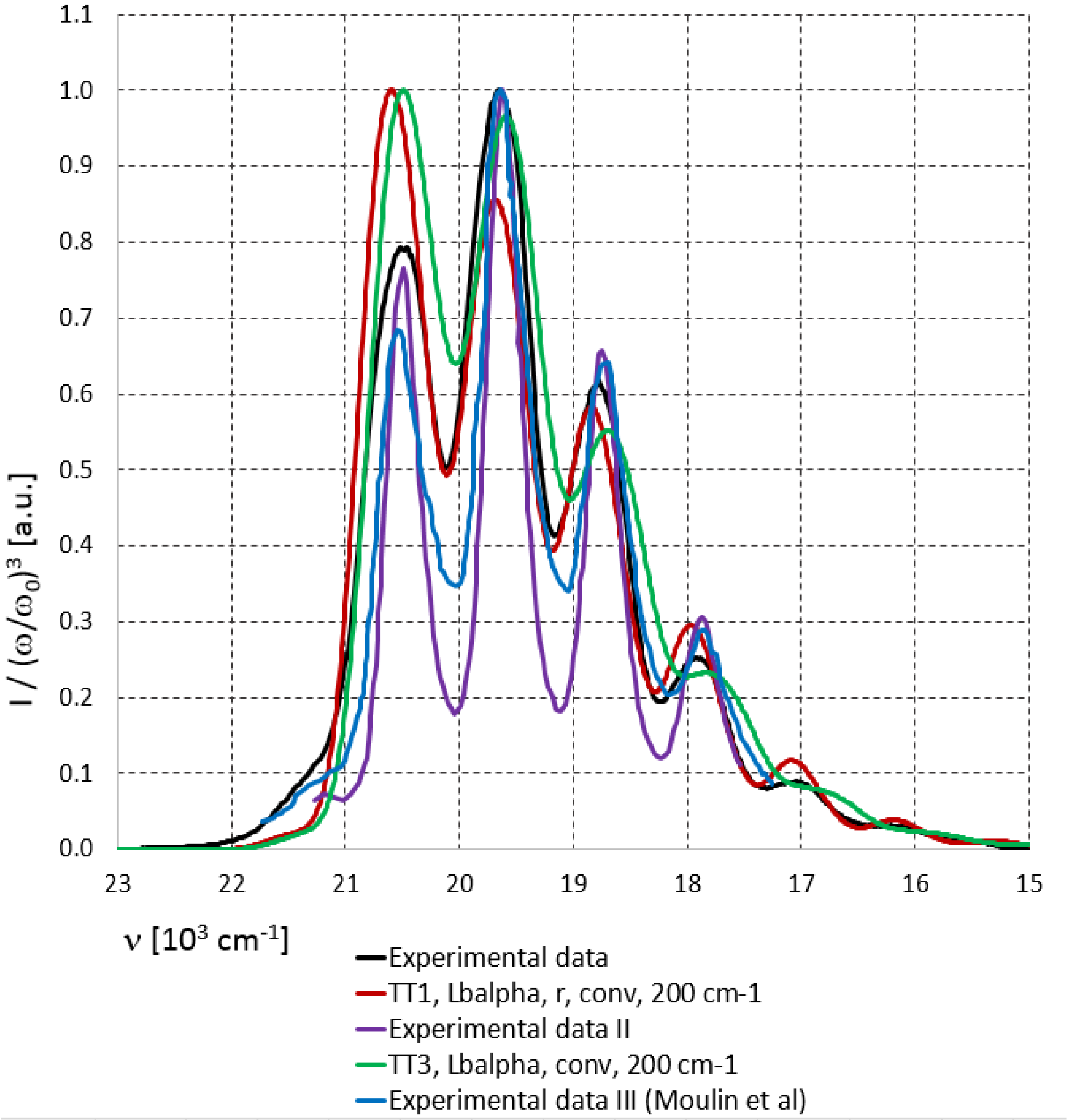

**Fig. 16:** Spectral predictions (TT1/TD-LBα) with inclusion of electronic transition coefficients variation among snapshots. Smooting parameter $\sigma$ = 200 cm$^{-1}$ (TT3: green, TT1 with electronic transition coefficient variation included: red).

The electronic transition moment variation among snapshots ($\mu = \mu(R)$) does significantly affect spectrum envelope shape with result for smaller $\Delta R$ (more rapid decrease of peak intensity with the peak number, red curve in Fig. 16, method "rLBα" in Tab. 6) – in comparison to simulation with strict $\mu = \mu_0$ condition do.

Without smoothing, the peak splitting for $\mu = \mu(R)$ case is even stronger as bimodality is promoted to both $T_{00}$ and $\mu$ distributions. On the other side, the TT3 settings simulations aren't affected by this artefact as the $N_1$ region is larger. However, they are affected by poorer snapshot statistics, so the aforementioned benefit cannot be fully appreciated in this study.

### 3.4. Numerical comparison to experimental data

The following Tab. 6 summarize only main peak position ($T_{00,eff}$), peak spacing ($\omega_{gs,eff}$) and Franck-Condon profile parameter $\Delta R_{eff,n}$ all derived from 7 (or 5 in case of (ii)-(iv) experimental spectra) individual gaussian peak parameters (peak maxima $\nu_{max,j}$, peak maximum intensity $I_j$ and peak width parameter $\sigma_j$, for $j = -\Delta n = -1, 0, 1 \ldots 5$ (index interval choice denoting difference in vibrational quantum numbers $\Delta n = n - n'$)). For the individual peak parameters, from which above mentioned spectroscopic parameters have been derived, please see the Supplementary Information.

| | | | $T_{00,eff}$ | $\delta(T_{00})$ | $\omega_{gs,eff}$ | $\delta(\omega_{gs})$ | $\Delta R_{eff,3}$ | $\Delta R_{eff,2}$ | $\Delta R_{eff,1}$ |
|---|---|---|---|---|---|---|---|---|---|
| settings | ref | method \ unit | $cm^{-1}$ | $cm^{-1}$ | $cm^{-1}$ | $cm^{-1}$ | pm | pm | pm |
| TT1 | | HF | 20615 | -40 | 882 | -6 | 9.91 | 9.37 | 7.49 |
| | | **LBα** | **20506** | **-45** | **919** | **31** | **11.43** | **11.1** | **11.98** |
| | | **rLBα** | **20573** | **22** | **874** | **-14** | **10.27** | **9.88** | **9.73** |
| | | LB94 | 20003 | -28 | 894 | 6 | 10.82 | 10.49 | 9.63 |
| | | CAM | 20014 | -50 | 842 | -46 | 9.15 | 7.86 | 7.91 |
| | | B3LYP | 18357 | -46 | 873 | -15 | 9.21 | 7.79 | 6.42 |
| **TT3** | | **LBα** | **20583** | **-72** | **891** | **-19** | **12** | **11.66** | **11.04** |
| | | LB94 | 20107 | -56 | 895 | -15 | 11.98 | 11.61 | 10.91 |
| | | CAM | 20129 | -60 | 894 | -16 | 11.99 | 11.64 | 10.95 |
| | | B3LYP | 18399 | -67 | 890 | -20 | 12.01 | 11.64 | 11.07 |
| Exper | i | JV[24] | 20511 | | 863 | | 10.48 | 10.07 | 9.21 |
| | ii | AV,norm[25] | 20474 | | 865 | | 5.94 | 6.21 | 6.26 |
| [158] | iii | Moulin et. al. | 20490 | | 875 | | 6.00 | 6.26 | 6.45 |
| [10] | iv | JV fS | 20466 | | 863 | | 6.20 | 6.49 | 6.56 |
| | | **Average** | **20485** | | **867**[26] | | **7.2** | **7.3** | **7.1** |
| | | Stand.dev. | 20 | | 6 | | 2.2 | 1.9 | 1.4 |

**Table 6:** Spectroscopic parameters of simulated and experimental luminescence spectra from gaussian peak fit. $\delta(X) \equiv X_{eff} - X$ (where quantity $X$ has been determined as median – almost identical to the average – over optimization-based analyzed CMD trajectory snapshots) and $\Delta R_{eff,n}$ stands for $\Delta R_{eff}$ fit from first $n$+1 peak area. rLBα stands for SAOP LBα corrected B3LYP functional and "r" for weighting individual snapshot spectra by electronic transition probability determined in excited state equilibrium geometry.

[24] Our measurement, ionic strength $I_m = 1.0$ mol·kg$_w^{-1}$, $pH = 2$, total unranium concentration $c_U = (1.0 \pm 0.1)\cdot 10^{-4}$ mol·dm$^{-3}$, temperature $t = (25.0 \pm 0.1)$ °C.

[25] Aleš Vetešník, personal communication, ionic strength $I_m = 1.0$ mol·kg$_w^{-1}$, $pH = 2$, total unranium concentration $c_U = (1.0 \pm 0.1)\cdot 10^{-6}$ mol·dm$^{-3}$ , temperature $t = (25.0 \pm 0.1)$ °C. The experimentals have been identical as for this article.

[26] Literature [98] experimental value is 870 cm$^{-1}$ [57]. Bell [159] reports $T_{00} = (20\,502.0 \pm 0.7)$ cm$^{-1}$ for fifth peak (Tab. III, fluorescence spectrum, ionic strength = 3 mol·kg$_w^{-1}$, $t = 28$°C) and $\omega_{gs} = 868$ cm$^{-1}$ spacing towards fourth peak (0'→1), by linear fit of four peaks of I(ν)/ν$^3$ function of -25 cm$^{-1}$ and -11 cm$^{-1}$ for $T_{00}$ and $\omega_{gs}$ can be observed.

Experimental luminescence spectra used for comparison in Tab. 6 and Fig. 16 are referenced by i-iv, where "i" accounts for measurement done in this study for a rather high ionic strength $I_m$ = 4.0 mol·kg$_W^{-1}$, "ii" for the measurement performed sooner with similar experimental set up, but for a lower ionic strength by A. Vetešník (see footnote 25) with correction to "machine transfer function" (most contribution due to the ICCD detection response variation with wave-number) derived from [158] luminescence spectrum referred as "iii". The last, "iv", spectrum has been adopted from $UO_2^{2+}$ - $SO_4^{2-}$ - $H_2O$ TRLFS speciation study ($I_m$ = 0.3 mol·kg$_W^{-1}$, $t$ = 18°C). It is probable that shifts in „i“ with respect to the other spectra are due to the higher ionic strength and to a lesser extend due to the omitted "machine transfer function" correction[27] or temperature difference (other spectra correspond to $t$ = 25°C).

To complete data in Tab. 6, frequency (peak spacing) parameter $\omega_{gs}$ determined from normal mode statistics along CMD sampling trajectory have been (926 ± 7) cm$^{-1}$ for TT1 (and $\omega_{gs}$ = (889 ± 7) cm$^{-1}$ after scaling by 0.96 factor [173] applied here) settings and $\omega_{gs}$ = (910 ± 6) cm$^{-1}$ for TT3 settings (scaling by factor 0.96 [173], not provided for this settings, would lead to (874 ± 5) cm$^{-1}$). As $\delta(T_{00})$ and $\delta(\omega_{gs})$ corresponds to a rather small and in most cases negative fraction, for a preliminary insight on this chemical system, $T_{00}$ and $\omega_{gs}$ determination from statistical data, yet avoiding Franck-Condon profile computation (in the TT3 settings (Tab. 1) very time-consuming) gives a good correlate for an "effective" $T_{00,eff}$ and $\omega_{gs,eff}$ parameters (determined by simulated spectra shift) comparable directly to the values determined by experimental luminescence spectra fit. That is not a case for $\Delta R$, where snapshot statistics provide (4.56 ± 0.07) pm for TT1 and (4.55 ± 0.09) pm for TT3.

### 3.5. The excitation-related U-O elongation, ΔR

The $\Delta R$ underestimation of $\Delta R_{eff}$ by 31% to 164% (according to particular determination and simulated spectrum) is due to the larger set of normal modes contributing to each peak intensity. The individual $\Delta R$ determined for vibrational modes of ligand-uranyl bonds, uranyl bending, anti-symmetric stretch, ligand localized modes or even modes connected to the solvent are at least of order of magnitude smaller than $\Delta R$ for uranyl symmetric stretching mode each, but together (there are 48 vibrational modes for $[UO_2(H_2O)_5]^{2+}$) matter. The simple summation over modes (n.b. Duschinsky effect [77]-[81]) cannot account for a correct $\Delta R_{eff}$ , this parameter has to be determined through fitting of (by means of Franck-Condon profile computation) simulated spectra.

The $\Delta R_{eff,n}$ values have been determined from area under the peaks, but evaluation based on peak maxima follows similar trends (for comparison TT1/rLBα peak maxima based $\Delta R_{eff,n}$ values are 8.89 pm, 9.50 pm and 9.83 pm for $n$ = 1, 2 and 3 respectively, for TT1/HF 10.25 pm, 10.91 pm and 11.27 pm and for TT3/LBα 10.11 pm, 11.17 pm and 11.64 pm, for experimental data "iii" $\Delta R_{eff,n}$ fitted from peak maxima intensities are 6.92 pm, 6.73 pm and 6.37 pm for $n$ = 1, 2 and 3 respectively).

### 3.6. Hot band and $\omega_{es}$ parameter

Aside up to six cold-bands, 0'→$n$ ($n$ = 0, 1, ..., 5) with average spacing $\omega_{gs}$, uranyl(VI) luminescence spectrum feature one hot-band 1'→0 (violet curve in Fig. FSI5 and Fig. FSI6 in Supplementary Information) differing by one quantum in the uranyl group symmetric stretching mode in the excited electronic state (i.e. the corresponding vibrational frequency is denoted $\omega_{es}$). From luminescence spectrum recorded and to individual gaussian peak decomposed by Bell [159], [160] $\omega_{es}$ = (768 ± 72) cm$^{-1}$ can be deduced. Four experimental spectra presented here as (i)-(iv) corresponds to much lower average $\omega_{es}$ = (676 ± 143) cm$^{-1}$ (uncertainty estimated as half of range). Among four most accurate spectral simulations – TT1/HF, TT1/LBα, TT1/rLBα and TT3/LBα only the last one provide $\omega_{es,eff}$ (determined by gaussian spectrum fit and counting band maxima difference for $\omega_{es,eff}$) comparable with experimental value, $\omega_{es,eff}$ = (772 ± 10) cm$^{-1}$ (for), others overestimate the quantity by 80, 180 and 180 cm$^{-1}$ (for TT1/HF, TT1/LBα and TT1/rLBα respectively, with respect to 768 cm$^{-1}$ determined from [159]). From normal mode analysis in excited electronic state, for TT3 settings $\omega_{es}$ = (804 ± 8) cm$^{-1}$ (from 9 snapshots) and for TT1 settings $\omega_{es}$ = (819 ± 6) cm$^{-1}$ (from 60 snapshots, after 0.96 scaling factor reduction, $\omega_{es}$ = (786 ± 6) cm$^{-1}$). For vibrational frequency in excited electronic state absorption spectra (which are also well vibrationally resolved by symmetric stretching mode for most aqueous uranyl(VI) complex species) should be used instead as there this parameter characterize spacing of cold-bands. Unfortunately, the absorption prefers higher excited states and average band spacing refer to their symmetric stretching vibrational frequencies instead.

[27] The (iv) hasn't been corrected and is still almost identical to (ii) (and therefore for simplicity was not plotted in Fig. 14).

## *4. Other complex species - $[UO_2(CO_3)_3]^{4-}$, $[UO_2(\kappa^2\text{-}SO_4)_2(H_2O)]^{2-}$ and $[UO_2(CO_3)_3Mg(H_2O)_n]^{2-}$ excitation energy*

In order to prove predictability of our uranyl complex luminescence spectra simulation protocol, it should be applied to set of different uranyl complex species. First steps in this direction have been made for bis(sulphate), tris(carbonate) complex and ternary magnesium tris(carbonate) complex of uranyl(VI). Instead of luminescence spectra, we present estimation for spectroscopic parameters $T_{00}$ (Tab. 7), $\omega_{gs}$ and $\Delta R$ (Tab. 8) from which can be the aforementioned spectra reconstructed (assuming the peak widths to be $\sigma$ = 200 cm$^{-1}$ and that $T_{00}^{eff} = T_{00}$, $\omega_{gs}^{eff} = \omega_{gs}$, $\Delta R^{eff} = \Delta R$). The computation protocol followed Fig. 3 except that CMD sampling has been omitted (p1 and p12 missing), and we seek large $N_1$ at least in the ground state optimization (to match amorphous ice phase in cryo-TRLFS experiments)

| $[UO_2(CO_3)_3]^{4-}$ | | | | |
|---|---|---|---|---|
| Method | $T_v^{(vac)}$ | $T_v^{(hyd)}$ | $T_a^{(hyd)}$ | $T_{00}^{(hyd)}$ |
| B3LYP | 19.56 | 18.94 | 18.35 | 18.20 |
| .XALDA | 21.44 | 20.81 | 19.46 | 19.31 |
| CAM | 20.53 | 19.91 | 20.02 | 19.85 |
| **.XALDA** | **21.56** | **20.94** | **21.00** | **20.85** |
| **LB94** | **22.21** | **21.58** | **20.95** | **20.80** |
| **LBα** | **23.32** | **22.70** | **21.33** | **21.18** |
| TDHF | 20.65 | 20.02 | 21.78 | 21.63 |
| **Exper.1** | | | | **20.80** |

**Table 7:** Theoretical and experimental values for excitation energies ($T_v$ vertical, $T_a$ adiabatic, $T_{00}$ phonon-less) *In Vacuo* (vac) and including hydration (hyd).

For above mentioned species ($[UO_2(CO_3)_3]^{4-}$, $[UO_2(\kappa^2\text{-}SO_4)_2(H_2O)]^{2-}$ and $[UO_2(CO_3)_3Mg(H_2O)_n]^{2-}$), the complex in question has been pre-optimized in ground state *in Vacuo*[28], 55 explicit water molecules have been added through Packing software [S7], hydration pre-optimized within (classical) Universal Force Field (UFF) in Turbomole [S8] (with central complex atomic position frozen) and subsequently optimized in ground electronic state with scalar quasi-relativistic ECP/B3LYP-D3/def-SVP in Turbomole[29] [S8]. From ground electronic state equilibrium geometry the lowest lying excited electronic state geometry optimization has been started with following constrains[30]:

1. for $[MgUO_2(CO_3)_3]^{2-}$ with all atomic positions aside to uranyl group frozen
2. for $[UO_2(CO_3)_3]^{4-}$ with solvent water molecule positions fixed
3. for $[UO_2(\kappa^2\text{-}SO_4)_2(H_2O)]^{2-}$ without freezing limitations

For the central complex (in case of $[MgUO_2(CO_3)_3]^{2-}$ supplemented by four water molecules[31]) spin-orbit resolved SORECP/ATD-CAM-B3LYP/def-TZVPP computation in Dirac [S1] has been performed for both ground and excited electronic state equilibria to determine main part of $T_{00}$ (2). For this determination the surrounding (i.e. those not ligated to any metal atom) water molecules have been represented as point charges only.

The above mentioned geometry optimization procedue in the ground electronic state has been repeated four times and structure with lowest SORECP/TD-CAM-B3LYP/XALDA/def-TZVPP determined electronic ground state energy of the central complex of interest has been chosen for a subsequent analysis.

To account for solvent interaction more accurately (4), hydration shift $H^{SF}_{pch}(T_a)$ has been computed from explicit hydrated vs. point-charge modelled solvent within ECP/B3LYP-D3/def-SVP in Turbomole [S8]. ΔZPE has been estimated as -150 $cm^{-1}$ in all cases, value adopted from the previous uranyl aquo complex study (in TT1 settings, section 3 of Results).

[28] except for ternary $[MgUO_2(CO_3)_3]^{2-}$ where 12 waters have been added even in this first phase to assure magnesium position in ligation plane

[29] In all above mentioned steps, more structures have been studied and among them the one with lowest energy have been chosen, the study is still to be considered as a preliminary and auxiliary for the purpose of this article.

[30] Implemented to cut computational time with rather small loss of accuracy as the main structural change between electronic states in question is uranyl-located.

[31] found to be directly coordinated to $Mg^{2+}$ (having octahedral coordination surrounding, see Fig. FSI12 and Fig. FSI13 in Chapter 3 of Supplementary Information, [163]).

| Molecule[32] | $\omega_{gs}$ | $T_{00}$ | $\Delta R$ | $T_{00,exp}$ | $\omega_{gs,exp}$ | $\Delta R_{exp}$ | ref |
|---|---|---|---|---|---|---|---|
| $[UO_2(CO_3)_3]^{4-}$ | 801[33] | 20.85 | 4.3 | 20.66<br>20.80 | 812<br>792[34] | 10.4 | [5]<br>[162] |
| $[MgUO_2(CO_3)_3]^{2-}$ | 820 | 20.90 | 4.4 | 20.78<br>20.63 | 808<br>827 | 10.4 | [5]<br>[162] |
| $[UO_2(SO_4)_2]^{2-}$ | 861 | 20.41 | 4.0 | 20.27<br>20.17 | 854<br>865 | 6.8 | [10]<br>[8] |
| $[UO_2]^{2+}$ | 894 | 20.53 | 4.6 | 20.49<br>20.50<br>20.51 | 875<br>855<br>875 | 6.3<br><br>11.1[35] | [158]<br>[3]<br>[10] |

**Table 8:** Theoretical and experimental values for luminescence spectroscopic parameters for few chosen uranyl complex species (common in natural water samples).

As all values presented in this sub-chapter come from optimization of few-water molecules explicit solvated clusters *in Vacuo* rather than from ($t$ = 25°C) CMD-snapshots (as has been done for uranyl(VI) aquo complex), results for $[UO_2(H_2O)_5]^{2+} \cdot 20\ H_2O$ (ref. L in Tab. 10 in [10]) has been appended for comparison ($T_{00}$ has been determined here with SORECP/TD-CAM-B3LYP/def-TZVPP as for other species complex).

Experimental parameters for first two species listen in Tab. 8 has been derived from cryo-TRLFS spectra rather than those acquired under room temperature to better compare with methodology seeking global energy minimum with respect to atomic coordinates (rather than analyzing CMD snap-shots).

The $T_{00,exp}$ has been determined with one hot-band assumption, as opposed to the two hot-bands preferred in [5] (and therefore here presented value differs from that in Tab. 17 in [5] by one vibrational quantum (~ 800 $cm^{-1}$) ). The one hot-band model for $UO_2^{2+}$ - $CO_3^{2-}$ - $Mg^{2+}/Ca^{2+}$ system luminescence spectra is in better agreement with recent spin-orbit resolved computations. Literature [161], [8], [158] presents peak maxima in nm only, we inverted them in $cm^{-1}$ and provided linear fit, to determine $\Delta R_{exp}$ transferring Figures of spectra to [x;y] and Franck-Condon fit would be needed – hasn't been done yet.

Please see the Chapter 3 in Supplementary Information for further details and discussion.

[32] Water molecules (both ligated and solvent) omitted for brevity.
[33] An average of frequencies (747 $cm^{-1}$ and 854 $cm^{-1}$) of two modes with significant contribution of symmetric stretching coordinate.
[34] Fit from [162] results in $\omega_{gs}$ = (792 ± 10) $cm^{-1}$ (and $T_{00}$ = (20 800 ± 20) $cm^{-1}$), but $\omega_{gs}$ ~ 812 $cm^{-1}$ from PARAFAC determined (individual components assigned to concentration profiles geochemically modelled in PHREEQC [5]) is consistent with Ikeda et al theoretical prediction (810.2 $cm^{-1}$ [164]).
[35] Values determined from different experimental spectra vary (also see Tab. 6). The most probable reason would be different ionic strength.

# Discussion and Conclusions

## *1. Summary*

Simulation protocol for vibrationally resolved luminescence spectra based on Franck-Condon profile computation (within double-harmonic approximation) with CMD trajectory sampled configuration space and TD-DFT (with XALDA approximation) has been successful in terms of qualitative agreement with the experimental spectral data for $[UO_2(H_2O)_5]^{2+}$, the model uranyl(VI) aquo complex (Fig. 16, Tab. 6). The most accurate functional have been from Statistical Average of Orbital Potential (SAOP) group, namely LB$\alpha$ asymptotically corrected B3LYP (in particular, when variation of electronic transition probability in excited state equilibrium with snapshot number is accounted for).

The asymptotically uncorrected B3LYP functional provides good spectral shape ($\Delta R$ and $\omega_{gs}$ parameters in a good agreement with experiment), but seriously underestimate phonon-less excitation energy $T_{00}$. The reason behind this should be the partial charge-transfer character of corresponding electron transition localized at uranyl central group (HOMO($\sigma_u$) $\leftarrow$ LUMO(U$f_{\delta,3/2}$)), B3LYP is known to underestimate the charge-transfer excitation energies due to -0.2 $r^{-1}$ asymptotics of exchange energy with respect to electron-electron distance $r$. The SAOP or CAM- correction improves situation. The complete exclusion of electron correlation in TD-HF gives almost as good description as the LB$\alpha$ variant (but is less reliable for other related systems, such as $D_{5h}$-$[UO_2(H_2O)_5]^{2+}$ *in Vacuo* (Tab. 3)).

Spin-orbit splitting has been included via suitable Spin-Orbit Relativistic Effective Core potential (SORECP, [143]-[146]) as included in the DIRAC (version 17) quantum chemistry package [S1]. The correction accounts for up to -1400 $cm^{-1}$ excitation energy shift ($UO_2^{2+}$, Results, Chapter 1), and almost cancel out with an important ~ +1300 $cm^{-1}$ terms from XALDA approximation, therefore the former, scalar quasi-relativistic studies [5,10] give almost correct values. The inclusion of spin-orbit interaction would allow estimating radiative contribution to luminescence life-times – for a preliminary study with future prospects, please see Chapter 4 in Supplementary Information.

The LB94/B3LYP and CAM-B3LYP functionals led to a less accurate, but still interesting results and the latter have been tested for simulation protocol predictivity on $[UO_2(CO_3)_3]^{4-}$, $[MgUO_2(CO_3)_3]^{2-}$, $[UO_2(SO_4)_2]^{2-}$ and $[UO_2]^{2+}$ (water molecules omitted in formulae, cluster optimization instead of CMD sampling has been used in the simulation protocol in this case, thus comparable better to a cryo-TRLFS data) complex species – the lowering of $\omega_{gs}$ with both $SO_4^{2-}$ and $CO_3^{2-}$ ligation and decrease of $T_{00}$ in former case and increase in latter case (all with respect to the aquo complex) has been qualitatively explained by CAM-B3LYP based simulation protocol as well.

The agreement between CAM-B3LYP-XALDA-based $T_{00}$ and $T_{00,exp}$ as well as the agreement between ATD-B3LYP-based $\omega_{gs}$ and $\omega_{gs,exp}$ in Tab. 8 is good, taking in account that rather the “effective” values of $T_{00,eff}$ and $\omega_{gs,eff}$ (based on experience with aquo complex, they are smaller by few to few tens of $cm^{-1}$, see $\delta$(.) quantities in Tab. 6), should be compared with experimental ones. The CAM-B3LYP has been chosen here as it is the best performing functional for $D_{5h}$-$[UO_2(H_2O)_5]^{2+}$ *in Vacuo* (Tab. 3), system studied when first computation on other species has been started.

The dependence on particular DFT functional and breakdown of any DFT-based simulation for multi-reference systems (e.g. excited state equilibrium geometries of uranyl *mono*sulphate or *mono*carbonate) lead us to think about future post-HF *ab initio* – based simulation protocol variants. As studied systems are large and individual species spectra usually differ very little – powerful approach would be needed to cover the respective computational demands.

## 2. The $\Delta R$ and $\sigma$ spectral parameters

To complete spectral information, $\Delta R$ and peak widths $\sigma_j$ should be discussed. The average peak width (193 $cm^{-1}$ Exper.(ii) – 245 $cm^{-1}$ Exper.(iv), three peaks considered, see Tab. TSI1.3 in Supplementary Information) correlate well to $T_{00}$ distribution variance in TT1 settings ($\sigma_T$ in last column of Tab. 3). Presented simulated spectra (Fig. 10-13, 15, 16) had to be artificially smoothed by discrete convolution with gaussian peak profile (the splitting resulting from artificial $T_{00}$ distribution bimodality (see Chapter 3.3 of Results section, Fig. 14) needs smoothing parameter up to $\sigma = 200$ $cm^{-1}$). Therefore, peak widths in simulated spectra can be attributed to the smoothing convolution instead. However, the previously mentioned correlation to $\sigma_T$ parameter provide solid hope that in case of simulation based on a larger set of snapshots (hundreds at least) would provide peak widths with lesser or no inference of smoothing convolution parameters. The $\Delta R_{eff}$ parameter has been slightly overestimated by all methods except for TD-HF and B3-LYP (which, however gives wrong value of a much more important $T_{00,eff}$ parameter). This and the artificial peak splitting problems ask for an improvement not only in the electronic part of simulation procedure, but also in further Franck-Condon profile part.

## 3. Hot band and eventual higher excited electronic states

Interesting feature of uranyl(VI) complex luminescence spectra is one (rather wide) hot-band located at ≈ 21 100 $cm^{-1}$ (474 nm). According to [14], 4% of total luminescence intensity can be assigned to this peak. Simulated spectra underestimate (Fig. 16, note the tiny, blue-most peak in all spectra) hot-band peak area. For $[UO_2(H_2O)_5]^{2+}$ in water, the de-excitation rates for higher electronic states are by factor 3 to 5 higher than for the lowest-lying and their energy is around 250 $cm^{-1}$ (SORECP/TD-DFT/TT1 settings) – 1200 $cm^{-1}$ (SORECP/TD-HF/TT1 settings) higher than the lowest excited state. Moreover, the energy difference between lowest excited electronic state and the first higher lying excited state have been determined in the a $^3\Delta_g$ equilibrium geometry (ECP/TD-B3LYP predicted) by post-HF *ab initio* methods – PPM and a preliminary study with FS-CCSD provided (890 ± 40) $cm^{-1}$ difference between the two lowest lying excited electronic states (four snapshots, TT3 settings, excited state eq. geometry) – value close to the uranyl group stretching vibrational frequency, i.e. peak spacing. GASCI with generalized active spaces specified in Tab. 3 predicts 989 $cm^{-1}$. Aforementioned electronic level spacing value suggests possibility that part of luminescence intensity in "hot-band" peaks comes from an ordinary cold-band 0'→0 transition, but from higher excited electronic state. Similarly, the "missing intensity" in the following peaks (0'→*n*) might come from 0'→(*n*+1) transitions from the higher excited electronic state. In future, we plan to extend simulation to multiple excited electronic states to test this hypothesis in detail.

While the TT1 settings led to a slightly more accurate (theory vs. experiment) simulation (Tab. 6, e.g., the agreement in $\Delta R_{eff}$, but with the help of smoothing of simulated spectra, which removes peak splitting), we believe that TT3 settings has potential to be improved beyond this level of agreement and would not suffer from peak splitting. Also, for the $\omega_{gs,eff}$ agreement TT3 settings need not the scaling factor of 0.96 to be applied (in contrast to TT1). That is because in TT3 settings larger number of vibrational modes are treated quantum-mechanically. Ligand connected modes are, however, less harmonic and therefore simulation should extend beyond double-harmonic approximation (first approach would be based on third and fourth order energy derivatives with respect to normal modes, then Vibrational SCF or even Vibrational post-HF methods and connected Franck-Condon profile evaluation would be used). Suggested extensions are a future prospect.

## 4. Further prospects

Moreover, peaks in experimental luminescence spectra show systematic deviations from gaussian shape and it would be an interesting challenge for theoretical simulation to explain them. Presented simulated spectra have gaussian shape forced by smoothing function (and raw simulated spectra (Fig. 10, 11 and 13) are too noisy for detailed peak shape discussion due to poor statistics (only 33 and 6 snapshots for TT1 and TT3 settings respectively)), but future increase in analyzed snapshot total number and/or extensions mentioned in previous paragraph would avoid the need for large smoothing parameter and provide insight to true peak shapes.

## Acknowledgements

We would like to thank Lukasz Cwiklik, PhD, DSc. (J. Heyrovský Institute of Physical Chemistry, CAS) for helpful hints and the CMD computation, to the Department of Nuclear Chemistry for support (including the possibility for TRLFS measurements) and the Czech Student Grant Agency (grant no. SGS16/250/OHK4/3T/14) for material support.

We woud like to greatly thank all other co-authors of [5] - Dr. Robin Steudtner and Dr. Andrea Kassahun for discussions. Similarly, we would like to thank to Dr. Trond Saue, Doc. Mgr. Jiří Pittner, Dr. rer. nat., DSc., doc. Ing. Pavel Soldán, PhD., Mgr. Jindřich Kolorenč, PhD. and Mgr. Aleš Vetešník, PhD.

Computational resources were provided by the CESNET LM2015042 and the CERIT Scientific Cloud LM2015085, provided under the programme "Projects of Large Research, Development, and Innovations Infrastructures.

## Software used

**[S1] DIRAC 16,** a relativistic *ab initio* electronic structure program, Release DIRAC16 (2016), written by L. Visscher, H. J. Aa. Jensen, R. Bast, and T. Saue, et al. http://diracprogram.org/doku.php

**[S2] ezSpectrum,** V.A. Mozhayskiy and A.I. Krylov, ezSpectrum, http://iopenshell.usc.edu/downloads, This work was conducted using the resources of the iOpenShell Center for Computational Studies of Electronic Structure and Spectroscopy of Open-Shell and Electronically Excited Species (http://iopenshell.usc.edu) supported by the National Science Foundation through the CRIF:CRF program.

**[S3] GROMACS,** Pronk, S., Pall, S., Schulz, R., Larsson, P., Bjelkmar, P., Apostolov, R., Shirts, M. R., Smith, J. C., Kasson, P. M., van der Spoel, D., Hess, B., Lindahl, E. GROMACS 4.5: a high throughput and highly parallel open source molecular simulation toolkit. *Bioinformatics* **29**(7): 845–854, (2013). http://www.gromacs.org/.

**[S4] MATLAB,** Release 8.7, The MathWorks, Inc., Natick, Massachusetts, United States. https://www.mathworks.com/products/matlab.html.

**[S5] Molden,** Schaftenaar G., Vlieg E., Vriend G.,"Molden 2.0: quantum chemistry meets proteins", *J Comput Aided Mol Des* (2017) **31**: 789; Schaftenaar G., Noordik J.H., "Molden: a pre- and post-processing program for molecular and electronic structures", *J Comput Aided Mol Design*, **14** (2000) 123-134. http://www.cmbi.ru.nl/molden/.

**[S6] Newton-X**, A package for Newtonian dynamics close to the crossing seam, http://www.newtonx.org/.

**[S7] Packing,** J. M. Martinez, L. Martinez, *J. Comput. Chem.*, **24**:819-825,(2003).

**[S8] Turbomole V7.1**, a development of University of Karlsruhe and Forschungszentrum Karlsruhe GmbH, 1989-2007, TURBOMOLE GmbH, since 2007; available from http://www.turbomole.com.

**[S9] Wolfram Mathematica**, Wolfram Research, Inc., Mathematica, Version 10.0.1, Champaign, IL (2014). http://www.wolfram.com/?source=nav.

## References

[1] Osman, A.A.A. (2014) Investigation of Uranium Binding Forms in Environmentally Relevant Waters and Bio-fluids. Ph.D. thesis, TU Dresden, 124pp.

[2] Lakowicz R. J.: Principles of Fluorescence Spectroscopy, Third Edition, Springer, 2006.

[3] Moulin C., Laszak I., Moulin V., Tondre C., Time-Resolved Laser-Induced Fluorescence as a Unique Tool for Low-Level Uranium Speciation, Applied Spectroscopy, Vol. 52, No. 4, 1998, p. 528-535.

[4] Wang, Z., J. M. Zachara, W. Yantasee, P. L. Gassman, C. X. Liu, and A. G. Joly. (2004) Cryogenic Laser Induced Fluorescence Characterization of U(VI) in Hanford Vadose Zone Pore Waters. Environmental Science & Technology 38:5591-5597.

[5] Višňák J., Steudtner R., Kassahun A., Hoth N.: Multilinear analysis of Time-Resolved Laser-Induced Fluorescence Spectra of U(VI) containing natural water samples, EPJ Web of Conferences, Volume 154, p. 01029 (2017), DOI: 10.1051/epjconf/201715401029.

[6] Meinrath G., Aquatic Chemistry of Uranium, Review Focusing on Aspects of Environmental Chemistry, Freiberg On-line Geoscience Vol.1., http://www.geo.tu-freiberg.de/fog/FOG_Vol_1.pdf .

[7] Geipel G., Brachmann A., Brendler V., Bernhard G., Nitsche H., Uranium(VI) Sulfate Complexation Studied by Time-Resolved Laser-Induced Fluorescence Spectroscopy (TRLFS), Radiochimica Acta 75:199-204.

[8] Vercouter T., Vitorge P., Amekraz B., Moulin Ch., Stoichiometries and Thermodynamic Stabilities for Aqueous Sulfate Complexes of U(VI), Inorg. Chem. 2008, 47, 2180-2189.

[9] Vetešník A., Semelová M., Štamberg K., Vopálka D.: Uranium(VI) sulfate complexation as a function of temperature and ionic strength studied by TRLFS In: Uranium, Mining and Hydrogeolology. Berlin: Springer-Verlag, 2008, p. 623-630. ISBN 978-3-540-87745-5.

[10] Višňák J., Sobek L.: Quantum chemical calculations and spectroscopic measurements of spectroscopic and thermodynamic properties of given uranyl complexes in aqueous solutions with possible environmental and industrial applications, EPJ Web of Conferences, 128, p. 02002, (2016). DOI: http://dx.doi.org/10.1051/epjconf/201612802002.

[11] Y. Yokoyama, M. Moriyasu, S. Ikeda, J. inorg, nucl. Chem., (1976), 38, pp. 132%1333. Pergamon Press.

[12] H. D. Burrows, S. J. Formosinho, M. G. Miguel, F.P. Coelho, J. Chem. Soc., Faraday Trans. 1, (1976),72, 163-171, 10.1039/F19767200163.

[13] R. Matsushima, S. Sakuraba, J. Am. Chem. Soc., (1971), 93 (26), pp 7143–7145, DOI: 10.1021/ja00755a004.

[14] H. D. Burrows, S. J. Formosinho. J. Chem. Educ, (1978), 55 (2) p 125., DOI 10.1021/ed055p125.

[15] M. E. D. G. Azenha, H. D. Burrows, S. J. Formosinho, M G. M. Miguel, J. Chem. Soc., Faraday Trans. I, (1989), 85(8), 2625-2634.

[16] Hitchcock F. L. (1927). "The expression of a tensor or a polyadic as a sum of products". Journal of Mathematics and Physics 6: 164–189.

[17] Rasmus B.: PARAFAC. Tutorial & applications. Chemometrics Group, Food Technology, Royal Veterninary & Agricultural University. Rolighedsvej 30, III, DK-1958 Frederiksberg C, Denmark, http://www.models.kvl.dk/~rasmus/presentations/parafac_tutorial/paraf.htm.

[18] Geladi P., Analysis of multi-way (multi-mode) data. Chemom. Intell. Lab. Syst., 7 (1989) 11.

[19] Smilde A.K., Three-way analyses. Problems and prospects. Chemom. Intell. Lab. Syst., 5 (1992) 143.

[20] Harshman, R. and Lundy, M. (1984). The PARAFAC model for three-way factor analysis and multidimensional scaling. In Research methods for multimode data analysis, chapter 5, pages 122-215. Praeger, New York.

[21] Harshman, R. and Lundy, M. (1994). PARAFAC: Parallel factor analysis. Computational Statistics and Data Analysis, 18:39-72.

[22] S. Leurgans, R.T. Ross and R.B. Abel, A decomposition for three-way arrays. SIAM J. Matrix Anal. Appl., 14 (1993) 1064.

[23] Ziegler T., Krykunov M., Autschbach J., Derivation of the RPA (Random Phase Approximation) Equation of ATDDFT (Adiabatic Time Dependent Density Functional Ground State Response Theory) from an Excited State Variational Approach Based on the Ground State Functional, J. Chem. Theory Comput. 2014, 10, 3980-3986.

[24] Hohenberg, Pierre; Walter, Kohn (1964). "Inhomogeneous electron gas". Physical Review. 136 (3B): B864–B871.

[25] Levy, Mel (1979). "Universal variational functionals of electron densities, first-order density matrices, and natural spin-orbitals and solution of the v-representability problem". Proceedings of the National Academy of Sciences. 76 (12): 6062–6065.

[26] Vignale, G.; Rasolt, Mark (1987). "Density-functional theory in strong magnetic fields". Physical Review Letters. 59 (20): 2360–2363.

[27] Hohenberg, P.; Kohn, W. (1964). "Inhomogeneous Electron Gas". Physical Review. 136 (3B): B864.

[28] Runge, Erich; Gross, E. K. U. (1984). "Density-Functional Theory for Time-Dependent Systems". Physical Review Letters. 52 (12): 997.

[29] M.A.L. Marques; C.A. Ullrich; F. Nogueira; A. Rubio; K. Burke; E.K.U. Gross, eds. (2006). Time-Dependent Density Functional Theory. Springer-Verlag. ISBN 978-3-540-35422-2.

[30] Carsten Ullrich (2012). Time-Dependent Density-Functional Theory: Concepts and Applications (Oxford Graduate Texts). Oxford University Press. ISBN 978-0199563029.

[31] A. D. Becke (1988). "Density-functional exchange-energy approximation with correct asymptotic behavior". Phys. Rev. A. 38 (6): 3098–3100. Bibcode:1988PhRvA..38.3098B. doi:10.1103/PhysRevA.38.3098. PMID 9900728.

[32] Chengteh Lee; Weitao Yang; Robert G. Parr (1988). "Development of the Colle-Salvetti correlation-energy formula into a functional of the electron density". Phys. Rev. B. 37 (2): 785–789. Bibcode:1988PhRvB..37..785L. doi:10.1103/PhysRevB.37.785.

[33] S. H. Vosko; L. Wilk; M. Nusair (1980). "Accurate spin-dependent electron liquid correlation energies for local spin density calculations: a critical analysis". Can. J. Phys. 58 (8): 1200–1211. Bibcode:1980CaJPh..58.1200V. doi:10.1139/p80-159.

[34] Becke, Axel D. (1993). "Density-functional thermochemistry. III. The role of exact exchange". J. Chem. Phys. 98 (7): 5648–5652. Bibcode:1993JChPh..98.5648B. doi:10.1063/1.464913.

[35] K. Kim and K. D. Jordan (1994). "Comparison of Density Functional and MP2 Calculations on the Water Monomer and Dimer". J. Phys. Chem. 98 (40): 10089–10094. doi:10.1021/j100091a024.

[36] P.J. Stephens, F. J. Devlin, C. F. Chabalowski and M. J. Frisch (1994). "*Ab Initio* Calculation of Vibrational Absorption and Circular Dichroism Spectra Using Density Functional Force Fields". J. Phys. Chem. 98 (45): 11623–11627. doi:10.1021/j100096a001.

[37] Yanai T., Tew D.P., Handy N.C., Chemical Physics Letters, 393, 51–57, (2004).

[38] Bernasconi L., The Journal of Chemical Physics 132, 184513 (2010).

[39] Gritsenko O.V. et al, Chemical Physics Letters, 302, (1999), 199–207.

[40] Gritsenko O.V. et al, International Journal of Quantum Chemistry, Vol. 76, 407-419 (2000).

[41] Schipper P.R.T. et al, The Journal of Chemical Physics 112, 1344 (2000); doi: 10.1063/1.480688.

[42] Saue T., Helgaker T., Four-component relativistic Kohn-Sham theory. Journal of Computational Chemistry, 23(8):814–823, 2002. URL: http://dx.doi.org/10.1002/jcc.10066, doi:10.1002/jcc.10066.

[43] Olsen J.M.H., Steinmann C., Rudd K., Kongsted J., J. Phys. Chem. A2015, 119, 5344−5355, DOI: 10.1021/jp510138k.

[44] Boereboom J.M., Fleurat-Lessard P., Bulo R.E., J. Chem. Theory Comput.2018, 14, 1841−1852.

[45] Brancato G., Rega N., Barone V., THE JOURNAL OF CHEMICAL PHYSICS128, 144501 2008.

[46] Warshel A., Karplus M., J. Am. Chem. Soc., 1972, 94, 5612–5625.

[47] Warshel A., Levitt M., J. Mol. Biol., 1976,103, 227–249.

[48] Gao J., Acc. Chem. Res., 1996, 29, 298–305.

[49] Lin H., Truhlar D., Theor. Chem. Acc., 2007, 117, 185–199.

[50] Senn H. M., Thiel W., Angew. Chem., Int. Ed., 2009, 48, 1198–1229.

[51] Tomasi J., Persico M., Chem. Rev., 1994, 94, 2027–2094.

[52] Cramer C. J., Truhlar D. G., Chem. Rev., 1999,99, 2161–2200.

[53] Orozco M., Luque F. J., Chem. Rev., 2000, 100, 4187–4226.

[54] Tomasi J., Mennucci B., Cammi R., Chem. Rev., 2005, 105, 2999–3093.

[55] Pernpointner M., The relativistic polarization propagator for the calculation of electronic excitations in heavy systems. The Journal of Chemical Physics, 140:084108, 2014. URL: http://dx.doi.org/10.1063/1.4865964, doi:10.1063/1.4865964.

[56] Pernpointner M., Visscher L., Trofimov A.,B., Influence of Spin-Orbit Coupling on Electronic Transition Moments obtained by the Four-component Polarization Propagator. Journal of Chemical Theory and Computation, to be published, 000:000000, 2017. URL: http://, doi:xxx.

[57] B. Siboulet et al., *Chemical Physics* **326** (2006) 289–296.

[58] Hagberg D., Karlstroem G., Roos B.O., Gagliardi L., J Am Chem Soc 2005, 127, 14250-14256.

[59] Li B., Matveev A.V., Krüger S., Rösch N., Computational and Theoretical Chemistry 1051 (2015) 151–160.

[60] Cao Z., Balasubramanian K., THE JOURNAL OF CHEMICAL PHYSICS123, 114309 2005.

[61] Drobot B., Steudtner R.,Raff J.,Geipel G.,Brendler V.,Tsushima S.: Chem. Sci., 2015, 6, 964.

[62] Rodríguez-Jeangros N., Seminario J.M., J Mol Model (2014) 20:2150.

[63] D. Vopálka et al, Report to the final control day with respect to the contract between Radioactive Waste Repository Authority (RAWRA) and Czech Technical University (FNSPE) (Phase 4) (n. 4007016), Prague, July 2009 (Chapter 2.2 (author: A. Vetešník), Chapter 2.3 (J. Višňák), Chapter 3 (J. Šebera)) (In Czech).

[64] Tsushima S., Uchida Y., Reich T., A theoretical study on the structures of UO2(CO3)34-, Ca2UO2(CO3)30, and Ba2UO2(CO3)30, Chem Phys Lett 357 (2002) 73-77.

[65] Tirler A.O., Hofer T.S., Structure and dynamics of the uranyl tricarbonate complex in aqueous solution: insights from quantum mechanical charge field molecular dynamics., J Phys Chem B. 2014 Nov 13;118(45):12938-51. doi: 10.1021/jp503171g.

[66] Marchenko A., Truflandier L.A., Autschbach J., Uranyl Carbonate Complexes in Aqueous Solution and Their Ligand NMR Chemical Shifts and 17O Quadrupolar Relaxation Studied by *ab Initio* Molecular Dynamics, Inorg Chem, 2017 Jul 3;56(13):7384-7396. doi: 10.1021/acs.inorgchem.7b00396.

[67] Tirler A.O., Weiss A.K., Hofer T.S., A comparative quantum mechanical charge field study of uranyl mono- and dicarbonate species in aqueous solution., J Phys Chem B. 2013 Dec 19;117(50):16174-87. doi: 10.1021/jp407179s.

[68] van Wezenbeek E.M., Dissertation, Free University of Amesterdam. 1991.

[69] Cornehl, H.H., Heinemann, Ch., Marcalo, J., Pires de Matos, A. & Schwarz, H. (1996): The 'Bare' Uranyl(2+) Ion, UO2 2+, *Angew. Chemie. Int. Ed. Engl.*, **35**, 891 – 894.

[70] Bast R., Jensen H. J. AA., Saue T., Relativistic Adiabatic Time-Dependent Density Functional Theory Using Hybrid Functionals and Noncollinear Spin Magnetization, International Journal of Quantum Chemitsry, Vol. 109, 2091-2112 (2009).

[71] Bast R., Quantum chemistry beyond the charge density, PhD thesis, Louis Pasteur University, Strasbourg I, (2008).

[72] Pierloot K., van Besien E.: Electronic structure and spectrum of UO2(2+) and UO2Cl4(2-), J Chem Phys **123**, 204309 (2005).

[73] Pierloot, K.; van Besien, E.; van Lenthe, E.; Baerends, E. J. J Chem Phys 2007, 126, 194311.

[74] Réal, F.; Vallet, V.; Marian, C.; Wahlgren, U. J Chem Phys 2007,127, 214302.

[75] Infante I., Visscher L., QM/MM Study of Aqueous Solvation of the Uranyl Fluoride [UO2F42] Complex

[76] Rabinowitch, E., Belford, R.L., Spectroscopy and Photochemistry of Uranyl Compounds, A Pergamon Press Book, The Macmillan company, New York, 1964.

[77] Kupka H.J., Transitions in Molecular Systems, ISBN:978-3-527-41013-2, (2010), Wiley VCH, ISBN:978-3-527-41013-2.

[78] Duschinsky, F. (1937) On the interpretation of electronic spectra of polyatomic molecules. Translated by Christian W. M€ uller.Acta Physicochim.U.R.S.S., 7, 551.

[79] Craig, P. and Small, G.J. (1969) Totally symmetric vibronic perturbations and the phenanthrene 3400-A spectrum. J.Chem. Phys., 50, 3827.

[80] Small, G.J. (1971) Herzberg–Teller vibronic coupling and the Duschinsky effect.J. Chem. Phys., 54, 3300.

[81] Sharf, B. and Honig, B. (1970) Comments on vibronic intensity borrowing.Chem. Phys. Lett., 7, 132.

[82] Tsushima S., Dalton Trans., 2011, 40, 6732.

[83] Cotton, F.A., G. Wilkinson, and P.L. Gaus. 1987. Basic Inorganic Chemistry, 2nd ed. Wiley, New York, NY. p. 219.

[84] Vopálka et. al, Czech Radioactive Waste Repository Authority (RAWRA) Report: Uranium speciation study in system $UO_2^{2+}$ - $SO_4^{2-}$ by means of the TRLFS method (In Czech).

[85] C. Eckart, G. Young, (1936). 1 (3): 211–8.doi:10.1007/BF02288367.

[86] M. R. Hestenes, (1958). J Soc Ind Appl Math. 6 (1): 51–90. JSTOR 2098862. MR 0092215. doi:10.1137/0106005.

[87] G. H. Golub, W. Kahan, (1965). 2 (2): 205–224. JSTOR 2949777. MR 0183105. doi:10.1137/0702016.

[88] T. W. Anderson, An Introduction to Multivariate Statistical Analysis, (Wiley, New York, 1958).

[89] K.V. Mardia, J.T. Kent, J.M. Bibby (1979). Multivariate Analysis. Academic Press,. ISBN 0124712525. (M.A. level "likelihood" approach).

[90] R. A. Johnson, D. W. Wichern, (2007). Applied Multivariate Statistical Analysis (Sixth ed.). Prentice Hall. ISBN 978-0-13-187715-3.

[91] O. Y. Rodionova, A. L. Pomerantsev, Russ Chem Rev 75 (4) 271-287 (2006).

[92] R. G. Brereton, Applied Chemometrics for Scientists, University of Bristol, Wiley, 2007, ISBN-13: 978-0-470-01686-2.

[93] J. N. Miller, J. V. Miller, Statistics and Chemometrics for Analytical Chemistry, Pearson Edu. L., Sixth edition 2010, ISBN: 978-0-273-73042-2.

[94] M. Hazewinkel, ed. (2001), "Maximum-likelihood method", Encyclopedia of Mathematics, Springer, ISBN 978-1-55608-010-4.

[95] Pomogaev V., Tiwari S. P., Rai N., Goff G. S., Runde W., Schneider W.F., Maginn E.J., Development and application of effective pairwise potentials for $UO_2^{n+}$, $NpO_2^{n+}$, $PuO_2^{n+}$, and $AmO_2^{n+}$ (n = 1, 2) ions with water, Phys. Chem. Chem. Phys., 15, 15954-15963 (2013).

[96] Jorgensen W.L., Chandrasekhar J., Madura J. D., Impey R. W., Klein M. L., J. Chem. Phys., 1983, 79, 296.

[97] Chang J.-L., J. Mol. Spectrosc., 232, 102-104, (2005).

[98] Ismail N., Etude théorique de l'ion uranyle et de ses complexes et dérivés, Université Paul Sabatier de Toulouse, 2000.

[99] Denning R.G., Journal of Luminescence 128 (2008) 1745–1747.

[100] Virgil E. Jackson, Raluca Craciun, and David A. Dixon, Kirk A. Peterson, Wibe A. de Jong, Prediction of Vibrational Frequencies of UO22+ at the CCSD(T) Level, J. Phys. Chem. A2008,112,4095-4099.

[101] Višňák J., Kuba J., Vetešník A., Bok J., In Lectures of colloquium on Radioanalytical methods, IAA 14, (2014).

[102] Deglmann P., Furche F., Ahlrichs R., Chem. Phys. Lett. 362:511 (2002).

[103] Deglmann P., Furche F., J. Chem. Phys. 117:9535 (2002).

[104] Ahlrichs R., Baer M.,Haeser M., Horn H., Koelmel C., Electronic structure calculations on workstation computers: the program system TURBOMOLE, Chem. Phys. Lett. 162: 165 (1989).

[105] Treutler O., Ahlrichs R., Efficient Molecular Numerical Integration Schemes, J. Chem. Phys. 102: 346 (1995).

[106] Born M., Oppenheimer R., Ann. Phys. 84, 457 (1927).

[107] Tennyson J., WIREs Comput Mol Sci 2011. Doi: 10.1002/wcms.94.

[108] N. Kh. Bikbaev, A. I. Ivanov, G. S. Lomakin, and O. A. Ponornarev, Theory of nonradiative processes in the "non-Condon" approximation, Zh. Eksp. Teor. Fiz. 74, 2154-2166 (June 1978). [Sov. Phys. JETP 47(6), June 1978 0038-5646/78/061121-0].

[109] K. Huang and A. Rhys, Proc. R. Soc. London Ser. A 204, 406 (1950).

[110] R. Kubo and Y. Toyozawa, Prog. Theor. Phys. 13, 160 (1955).

[111] Yu. E. Perlin, Usp. Fiz. Nauk 80, 553 (1963) [Sov. Phys. Usp. 6, 542 (1964)]

[112] W. B. Fowler and D. L. Dexter, PhyS. Rev. 128, 2154 (1962).

[113] N. N. Kristofel' and G. S. Zavt, Fiz. Tverd. Tela (Leningrad) 5, 1279 6963) [Sov. Phys. Solid State 5, 932 (1963)].

[114] Joachain C. J., Quantum collision theory, North-Holland Publishing, 1975, ISBN 0-444-86773-2 (Elsevier), p. 340.

[115] Franck J., Transactions of the Faraday Society, 21: 536–542, (1926).

[116] Condon E., Phys Rev, 28: 1182–1201, (1926).

[117] Condon E., Physical Review 32, 858 (1928).

[118] Coolidge A.S., James H.M., Present R.D., J. Chem. Phys, 4: 193–211, (1936).

[119] Atkins P.W., Friedman R.S., Molecular Quantum Mechanics, Oxford University Press, (1999).

[120] Islampour R., et al., J. Mol. Spectrosc. 194, 179–184, (1999).

[121] Malmqvist P.-A., Forsberg N., Chem. Phys., 228, 227–240, (1998).

[122] Rudberg E., Diploma work in Physical Chemistry, Royal Institute of Technology, Sweden (2003).

[123] Barone V., Bloino J., Biczysko M., Santoro F., Fully Integrated Approach to Compute Vibrationally Resolved Optical Spectra: From Small Molecules to Macrosystems, J. Chem. Theory Comput.2009,5,540–554.

[124] Hsue Ch.-S., Chinese Journal of Physics, Vol. 32, No. 6-I, 1994.

[125] Hilborn, R. C. (2002). Einstein coefficients, cross sections, f values, dipole moments, and all that.

[126] Mulder B.J., J Phys Chem Solids, Vol. 29, Is. 1, 182-184, 1968.

[127] van Meer R., Gritsenko O. V., Baerends E. J., Physical Meaning of Virtual Kohn-Sham Orbitals and Orbital Energies: An Ideal Basis for the Description Of Molecular Excitations, J. Chem. Theory Comput. 2014, 10, 4432-4441.

[128] van Leeuwen, R. and Baerends, E. J., Exchange-correlation potential with correct asymptotic behavior, Phys. Rev. A, Vol. 49, Is. 4, 2421-2431, 1994.

[129] Kullie O., Saue T., Range-separated density functional theory: A 4-component relativistic study of the rare gas dimers He2, Ne2, Ar2, Kr2, Xe2, Rn2 and Uuo2. Chemical Physics, 395(0):54 – 62, 2012. Recent Advances and Applications of Relativistic Quantum Chemistry.

[130] Si̇nanoğlu, Oktay (1962). "Many-Electron Theory of Atoms and Molecules. I. Shells, Electron Pairs vs Many-Electron Correlations". The Journal of Chemical Physics. 36 (3): 706.

[131] Čížek, J. (1966). "On the Correlation Problem in Atomic and Molecular Systems. Calculation of Wavefunction Components in Ursell-Type Expansion Using Quantum-Field Theoretical Methods". The Journal of Chemical Physics. 45 (11): 4256.

[132] Jeziorski, B.; Monkhorst, H. (1981). "Coupled-cluster method for multideterminantal reference states". Physical Review A. 24 (4): 1668.

[133] Shavitt, Isaiah; Bartlett, Rodney J. (2009). Many-Body Methods in Chemistry and Physics: MBPT and Coupled-Cluster Theory. Cambridge University Press. ISBN 978-0-521-81832-2.

[134] Nakai, Hiromi; Sodeyama, Keitaro (2003). "Many-body effects in nonadiabatic molecular theory for simultaneous determination of nuclear and electronic wave functions: *Ab initio* NOMO/MBPT and CC methods". The Journal of Chemical Physics. 118 (3): 1119.

[135] Dyall K.G., Faegri K., (2007): Introduction to relativistic quantum chemistry, Oxford Univ. Press.

[136] Grimme S., Antony J., Ehrlich S., Krieg H., J. Chem. Phys, 132 (2010), 154104.

[137] Olsen J.M., Aidas K.S., Kongsted J., Excited states in solution through polarizable embedding. Journal of Chemical Theory and Computation, 6(12):3721–3734, 2010.

[138] Olsen J.M.H., Kongsted J., Molecular properties through polarizable embedding. In Advances in Quantum Chemistry, pages 107–143. 2011.

[139] Hedegård E.D., Bast R., Kongsted J., Olsen J.M.H., Jensen H.J.Aa., Relativistic Polarizable Embedding. Journal of Chemical Theory and Computation, 13(6):2870–2880, 2017.

[140] J. Bigeleisen, J. Amer. Chem. Soc. 118, 3676 (1996).

[141] Y. Fujii, M. Nomura, M. Okamoto, H. Onitsuka, F. Kawakami and K. Takeda, Z. Naturforsch. 44a, 395 (1989).

[142] Suzen, Mehmet (2013). Generate Weighted Histogram (https://www.mathworks.com/matlabcentral/fileexchange/42493-generate-weighted-histogram), MATLAB File Exchange. Retrieved Jul 08, 2013.

[143] Bull.Korean.Chem.Soc. 33, p.803 (2012)

[144] Kuechle W., Dolg M., Stoll H., Preuss H., J. Chem. Phys. 100, 7535 (1994).

[145] Cao X., Dolg M., Stoll H., J. Chem. Phys. 118, 487 (2003)

[146] Cao X., Dolg M., J. Molec. Struct. (Theochem) 673, 203 (2004).

[147] Schäfer A., Huber C., Ahlrichs R., J. Chem. Phys. 100, 5829 (1994).

[148] Weigend F., Häser M., Patzelt H., Ahlrichs R.; Chem. Phys. Letters 294, 143, (1998).

[149] Weigend F., Köhn A., Hättig C.; J. Chem. Phys. 116, 3175 (2002).

[150] Dunning T.H.; JCP 90, 1007 (1989).

[151] Dolg M., Cao X., J. Phys. Chem. A 113, 12573(2009). http://www.tc.uni-koeln.de/cgi-bin/pp.pl?language=en,format=gaussian,element=U,ecp=ECP60MDF,job=getbset,bset=ECP60MDF.

[152] Groenewold, G.S., Gianotto, A.K., Cossel K.C., Van Stipdonk M.J., Moore D.T., Polfer N., Oomens J., de Jong W.A., Visscher L., J. AM. CHEM. SOC. 2006,128, 4802-4813.

[153] Fossgaard O., Gropen O., Corral Valero M., Saue T., On the performance of four-component relativistic density functional theory: Spectroscopic constants and dipole moments of the diatomics HX and XY (X,Y=F, Cl, Br, and I). The Journal of Chemical Physics, 118(23):10418–10430, 2003. URL: http://scitation.aip.org/content/aip/journal/jcp/118/23/10.1063/1.1574317, doi:10.1063/1.1574317.

[154] Dyall K.G., Theor. Chem. Acc. (2016) 135:128.

[155] Dyall K.G., Theor. Chem. Acc. (2007) 117:491. Available from the Dirac web site, http://dirac.chem.sdu.dk.

[156] Visscher L., Approximate molecular relativistic Dirac-Coulomb calculations using a simple Coulombic correction. Theoretical Chemistry Accounts, 98(2-3):68–70, 1997.

[157] Bernadotte S., Atkins A.J., Jacob C.R., J Chem Phys 137, 204106 (2012).

[158] Couston L., Pouyat D., Moulin Ch., Decambox P., "Speciation of Uranyl Species in Nitric Acid Medium by Time-Resolved Laser-Induced Fluorescence," Appl. Spectrosc. 49, 349-353 (1995).

[159] Bell J. T., Biggers R. E., The absorption spectrum of the uranyl ion in perchlorate media, Part I. Mathematical Resolution of the overlapping band structure and studies of the environmental effects, J Mol Spec, 1965.

[160] Bell J.T., Biggers R.E., JOURNAL OF MOLECULAR SPECTROSCOPY 25, 312-329 (1968) (Fig. 2, difference 21270-20502 levels)

[161] S. V. Lotnik, L.A. Khamidullina, V.P. Kazakov, (2003) Radiochemistry 45:499-502.

[162] Geipel G., Amayri S., Bernhard G., Spectrochimica Acta Part A 71 (2008) 53–58

[163] Bock W. Ch., Kaufman A., Glusker J. P., Coordination of Water to Magnesium Cations, Inorg. Che. 1994, 33 419-427.

[164] A. Ikeda et al, Inorg. Chem. (2007), 46, 4212−4219.

[165] Schafer A., Horn H., Ahlrichs R., J. Chem. Phys. 97, 2571 (1992).

[166] Darmanyan A.P., Khudyakov I.V., Photochem. Photobiol. 52 (1990) 293.

[167] Natrajan L. S., Coordination Chemistry Reviews 256 (2012) 1583-1603. (page 1591, second paragraph in 2.3.)

[168] Formosinho S. J., Burrows H. D., Miguel M. da G., Azenha M.E.D.G., Saraiva M.I., Ribeiro A.C.D.N., Khudyakov I.V., Gasanov R.G., Bolte M., Sarakha M., Photochem. Photobiol. Sci., 2003, 2, 569-575.

[169] Elliott P., Burke K., Furche F., Excited states from time-dependent density functional theory, Reviews in Computational Chemistry, Vol. 26, Is. 51, ISBN 978-0470-38839-6, ISSN 1069-3599, 2009, edited by Lipkowitz K.B., Cundari T.R., p.91-159.

[170] Rappoport D., Furche F.; J. Chem. Phys. 133, 134105 (2010).

[171] D. Papoušek, M.R. Aliev: MOLECULAR VIBRATIONAL - ROTATIONAL SPECTRA, Academia 1982.

[172] Tsushima, S., Dalton Transactions 40(25):6732-7.

[173] J. Merrick, D. Moran, L. Radom, An Evaluation of Harmonic Vibrational Frequency Scale Factors, J. Phys. Chem. A 2007, 111, 11683-11700.

# Supplementary information

Table of contents

## *1. Simulated luminescence spectrum for $UO_2^{2+}$ in Vacuo*

The phonon-less excitation energy $T_{00}$ for X $0_g^+ \leftarrow$ a $1_g$ transition, 18067 cm$^{-1}$, has been computed for this simulation through PES scan with respect to the symmetric stretching mode coordinate for the SORECP/TD-DFT/CAM-B3LYP/extended triple dzeta basis set method (without XALDA) with DIRAC software [S1]. The normal mode analysis have been adopted from the ECP/TD-DFT/B3LYP-D3 scalar quasi-relativistic computation of X $^1\Sigma_g^+$ and a a $^3\Delta_g$ (which is the second excited triplet in energy, for bare uranyl. The a $^3\Phi_g$ has lower energy, but transition is more strictly symmetry forbidden for the X $^1\Sigma_g^+ \leftarrow$ a $^3\Phi_g$ transition. For hydrated uranyl, counter-part for $^3\Delta_g$ has lower energy than $^3\Phi_g$ counter-part), normal mode outputs has been processed by ezSpectrum for Franck-Condon profile computation with 7 and 10 maximum vibrational quanta in the initial and target electronic state.

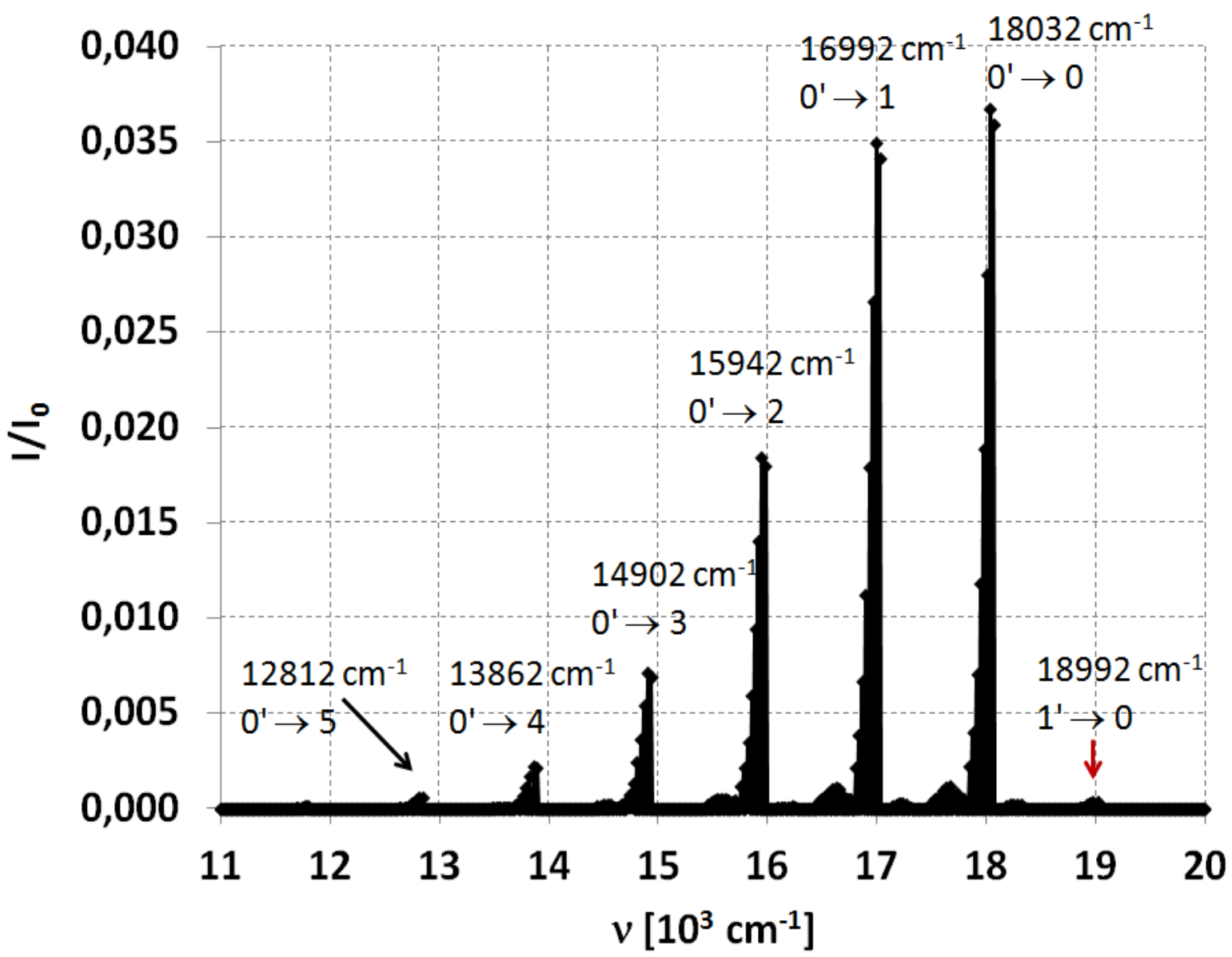

**Figure FSI1:** Simulated $^{238}U^{16}O_2^{2+}$ ($^3\Delta_g \rightarrow {}^1\Sigma_g^+$) lum. spectrum from normal mode calculation fully exploiting point group symmetry (the spectrum $T_{00}$ shift has been adapted from spin-orbit resolved $1_g \rightarrow$ X $0_g^+$ transition). [36]

For plotting (Fig. FSI1) histogram bin widths has been set as dν = 10 cm$^{-1}$, slightly wider than width determined by simplified rotationally-resolved simulation (giving dν = 4 cm$^{-1}$ half-width of peak progression differing by rotational numbers only). While dominant peaks correspond to symmetric stretching mode resolution, the small "bumps" between them to resolution by bending modes – which also explain broadening for the main, sym. stretching, progression peaks (marked by m'→n in Fig. FSI1). Lower vibrational frequency for bending modes (169 cm$^{-1}$ and 137 cm$^{-1}$ in ground and excited electronic states respectively) allow for hot-bands in their case, also broadening peaks. The FWHM for 0'→0 peak of ~ 64 cm$^{-1}$ is small compared to 450-650 cm$^{-1}$ for ($FWHM = 2\sqrt{2\ln 2}\,\sigma$, Tab. TSI1.3) hydrated uranyl in water).

The absence of Duschinsky effect allows vibrationally resolved luminescence spectrum for single electronic transition in $UO_2^{2+}$ to be estimated by convolution of four profiles,

[36] For all spectral simulations, the most abundant isotopologues have been considered.

$$I(\nu) = NL * I_{sym} * I_b * I_{solv}, \tag{SI0.1}$$

where $I$ is the luminescence intensity, $\nu$ wave-number, $NL$ is the natural line-shape (shifted to origin, eventually including rotational resolution), $I_{sym}$ is luminescence spectrum resolved by symmetric stretching mode only (Fig. FSI2), $I_b$ is the luminescence spectrum resolved by the two degenerated bending modes (Fig. FSI3) and $I_{solv}$ correspond to environmental spreading.

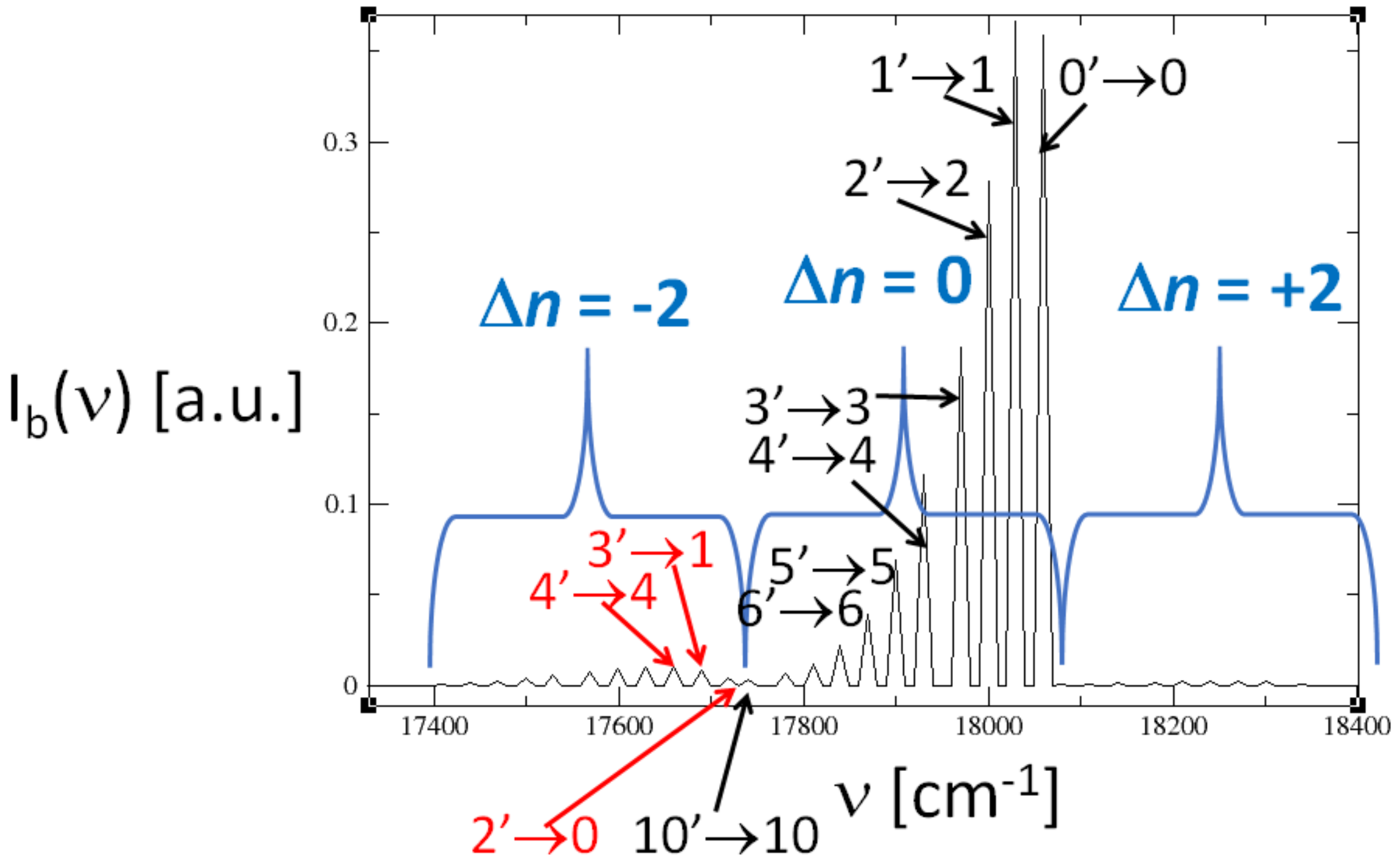

**Figure FSI2:** Bending modes resolution of $X^1\Sigma_{g,0+} \leftarrow a^3\Delta_{g,1}$ electronic transition in $^{238}U^{16}O_2^{2+}$.

In Fig. FSI2 please note that the most intense peak doesn't correspond to the (0‘,0‘)→(0,0) transition (18067 cm$^{-1}$, the blue-most among intense peaks), but is slightly shifted (by 32 cm$^{-1}$, i.e. difference between vibrational frequency of bending mode in excited (169 cm$^{-1}$) and ground (137 cm$^{-1}$) electronic states). This well correlates with $T_{00,eff} < T_{00}$ (even for solvated uranyl) by similar energy amount (in most cases).

The peaks in Fig. FSI2 are grouped by blue braces into $\Delta n$ = -2 (red-most pictured, range 17409 cm$^{-1}$ to 17729 cm$^{-1}$), $\Delta n$ = 0 (middle group with intense peaks, range 17747 cm$^{-1}$ to 18067 cm$^{-1}$ ($T_{00}$)) and $\Delta n$ = +2 (blue-most pictured, smallest peaks, range 18086 cm$^{-1}$ to 18341 cm$^{-1}$), where $\Delta n = n' - n$, the quantum vibrational numbers $n'$ and $n$ corresponds to total quanta in both degenerated bending modes in excited and ground electronic state respectively. Each group consists of transitions (denoted $n' \rightarrow n$) $k' \rightarrow k + \Delta n$, where $k \in \{0; 1; 2; ...; 10\}$, with $k$ = 0 peak corresponding to the lowest energy. With $k_{max}$ = 10 groups yet do not overlap. In harmonic approximation, which is probably much less applicable for bending mode than for the stretching mode (with consequences to luminescence peak shapes) the peak spacing is equidistant in each group (binning size and peak spacing are not divisible and so it doesn't look so in the Figure). The $\Delta n$ = -2 group is a hot-band and is corresponsible for the little “bumps” to the red side of each main luminescence peak in Fig. FSI1 (and even smaller "bump" to the blue side can be associated with $\Delta n$ = +2 group in Fig. FSI2). The "bumps" can be correlated with known problem to correctly fit uranyl(VI) luminescence spectra by gaussian profiles centred at main peak positions only (there is more intensity between peaks than expected).

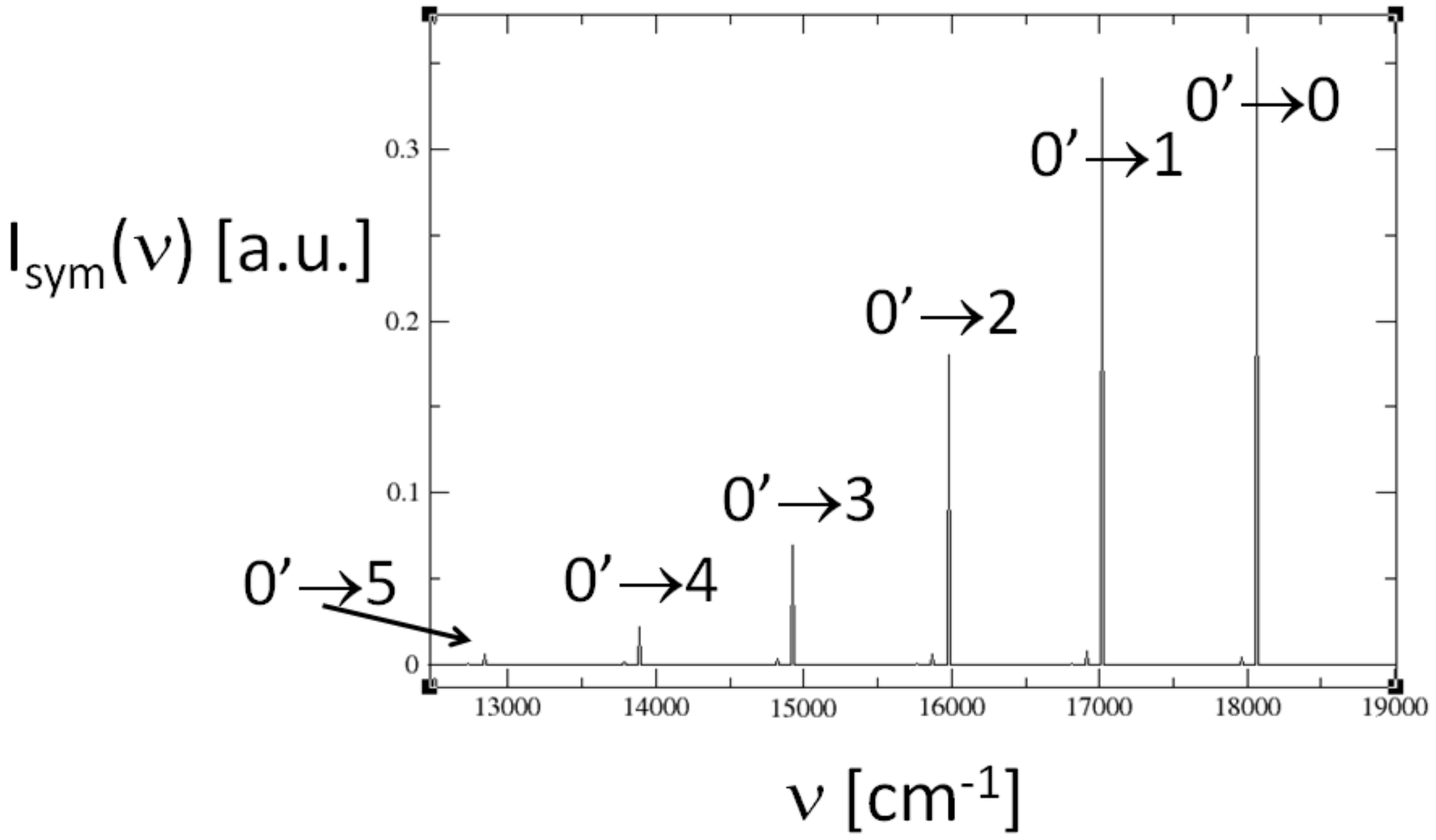

**Figure FSI3:** Train of peaks from symmetric stretching vibrational mode resolution of $X^1\Sigma_{g,0+} \leftarrow a^3\Delta_{g,1}$ electronic transition in $^{238}U^{16}O_2^{2+}$.

While bending mode associated resolution provide peak shapes, the main peak maxima position and relative intensities can be explained through symmetric stretching mode resolution (Fig. FSI3, $I_{sym}$ in (SI0.1)). The peaks in Fig. FSI3 could be grouped by $\Delta n \in \{0; -1; -2; ... -5\}$, but only one ($k$ = 0) peak dominate (for the room temperature) inside each group (there is another one tiny with $k$ = 1 visible on the redder side of each main peak).

## *2. Luminescence spectra fitting procedure*

The main part of computations in this study is described by following

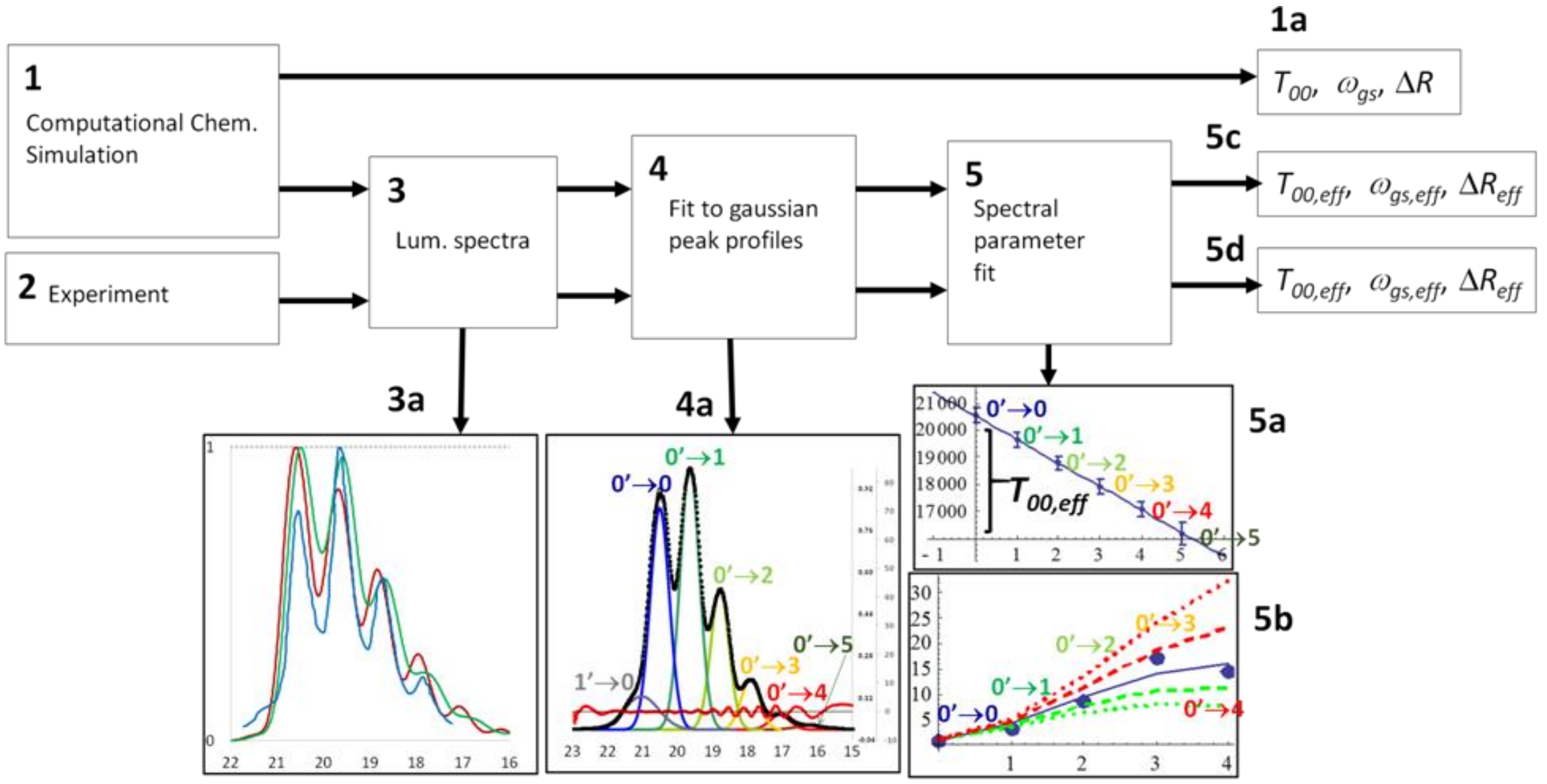

| exper (i) | 0'→5 | 0'→4 | 0'→3 | 0'→2 | 0'→1 | 0'→0 | 1'→0 |
|---|---|---|---|---|---|---|---|
| $I_j$ [] | 0.01615 | 0.05560 | 0.19076 | 0.53579 | 1.01105 | 0.84692 | 0.12648 |
| $\nu_{max,j}$ [cm$^{-1}$] | 16215 | 17080 | 17922 | 18788 | 19659 | 20521 | 21042 |
| $\lambda_{max,j}$ [nm] | 616.7 | 585.5 | 558.0 | 532.3 | 508.7 | 487.3 | 475.2 |
| $\sigma_j$ [cm$^{-1}$] | 416 | 288 | 269 | 262 | 274 | 274 | 494 |

4b

**Figure FSI4:** Luminescence spectra fits.

Fig. FSI4 – luminescence spectra from both simulation (1, the box "1" here correspond to the whole diagram from Fig. 3 of the main article) and experiment (2) are decomposed into individual gaussian peaks (4, according to formula (SI1.1) as depicted in box 4a for "(i)" experimental luminescence spectrum, corresponding parameters are summarized in table 4b below).

$$Z(\nu) = \sum_{j=-1}^{5} I_n \exp\left( - \frac{(\nu - \nu_{\max,j})^2}{2\sigma_j^2} \right), \qquad \text{(SI1.1)}$$

Parameters of (4a) fit are then themselves subject to another fits (5) – peak maxima wave-numbers $\nu_{max,j}$ are fitted with respect to model

$$\nu_{\max,j} = T_{00,eff} - \omega_{gs,eff} \cdot j \qquad (j \geq 0), \qquad \text{(SI1.2)}$$

$$\nu_{\max,j} = T_{00,eff} - \omega_{es,eff} \cdot j \qquad (j \leq 0), \qquad \text{(SI1.3)}$$

and peak area with respect to the model

$$\frac{S_j}{S_0} \;=\; \frac{\sigma_j\, I_j}{\sigma_0\, I_0} \;=\; \frac{\left|\left\langle \chi_{es}, j \middle| \chi_{gs}, 0 \right\rangle\right|^2}{\left|\left\langle \chi_{es}, 0 \middle| \chi_{gs}, 0 \right\rangle\right|^2} \;=\; f_j\left(\omega_{gs}, \omega_{es}; \Delta R_{eff}\right), \qquad \text{(SI1.4)}$$

where $S_j$ is the $j$-th peak area (for fit based on peak intensities $\sigma_j$ and $\sigma_0$ are omitted) and $\chi_{es}$ and $\chi_{gs}$ are vibrational parts of wave-functions in excited and ground electronic state respectively, bra and ket notation is completed by single mode vibrational quantum number, integration is over single normal mode coordinate. Double-harmonic approximation is used, $f_j$ exist in an analytical form (polynomial in $\Delta R_{eff}$) [124]. Resulting three parameters $T_{00,eff}$, $\omega_{gs,eff}$ and $\Delta R_{eff}$ can be compared theory vs. experiment ((7) vs. (SI1.2)) or correlated to the chemical-physical parameters $T_{00}$, $\omega_{gs}$ and $\Delta R$ connected to the simplified model (1a).

For the case of (SI1.1) decomposition Maximum Likeli-hood Method (MLM, [94]) based procedure has been applied, i.e. the following objective function,

$$\chi^2_{MLM}(\beta) \;=\; \sum_{k=1}^{N} \frac{\left(Z(\nu_k) - Z_k\right)^2}{\sigma_k^2} \;+\; \sum_{k=1}^{N} \ln\left(\sigma_k^2\right) \;+\; \Lambda \cdot P(\beta), \qquad \text{(SI1.5)}$$

where $(\nu_k, Z_k)_{k=1}^N$ are spectral data to be fitted, $\beta$ stands for $(\nu_{max,j}, I_j, \sigma_j)$ ($j$ = -1 to +5) peak parameters to be determined and one $\lambda$ parameter of variance model,

$$\sigma_k^2 \;=\; \lambda Z_k \;+\; \sigma_0^2, \qquad \text{(SI1.6)}$$

$P(\beta)$ is a penalization function enforcing constrains

$$\nu_{\max,j}^{(0)} - \sigma_j^{(0)} < \nu_{\max,j} < \nu_{\max,j}^{(0)} + \sigma_j^{(0)}, \qquad \text{(SI1.6b)}$$

where $\nu_{\max,j}^{(0)}$ (peak maxima by quadratic fit from 7 close most points) and $\sigma_j^{(0)}$ = 200 cm$^{-1}$ are initial guess for MLM optimization. The constrains prevent “peak coalescence” – i.e. situation where one peak in fitted spectrum is explained by two gaussian functions while another peak is omitted. The penalization function has been chosen as

$$P(\beta) \;=\; \sum_{j=-1}^{+5} \left( \frac{\nu_{\max,j} - \nu_{\max,j}^{(0)}}{\sigma_j^{(0)}} \right)^{2\cdot B}, \qquad \text{(SI1.7)}$$

where $B$ is big integer ($B$ has been set as $B$ = 10 here). Apparently as $B \to +\infty$ summands in (SI1.7) converge to zero inside allowed interval and to infinity outside. The $\Lambda$ prefactor and $B$ parameter have been chosen to be big enough to enforce “no coalescence”, yet small enough to provide minimal influence on optimized $\beta$ parameters. The positivity of $I_j$ and $\lambda$ has been forced by $I_j = c_j^2$ and $\lambda = (\lambda')^2$ parameterizations.

In the following tables, the red-most, 0'→5 peak, which is usually less accurately determined and omitted from subsequent $T_{00,eff}$, $\omega_{gs,eff}$, $\Delta R_{eff}$ fit, is not included.

| Method | 1'→0 | 0'→0 | 0'→1 | 0'→2 | 0'→3 | 0'→4 |
|---|---|---|---|---|---|---|
| TT1/HF | 465.6 | 484.8 | 506.8 | 531.4 | 556.5 | 583.1 |
| TT1/LBα | 466.2 | 487.8 | 510.3 | 535.8 | 562.6 | 594.6 |
| TT1/rLBα | 464.3 | 485.6 | 508.1 | 531.6 | 557.0 | 585.3 |
| TT3/LBα | 468.4 | 486.0 | 507.6 | 531.8 | 558.3 | 587.6 |
| Exper (i) | 468.7 | 487.3 | 509.1 | 532.6 | 558.4 | 585.6 |
| Exper (ii) | 472.6 | 488.4 | 510.0 | 533.5 | 559.4 | |
| Exper (iii) | 469.4 | 488.0 | 510.0 | 533.5 | 560.1 | |
| Exper (iv) | 472.7 | 488.7 | 509.9 | 533.7 | 559.0 | |

**Table TSI1.1:** Fitted peak maxima, $\lambda_{max,j} = 1/\nu_{max,j}$, in nm

Although the fitting procedure has been done in to-energy-proportional, $cm^{-1}$ scale, individual peak maxima in Tab. TSI1.1 are presented in nm for better comparison with literature data. TRLFS experimentalists usually focus on the 509 nm (determined rather as 510 nm in the above tabulated data) peak, which has the biggest intensity and corresponds to the 0'→1 vibronic transition. The main peak, i.e., initial peak of progression is, however, the 487-488 nm peak (0'→0). Their intensity ratio is in linear relation to $\Delta R_{eff,1,max}$ (the proportionality coefficient is, however, function of both $\omega_{gs}$ and $\omega_{es}$ – symmetric stretch frequency in ground and excited electronic states respectively) according to,

$$\Delta R_{eff,1,\max} = \sqrt{\frac{h}{2m}\frac{\omega_{gs}+\omega_{es}}{\omega_{es}\omega_{gs}^{1/2}}}\cdot\frac{I_1}{I_0}, \tag{SI1.8}$$

or alternatively for $\Delta R_{eff,1}$ parameter,

$$\Delta R_{eff,1} = \sqrt{\frac{h}{2m}\frac{\omega_{gs}+\omega_{es}}{\omega_{es}\omega_{gs}^{1/2}}}\cdot\frac{\sigma_1 I_1}{\sigma_0 I_0}. \tag{SI1.9}$$

| Method | 1'→0 | 0'→0 | 0'→1 | 0'→2 | 0'→3 | 0'→4 |
|---|---|---|---|---|---|---|
| TT1/HF | 0.01825 | 1.00000 | 0.81190 | 0.42649 | 0.16393 | 0.04198 |
| TT1/LBα | 0.01529 | 1.00000 | 0.95548 | 0.52792 | 0.21919 | 0.05024 |
| TT1/rLBα | 0.01555 | 1.00000 | 0.83661 | 0.55789 | 0.27708 | 0.09425 |
| TT3/LBα | 0.07612 | 1.00000 | 0.79679 | 0.37835 | 0.12831 | 0.03373 |
| Exper (i) | 0.07617 | 0.79246 | 1.00000 | 0.60807 | 0.24974 | 0.07938 |
| Exper (ii) | 0.08579 | 0.75870 | 1.00000 | 0.65359 | 0.30143 | |
| Exper (iii) | 0.09048 | 0.71876 | 1.00000 | 0.64452 | 0.27153 | |
| Exper (iv) | 0.08655 | 0.84763 | 1.00000 | 0.58337 | 0.29845 | |

**Table TSI1.2:** Fitted peak maxima intensities, $I_j$ (dimension-less), renormalized after fit, i.e. $I(n'\rightarrow m) = I'(n'\rightarrow m)/I'(0'\rightarrow 0)$, where intensities with prime are from the original fit.

| Method | 1'→0 | 0'→0 | 0'→1 | 0'→2 | 0'→3 | 0'→4 |
|---|---|---|---|---|---|---|
| TT1/HF | 220.8 | 250.0 | 273.9 | 317.8 | 294.1 | 550.5 |
| TT1/LBα | 198.0 | 275.5 | 328.9 | 297.2 | 359.5 | 294.5 |
| TT1/rLBα | 191.2 | 267.6 | 292.2 | 260.4 | 271.8 | 226.9 |
| TT3/LBα | 224.5 | 246.3 | 273.2 | 271.1 | 275.2 | 231.2 |
| Exper (i) | 313.7 | 292.0 | 272.8 | 262.8 | 270.5 | 269.7 |
| Exper (ii) | 150.5 | 187.6 | 191.6 | 201.2 | 235.4 | |
| Exper (iii) | 266.4 | 239.5 | 247.1 | 247.5 | 371.6 | |
| Exper (iv) | 275.4 | 187.5 | 199.3 | 213.6 | 179.0 | |

**Table TSI1.3:** Fitted peak width parameters, $\sigma_j$ in $cm^{-1}$.

Please note the small variance in peak width parameters within each row for the three most intense peaks - 0'→0, 0'→1 and 0'→2 (8 $cm^{-1}$ to 29 $cm^{-1}$ range for experimental and 27 $cm^{-1}$ to 69 $cm^{-1}$ range for theoretical). This correlates well with idea that variation in solvent positions affects mostly only $T_{00}$ of solvent, to lesser extend $\omega_{gs}$ and other normal-mode connected spectroscopic quantities are in fact unaffected. Therefore luminescence spectrum can be modelled as a convolution of "averaged" Franck-Condon histogram with $T_{00}$ distribution (Fig. 13-15 in the main article). The blue-most and red-most peaks (in particular in the experimental spectra) are less clearly determined (both by the position near detector range edges and by smaller overall intensity) and this puts their width parameters in question as well.

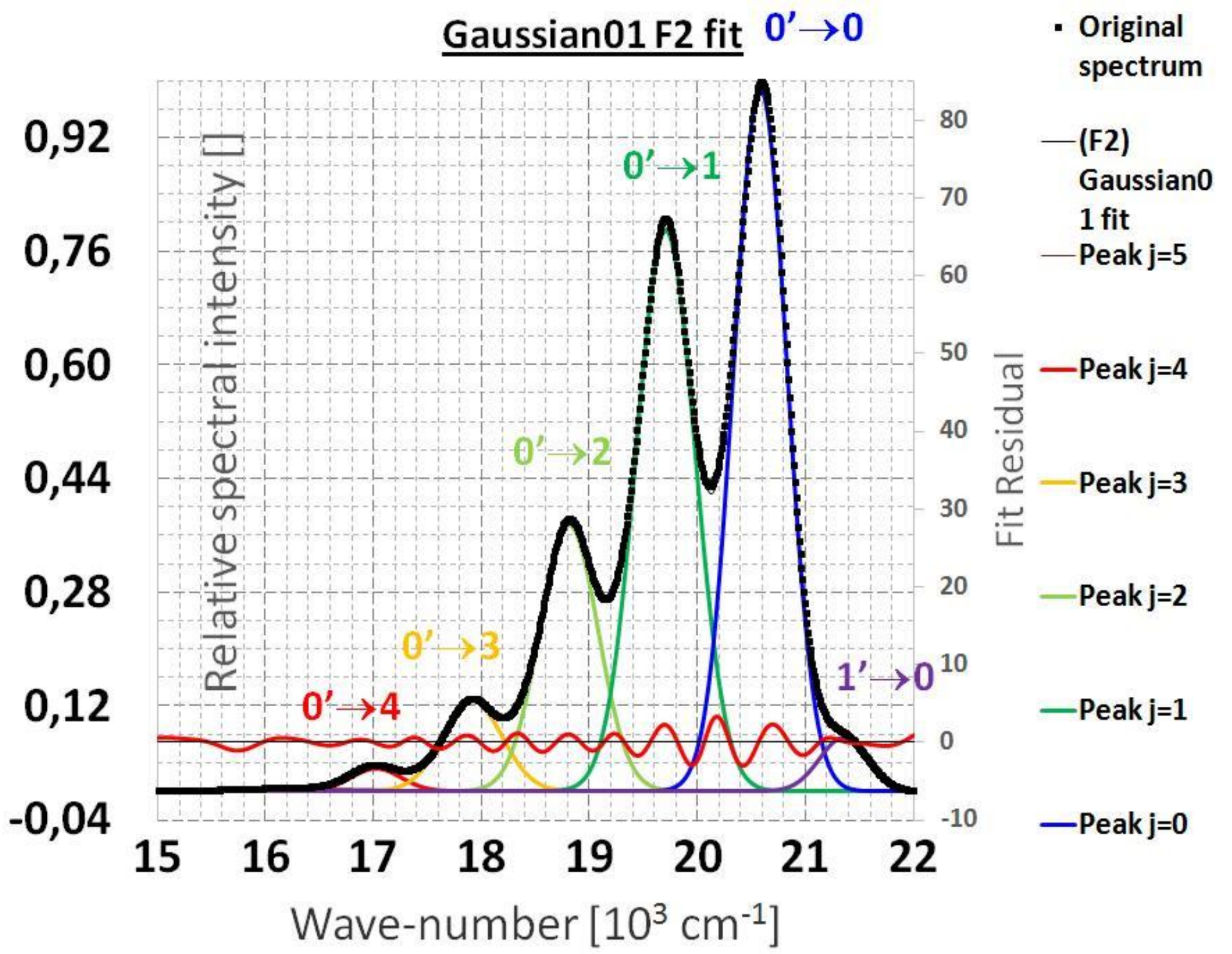

**Figure FSI5:** TT3/LBα simulated spectrum decomposition into gaussian peaks, normal scale. Fit residuals plotted (without logarithmic term and penalization) on auxiliary vertical axis – red curve. Values on vertical axis correspond to $I / (\omega/\omega_0)^3$ quantity.

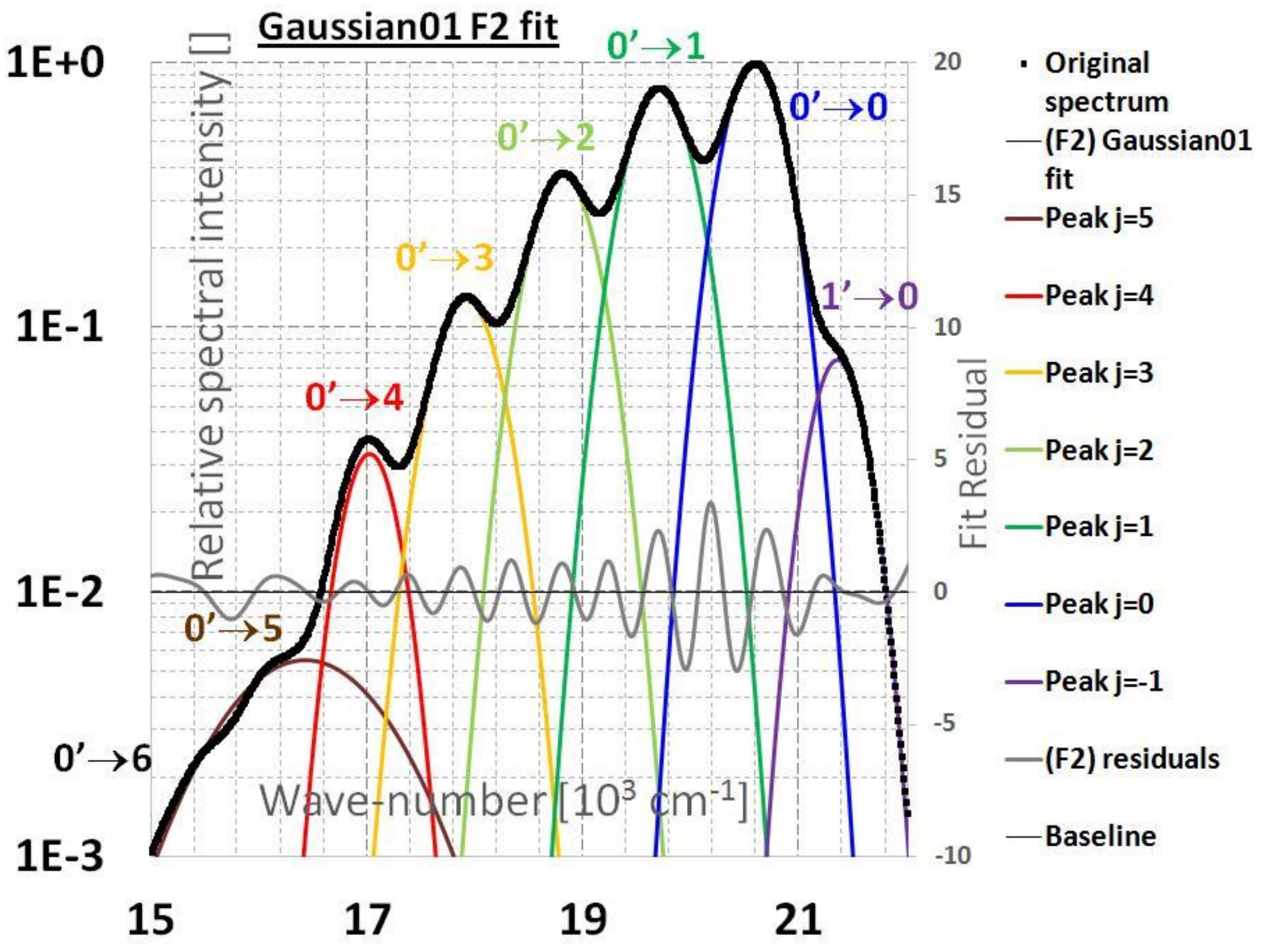

**Figure FSI6:** TT3/LBα simulated spectrum decomposition into gaussian peaks, logarithmic scale. Fit residuals plotted (without logarithmic term and penalization) on auxiliary vertical axis – grey curve. Values on vertical axis correspond to $I\ /\ (\omega/\omega_0)^3$ quantity.

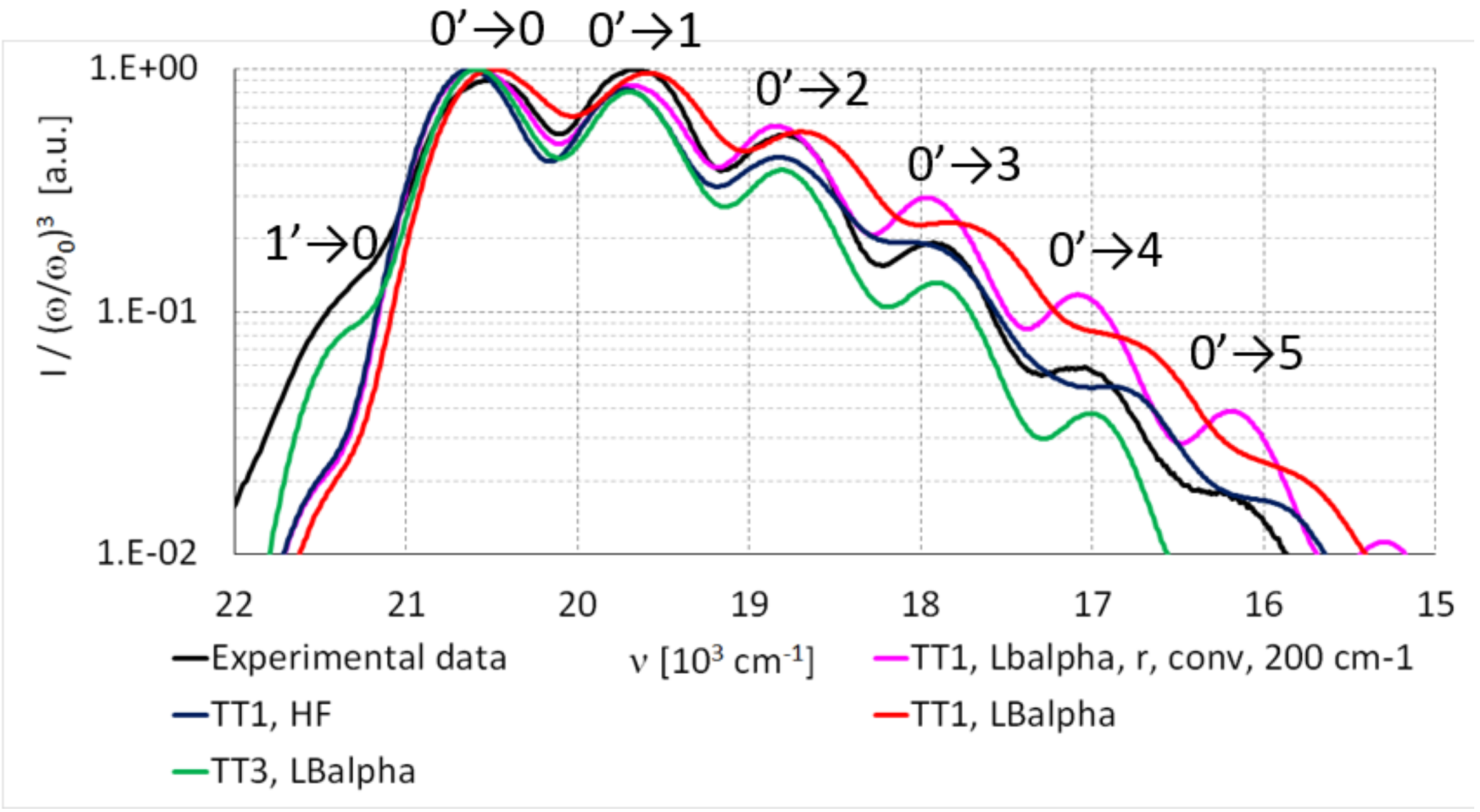

**Figure FSI7:** Comparison between experimental data measured within this study (black curve) and simulated spectra – TT1/HF (blue), TT3/LBα (green), TT1/LBα (red) and TT1/rLBα (magenta curve)

corresponds to $|\mu|^2$-weighted sum over snapshot Franck-Condon profiles ($\mu$ is the electronic transition moment in excited state equilibrium geometry). The logarithmic scale highlight simulation success in the "tail"

## *3. Molecular models*

### 3.1. The $D_{5h}$-$[UO_2(H_2O)_5]^{2+}$ complex *in Vacuo* for method comparison (Chapter 2, Results)

The excitation energies $T_a$, $T_v$ and $T_{de}$ in Tab. 3 (in Chapter 2 of Results in the main article) have been determined through formulae (5)-(7) with coordinates $R_{gs}$ and $R_{es}$ corresponding to the $D_{5h}$ point-group structures (Fig. FSI8, FSI9, Tab. TSI3.1). The respective coordinates corresponds to ground and first excited electronic state $D_{5h}$ constrained grometry optimization with ligated water molecule internal coordinates kept frozen from ground state optimization and difference between R(U-$O_w$) constrained as well (the respective difference between ground and electronic excited state equilibrium values is below 0.5%). Therefore, only the bond length of uranyl group, R(U-$O_{yl}$), changed by $\Delta R$ = 5 pm (tiny change depicted by orange lines in Fig. FSI9). The tiny difference yet corresponds for ~ 700 $cm^{-1}$ energy difference (roughly one vibrational quantum of the symmetric stretching uranyl group mode). The geometry optimization have been performed via scalar quasi relativistic ECP/B3LYP+D3/def-TZVPP method (*in Vacuo*) within Turbomole 7.1.

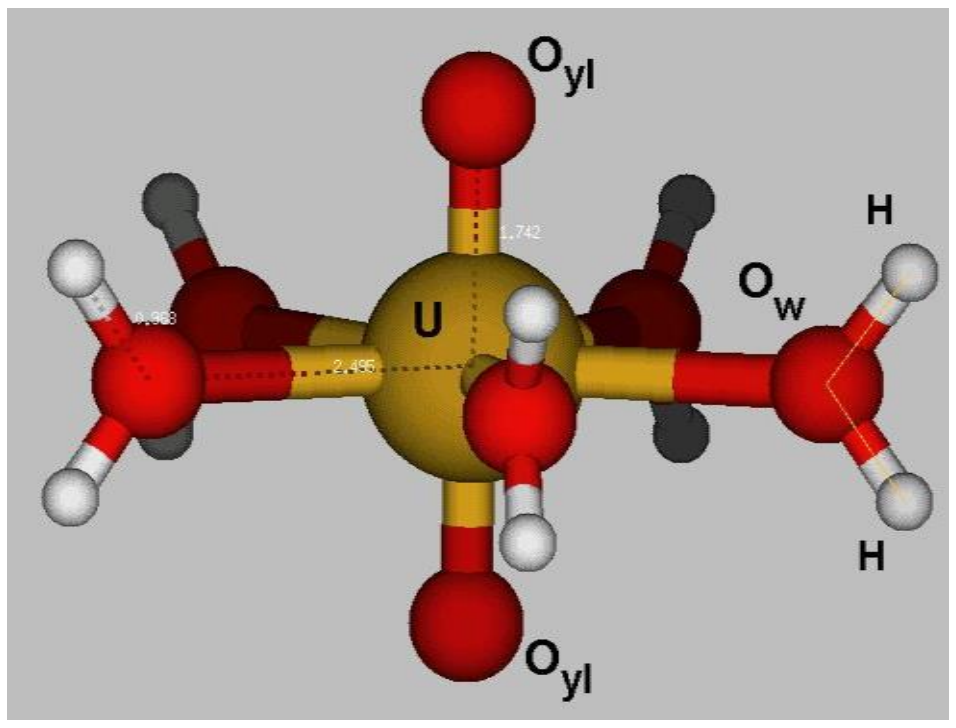

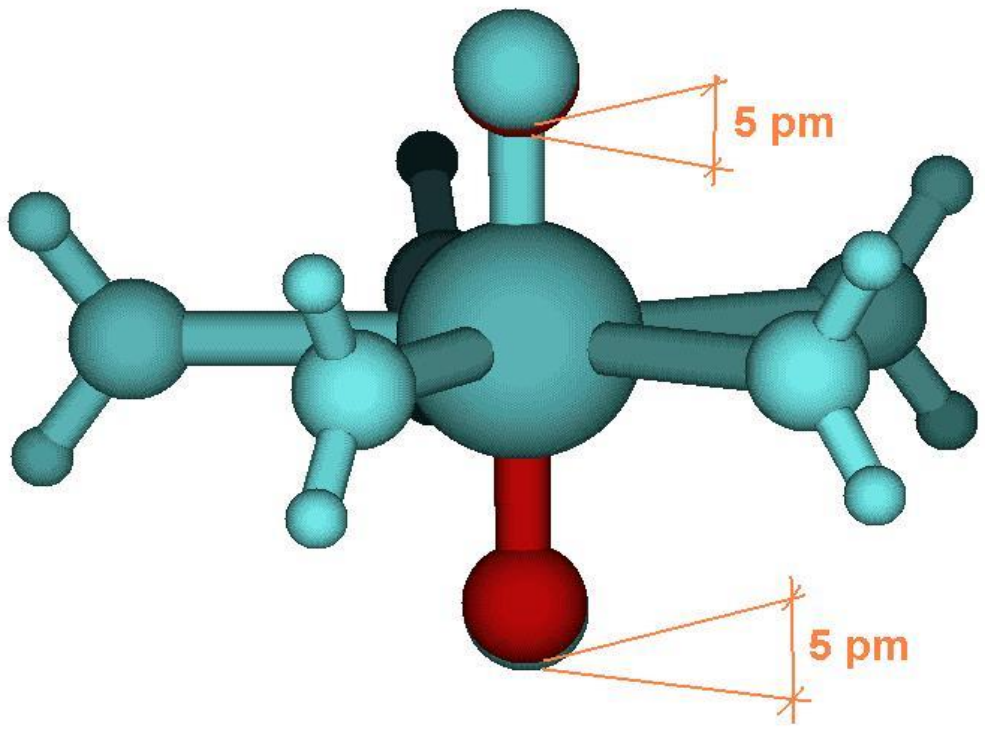

**Figure FSI8 (left):** The ground state $D_{5h}$-constrained equilibrium structure for $[UO_2(H_2O)_5]^{2+}$ **Figure FSI9 (right):** The ground and first excited electronic state equilibrium geometry models, corresponding to $R_{gs}$ and $R_{es}$ in (5)-(7) formulae.

| Variable | Ground state | Excited state |
|---|---|---|
| $R(U\text{-}O_{yl})$ | 174.243 pm | 179.243 pm |
| $R(U\text{-}O_w)$ | 249.531 pm | |
| $R(O_w\text{-}H)$ | 96.767 pm | |
| $\alpha(H\text{-}O_w\text{-}H)$ | 106.897° | |

**Table TSI3.1:** The ground and first excited electronic state equilibrium geometry ($D_{5h}$-constrained) models, corresponding to $R_{gs}$ and $R_{es}$ in (5)-(7) formulae. The with ligated water molecule internal coordinates in last two lines are delimited by bold line.

### 3.2. Complex species $UO_2(SO_4)_2^{2-}$, $UO_2(CO_3)_3^{4-}$ and $MgUO_2(CO_3)_3^{2-}$ for preditivity study (Chapter 4, Results)

In Chapter 4 of the Results section, uranyl(VI) bis(sulfate), tris(carbonate) and the ternary Magnesium-uranyl(VI) tris(carbonate) complex has been briefly studied to investigate predictive power of spectral simulation methodology (for SORECP/CAM-B3LYP/XALDA method), we append molecular models of aforementioned species for an illustration (the geometries correspond to electronic ground state equilibrium predicted by in-cluster energy minimization with all water molecules included explicitly, yet using only scalar quasi-relativistic ECP/TD-B3LYP/def-SVP (in Turbomole [S8], [104], without XALDA)). For the $T_{00}$ computation, the studied complex (cut out in Fig. FSI10-FSI12) has been accompanied by point charges instead of explicit water molecules, yet the bigger atomic basis set def-TZVPP and TD-DFT with CAM-B3LYP functional, XALDA approximation and spin-orbit resolved ECP has been used (in Dirac [S1], [143]-[146]).

The lowest lying (scalar quasi-relativistic) excited electronic state geometries differ only slightly by symmetric elongation of (both) uranyl U-O bonds (i.e. elongation with respect to the uranyl group symmetric stretching mode) by $\Delta R$ (Tab. 8). This is, in general, not true for complex species *in Vacuo*, where excited state optimization initialized from ground state equilibrium led (in case of tris(carbonate)) to ligand loss or similar significant deformations.

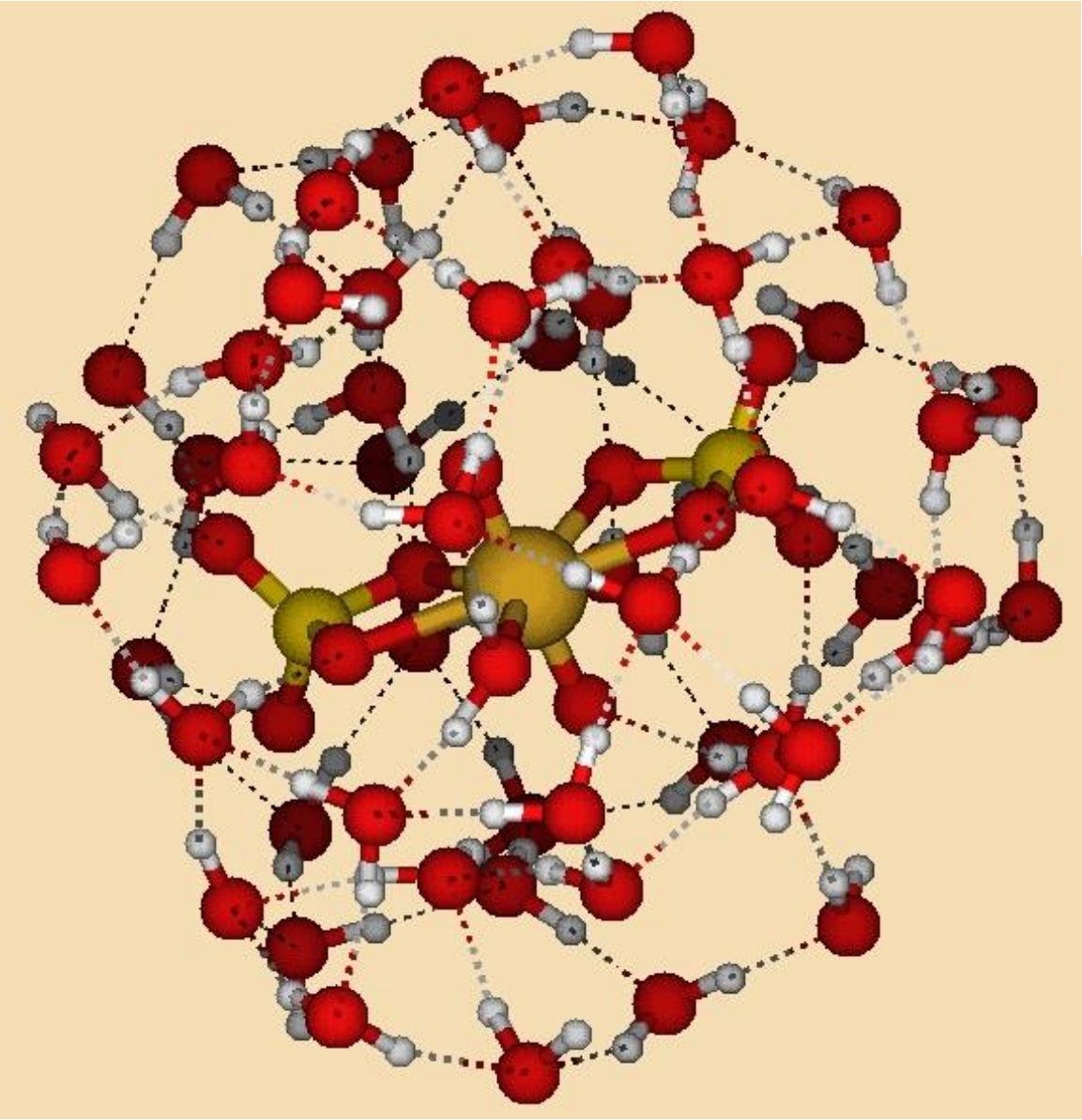

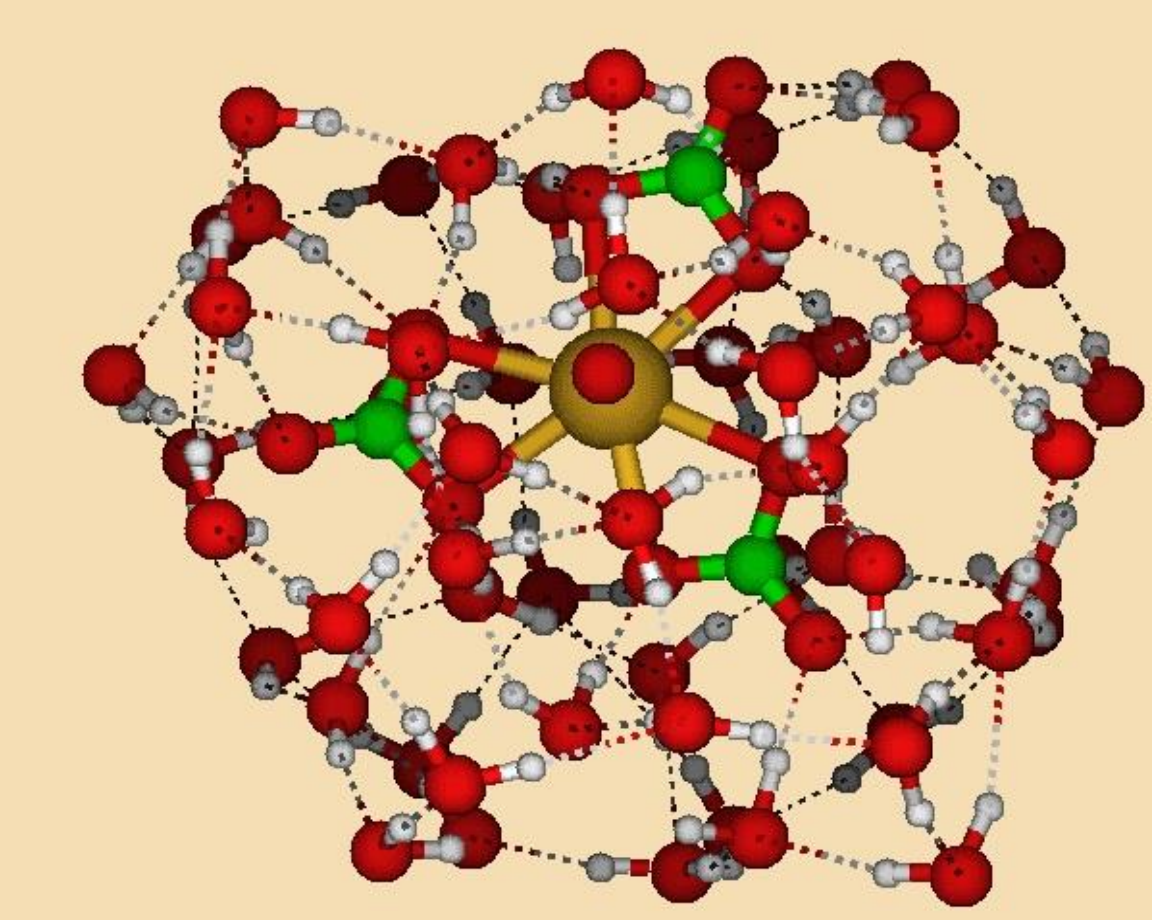

**Figure FSI10 (left):** Uranyl(VI) bis(sulfate) complex, $[UO_2(\kappa^2\text{-}SO_4)_2(H_2O)]^{2-}$ inside 55 water molecules cluster. **Figure FSI11 (right):** Uranyl(VI) tris(carbonate) complex, $[UO_2(\kappa^2\text{-}CO_3)_3]^{4-}$ (top view) inside 55 water molecule cluster. Uranium is plotted dark yellow and bigger than sulphur (yellow), oxygen is in red colour, carbon is green, hydrogen white. Hydrogen bonds are plotted by dashed lines. Molden package [S5] has been used for visualization.

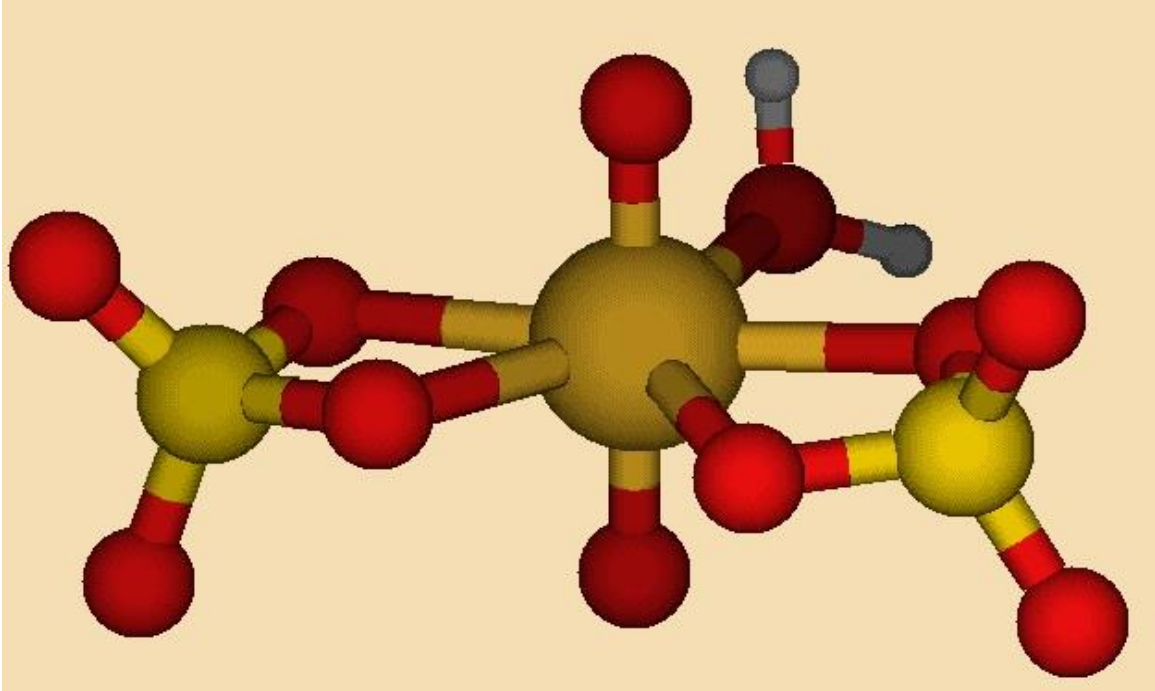

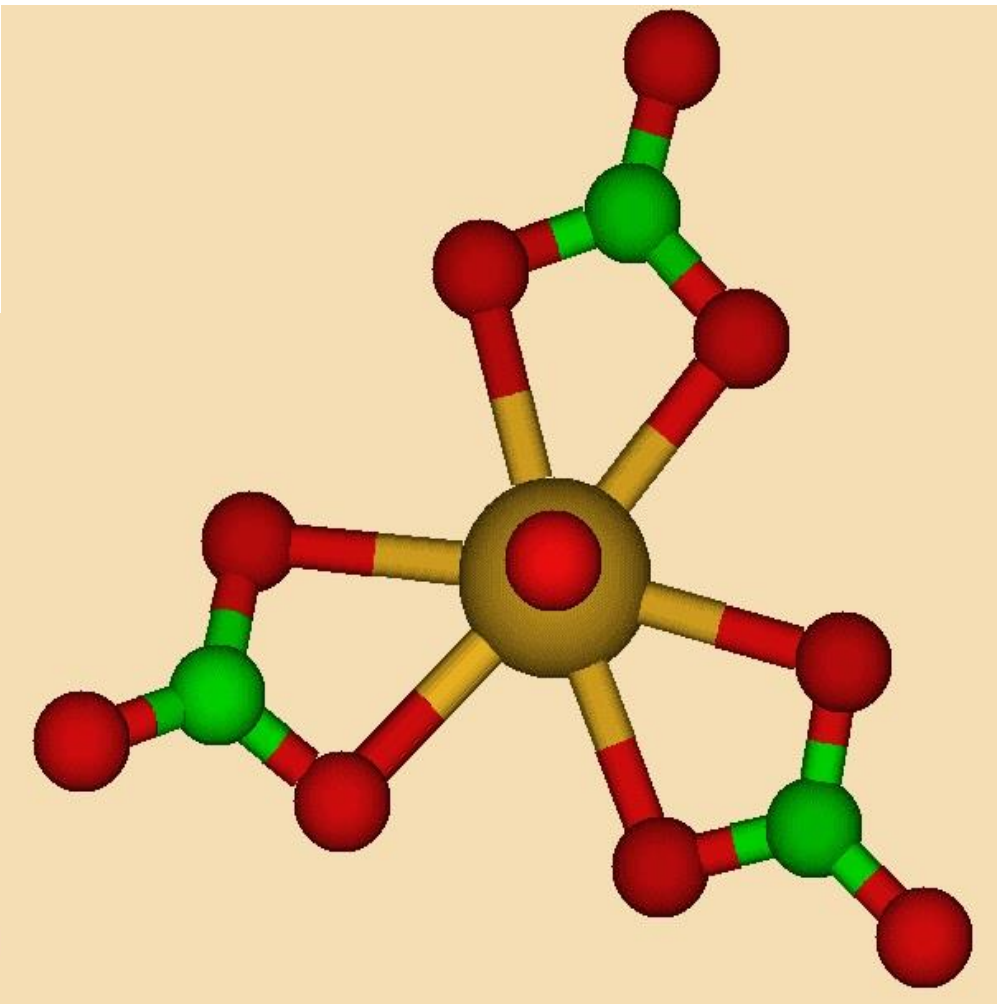

**Figure FSI12 (left):** Bis(sulfate), detail. **Figure FSI13 (right):** Tris(carbonate), detail.

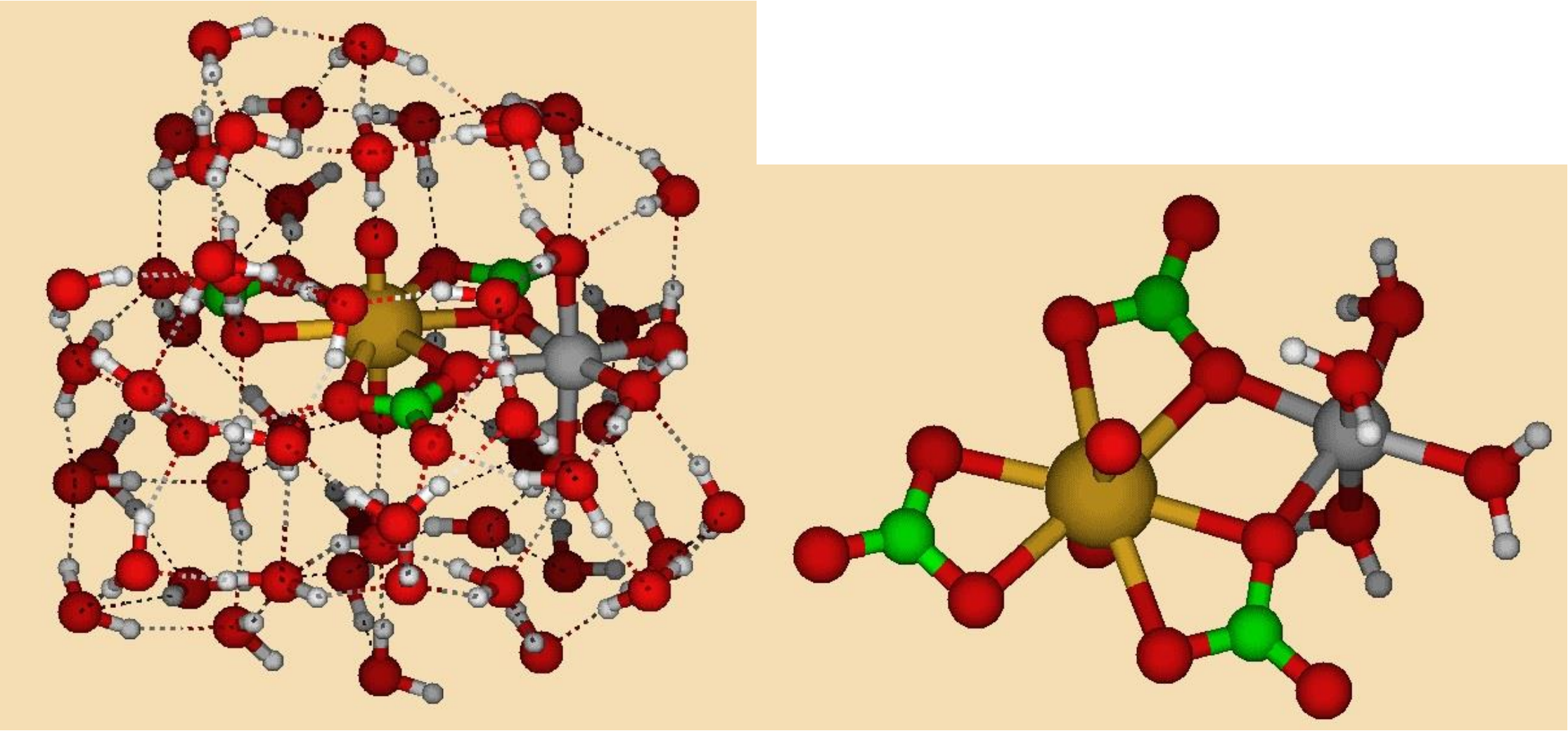

**Figure FSI14 (left):** Ternary, $[MgUO_2(CO_3)_3]^{2-}$ complex, **Figure FSI15(right)**: Detail. $Mg^{2+}$ ion is plotted grey.

Please note the octahedral coordination of $Mg^{2+}$, formula $[(CO_3)UO_2(\mu\text{-}CO_3)_2Mg(H_2O)_4]^{2-}$ would stress bridging by $CO_3^{2-}$ ligand between the two metals and four water molecules saturating $Mg^{2+}$ coordination number 6 (see [163] for comparison). The complex considered for $T_{00}$ spin-orbit resolved correction included the four water molecules coordinated directly to $Mg^{2+}$ atom. Without them, large variation of $T_{00}$ with respect to geometry, values significantly lower than experimental and even problems with ground state convergence and close-shell reference stability as observed, probably resulting from an unsaturated $Mg^{2+}$ coordination sphere.

## *4. Luminescence life-times*

Luminescence life-time $\tau_m$ (for species "*m*") can be estimated from de-excitation probability additivity formula

$$\frac{1}{\tau_m} = \frac{1}{\tau_m^{(0)}} + \lambda_{n.r.,m}, \qquad \text{(SI4.1)}$$

where $\tau_m^{(0)}$ stands for radiative de-excitation life-time, $\lambda_{n.r.}$ for non-radiative energy transfer rate. Chemical composition and temperature dependence of the latter, $\lambda_{n.r.}$ can be modelled as

$$\lambda_{n.r.,m} = \sum_Q k_Q[Q], \qquad \text{(SI4.2)}$$

where $Q$ denotes chemical species corresponsible for luminescence energy transfer (in particular quenching), $k_Q$ is the respective rate constant and $[Q]$ molar concentration. The Arrhenius law for $k_Q$ temperature could be expected,

$$k_Q = A_Q \exp\left(-E_{A,Q}/RT\right), \qquad \text{(SI4.3)}$$

with activation energy, $E_{A,Q}$ and frequency factor $A_Q$. While estimates for $\tau_m^{(0)}$ are presented in Table TSI4.1, quantum chemical study determining $E_A$ for $Q$ = $H_2O$ (to compare with model presented in [101]) is among future prospects, together with quantum-chemistry – statistical physics computation of $A_Q$ for the same quenching species.

| Settings | Method | System | $1/\lambda_v$ [ms] | $\tau_m^{(0)}$ [ms] | | |
|---|---|---|---|---|---|---|
| TT1 | HF | $[UO_2(H_2O)_5]^{2+}$ | 54 | 78 | ± | 4 |
| | B3LYP | | 9.1 | 10.2 | ± | 0.1 |
| | CAM | | 13.1 | 15.7 | ± | 0.2 |
| | LB94 | | 15.6 | 17.3 | ± | 0.3 |
| | LBα | | 11.6 | 13.2 | ± | 0.2 |
| TT3 | HF | $[UO_2(H_2O)_5]^{2+}$ | 53 | 80 | ± | 6 |
| | B3LYP | | 12.9 | 8.00 | ± | 0.05 |
| | CAM | | 12.5 | 16.4 | ± | 0.3 |
| | LB94 | | 13.1 | 16.4 | ± | 0.2 |
| | LBα | | 10.5 | 11.3 | ± | 0.1 |
| Cluster | HF | $[UO_2(H_2O)_5]^{2+}$ | 55.6 | 80.0 | | |
| | CAM | | 18.6 | 21.5 | | |
| | LBα | | 24.9 | 16.5 | | |
| Cluster | CAM | $UO_2(SO_4)_2^{2-}$ | 4.5 | 1.8 | | |
| | LBα | | 4.7 | 1.3 | | |
| Cluster | CAM | $UO_2(CO_3)_3^{4-}$ | 16.5 | 7.8 | | |
| | LBα | | 11.4 | 9.6 | | |

**Table TSI4.1:** Radiative luminescence life-times for studied uranyl(VI) complex species.

In Tab. TSI4.1, "Cluster" refers to geometry optimization of studied complex (ligated water molecules omitted in formula, i.e. $UO_2(SO_4)_2^{2-}$ is $[UO_2(H_2O)(\kappa^2\text{-}SO_4)_2]^{2-}$) surrounded by 55, 56 and 55 solvent water molecules. Geometry optimization has been done within scalar quasi-relativistic ECP/TD-B3LYP/def-SVP (in Turbomole, no XALDA) with explicit inclusion of solvent, in case of tris(carbonate), $UO_2(CO_3)_3^{4-}$, for excited state geometry optimization solvent atom positions has been fixed from ground state equilibrium value. Radiative transition probability has been computed with solvent atoms represented by point-charges only, yet by spin-orbit resolved SORECP/TD-DFT/XALDA/def-TZVPP (in Dirac [S1]). The $\lambda_v$ represents electronic transition probability rate corresponding to the ground electronic state equilibrium geometry (i.e. "vertical excitation"), $\tau_m^{(0)}$ is reciprocal of transition probability rate for excited electronic state equilibrium geometry. For $[UO_2(H_2O)_5]^{2+}$ the transition is more probable in ground state equilibrium geometry, but for bis(sulfate), $UO_2(SO_4)_2^{2-}$, and tris(carbonate), $UO_2(CO_3)_3^{4-}$, de-excitation from excited state equilibrium geometry is more probable. For TT1 and TT3 settings life-times has been determined by averaging based on formula,

$$\exp\left(-\frac{t}{\tau_m^{(0)}}\right) = \frac{1}{N}\sum_{j=1}^{N}\exp\left(-\frac{t}{\tau_{m,j}^{(0)}}\right), \tag{SI4.4}$$

for the "observational time" smaller than $t < 0.5$ ms (realistic as at ambient temperature with quenching life-times are in μs range and even under cryogenic conditions doesn't exceed 2.0 ms) results for $\tau_m^{(0)}$ are identical within statistical error with harmonic average for $\tau_{m,j}^{(0)}$ (as obvious when Taylor expansion with respect to $t$ to the first order is considered). Statistical errors has been determined by

$$\sigma\left(\tau_m^{(0)}\right) = \left(\tau_m^{(0)}\right)^2 \sigma\left(1/\tau_m^{(0)}\right) = \left(\tau_m^{(0)}\right)^2 \left|\frac{d\,y}{d\,x}\right| \sigma(x) = \left(\tau_m^{(0)}\right)^3 \frac{\sigma(f)}{t}, \tag{SI4.5}$$

where $x$ is (SI4.4) and $y = \tau_m^{(0)} = -(1/t)\ln(x)$. Shorter $\tau_m^{(0)}$ for bis(sulphate) and tris(carbonate) is consistent with experimentally observed small luminescence per unit concentration for the former. But as $1/\tau_m^{(0)}$ correspond to around $10^{-3}$ fraction of total de-excitation rate under room temperature, $\tau_m^{(0)}$ chemical trends need not to be same as for total luminescence life-times $\tau_m$.

Based on literature search [14], [166]-[168], we would probe as non-radiative de-excitation responsible chemical reaction the hydrogen abstraction by excited-state uranyl(VI) central group ($U^{VI}O_2^{2+}* + H\text{-}OH \rightarrow OU^{V}OH + *O^{-I}H \rightarrow U^{VI}O_2^{2+} + H_2O$).

## 5. *Diffuse atomic basis functions added to def-TZVPP on excitation energies*

As TD-DFT is known to be sensitive to diffuse atomic basis set function [29], [30], [169], their subsequent addition to def-TZVPP used in all calculations has been tried for TT3 settings optimized snapshot n. 130, spin-orbit resolved quasi-relativistic TD-HF and TD-DFT/XALDA computations (Tab. TSI5.1). In following table, $b = 0$ stands for def-TZVPP [5], [147], [148], $b = 1$ for def-TZVPPD [170] and $b = 2$ for def-TZVPPD [170] with single gaussian basis function with exponent 0.005 added for each angular moment (s to g) to uranium def-TZVPPD set.

| basis *b* | method | $T_{00}$(b) | D $T_{00}$(b) |
|---|---|---|---|
| 0 | HF | 20711.7311 | |
| | CAM | 20192.8630 | |
| | LBα | 20662.4476 | |
| 1 | HF | 20688.9207 | -22.8104 |
| | CAM | 20181.6440 | -11.2190 |
| | LBα | 20658.1395 | -4.3082 |
| 2 | HF | 20688.8976 | -0.0231 |
| | CAM | 20181.6299 | -0.0141 |
| | LBα | 20658.1458 | 0.0064 |

**Table TSI5.1:** Diffuse atomic basis set influence on $T_{00}$. All values in $cm^{-1}$. D $T_{00}$(b) = $T_{00}$(b) – $T_{00}$(b-1)

In conclusion, the influence is rather small in case of the aquo complex. D $T_{00}(1)$ only in the TD-HF case exceeds 20 $cm^{-1}$ of typical experimental spectral resolution for uranyl(VI) luminescence spectra.

## 6. Solvent model influence

Influence of water solvent oxygen atom partial charge change from the TIP3P value of -0.834 to -0.8476 of SPC/E model[37] (for which CMD computation have been done) have been studied for original TT1 settings "local equilibrium" structures. Adiabatic excitation energy in each snapshot have been shifted by (2.8 ± 0.8) $cm^{-1}$ higher (for both LBα/B3LYP and CAM-B3LYP tested here). Subsequently, "local equilibria" structures have been reoptimized with < 0.01 pm change of U-$O_{yl}$ uranyl bond length (and similarly negligible change of other geometry parameters) symmetric stretching mode vibrational frequency has been increased by 1 $cm^{-1}$ in average, but any spectral shape difference completely negligible with respect to any variance in snapshot collection choice. Naturally, more significant change of partial charges or solvent model in general would have significant effect and is of importance for further studies. Possible extension would be modelling solutions with non-zero ionic strength for better comparison with experimental data.

## 7. Master formula derivation

Starting from expression for spontaneous emission Einstein coefficient [125], which has been derived here for the case of dipole aproximation

$$A_{fi} \quad = \quad \frac{\omega_{fi}^3}{3\pi\varepsilon_0 \hbar c^3}\left|\int \psi_i^* \hat{\vec{\mu}} \psi_k dV\right|^2 , \tag{SI7.1}$$

where $\hat{\vec{\mu}}$ is the total electric dipole moment operator. We can evaluate the integral above (over all internal degrees of freedom of molecule in interest) considering Born-Oppenheimer approximation [106] and ignoring rotational degrees of freedom (either by summing over all transitions differing by rotational parts only (the case of bare uranyl *in Vacuo* spectrum – let us note that vibrational-rotational seprabation is well justified for this system due to the long U-O bond and therefore small ro-vibrational coupling constant $\alpha$) or considering the system to be embedded in condensed phase (the case of uranyl aquo complex in water), then

$$\begin{aligned}\sum_s \langle \psi_f | Z_s \vec{r}_s | \psi_i \rangle \quad &= \quad \vec{\varepsilon}^{\,*}_{\lambda,\vec{k}} \cdot \int_R \vec{\mu}_{e(f),e(i)}(\vec{R}) \cdot \overline{\chi}_f^{(rot,vib)} \cdot \chi_i^{(rot,vib)} \, d^{3N}\vec{R} \\ &+ \quad \delta_{e(f),e(i)} \cdot \left\langle \chi_{v,r(f);e(f)}^{(rot,vib)} \middle| \hat{T}_E^R \middle| \chi_{v,r(f);e(f)}^{(rot,vib)} \right\rangle\end{aligned} , \tag{SI7.2}$$

[37] The hydrogen atom partial charge have been modified accordingly, to keep water molecules electroneutral in any model used. "Frozen" (i.e., the $N_2$ and $N_3$ solvation shells) water molecule geometries have been close to SPC/E model in both cases.

where electronic transition moment can be expanded into series with respect to the normal mode coordinates (in initial electronic state equilibrium),

$$\vec{\mu}_{(e(f),e(i))}(\vec{R}) \;=\; \vec{\mu}^{0}_{(e(f),e(i))} \;+\; \left(\frac{\partial \vec{\mu}_{(e(f),e(i))}}{\partial Q_j}\right)^{0} Q_j \;+\; \frac{1}{2}\left(\frac{\partial^2 \vec{\mu}_{(e(f),e(i))}}{\partial Q_j \partial Q_k}\right)^{0} Q_j Q_k + \cdots, \qquad \text{(SI7.3)}$$

The upper index „0" denote evaluation for initial electronic state equilibrium positions of nuclei. The zero-th order term in (SI7.3) corresponds for Condon approximation, first order term to Herzberg-Condon. We have limited ourselves to the Condon aproximation (according to Papoušek and Aliev [171], each successive term contribution to rovibronic transition probabilities in (SI7.15), in general case, is smaller roughly by factor of 0.1 with respect to his lower-order predecessor). The spectrum is in Condon approximation proportional to the sum of all transition probabilities with the following form

$$P \;=\; K \cdot \left|\tilde{\vec{\mu}}^{0,body}_{e(f),e(i)}\right|^2 \left|\int_R \overline{\chi}^{(vib)}_f \cdot \chi^{(vib)}_i \, d^M \vec{R}\right|^2, \qquad \text{(SI7.4)}$$

where the integral in the last term is Franck-Condon factor [115]-[119], the proportionality constant $K$ is proportional to the cube of frequency therefore, experimental luminescence spectra $I(\omega)$ has been divided by dimension-less factor $(\omega/\omega_0)^3$ (where $\omega_0$ constant is set to 20 000 $cm^{-1}$) and eventually renormalized (by division by maximum value in resulting spectrum) before comparison to Franck-Condon profile corresponding to the Master formula (1).

Multiplying each Franck-Condon factor (FCF) by a well chosen peak profile and summing over all possible vibronic transitions leads to the Master formula as presented in the main article (1).

For harmonic approximation applied to vibrational motions in both electronic states the analytical formulae for FCF are known [120]-[123] for arbitrary number of normal modes (including Duschinsky rotations [77]-[81] between the normal mode sets corresponding to the two different electronic states). Let us remind works providing also analytical formula for the simplest, single mode case [97], [124],

$$\left|\int_R \overline{\chi}^{(vib)}_{0,\omega} \cdot \chi^{(vib)}_{n,\omega'} \, d^1\vec{R}\right|^2 \;=\; \frac{1}{2^n n!}\left(\sum_{k=0}^{\lfloor n/2 \rfloor} \binom{n}{2k}\left(\frac{4\omega}{\omega+\omega'}\right)^k (2k-1)!!\, H_{n-2k}\left(-\Omega\,\omega^{-1/2} d\right)\right), \qquad \text{( SI7.5)}$$

where

$$\Omega \;=\; \frac{\omega+\omega'}{\omega'\omega}, \qquad \text{(SI7.6)}$$

And $d = C \cdot \Delta R$, where

$$C^2 \;=\; \frac{\mu m_u c}{\hbar}, \qquad \text{(SI7.7)}$$

with lower indices of vibrational wave-functions in (SI7.5) corresponding to vibrational quantum number (at least one of the states is considered to be in the ground state) and vibrational frequency, $\Delta R$ is the difference between equilibria coordinates in the two electronic states, $m_u$ and $\mu$ in (SI7.7) are atomic mass unit (one Dalton in SI) and normal mode reduced mass in Daltons, respectively.

### 7B. Convolution and spectrum

In the special case of parallel normal modes in both electronic states (either by high level of problem symmetry in small molecules or by an approximation), the Franck Condon profile sum,

$$Y(\nu) = K' \cdot \sum_{n_1=0}^{n_{i,\max}} \sum_{n_2=0}^{n_{2,\max}} \cdots \sum_{n_M=0}^{n_{M,\max}} \sum_{m_1=0}^{m_{i,\max}} \sum_{m_2=0}^{m_{2,\max}} \cdots \sum_{m_M=0}^{m_{M,\max}} FCF(\{n_j\},\{m_j\}) \cdot V_{\Gamma_L,\Gamma_G}\left(\nu + \sum_{j=1}^{M} \omega'_j n_j - \sum_{j=1}^{M} \omega_j m_j\right), \tag{SI7.8}$$

(where $n_j$ and $m_j$ are vibrational quantum numbers for the target and initial electronic states respectively, $\omega_j'$ and $\omega_j$ are their frequencies, the {0',0',..,0'}→{0,0,...,0} transition have been centered at ν = 0 cm$^{-1}$ for further simplicity, $V_\Gamma(x)$ will be considered here Voigt profile with Gaussian width $\Gamma_G$ and Lorentzian width $\Gamma_L$), can be written as a convolution of Franck-Condon profiles connected to each single-mode harmonic oscillator,

$$Y = K' \cdot Y_1 * Y_2 * \cdots * Y_M, \tag{SI7.9}$$

where

$$Y_j = FCF(n_j, m_j) \cdot V_{\Gamma_{G,j},\Gamma_{L,j}}(\nu + \omega'_j n_j - \omega_j m_j), \tag{SI7.10}$$

considering properties of Fourier transform,

$$\Gamma_G^2 = \sum_{j=1}^{M} \Gamma_{G,j}^2, \tag{SI7.11}$$

$$\Gamma_L = \sum_{j=1}^{M} \Gamma_{L,j}, \tag{SI7.12}$$

for special case of $\Gamma_{G,j} \equiv \Gamma_{G,0} = konst.$ and $\Gamma_{L,j} \equiv \Gamma_{L,0} = konst.$, the final peak widths increase with $M^{1/2}$ and $M$ for Gaussian and Lorentz parts separately. The transformation of (SI7.9) to a simple product by Fourier transforms allows fitting of experimenal vibrationally resolved electron spectra to multimode models.

In the general case, normal modes are rotated with respect to each other (Duschinsky rotations) [77]-[81].